\documentclass[fleqn,usenatbib]{mnras}

\usepackage{newtxtext,newtxmath}
\usepackage[T1]{fontenc}
\usepackage{graphicx}
\usepackage{amsmath}
\usepackage{siunitx}
\usepackage{booktabs}
\usepackage{microtype}

\usepackage{orcidlink}  
\usepackage{lastpage}
\hypersetup{
    colorlinks=true,
    linkcolor=blue,
    citecolor=blue,
    urlcolor=blue
}

\title[ULIRGs and radio galaxies]{Finding the evolutionary link between ULIRGs and radio galaxies}

\author[Fida et al.]{Ayisha Fida$^{1}$\orcidlink{0009-0001-5850-2113},
Sumana Nandi$^{2}$\orcidlink{0000-0002-8142-4619}\thanks{E-mail: \href{mailto:sumana1981@gmail.com}{sumana1981@gmail.com}},
Mousumi Das$^{3}$\orcidlink{0000-0001-8996-6474},
K. Rubinur$^{4}$\orcidlink{0000-0001-5574-5104},
S. Shanbhog$^{5}$\orcidlink{0009-0004-0495-3877},
Preeti Kharb$^{6}$\orcidlink{0000-0003-3203-1613},
\newauthor
and Barenya Dev$^{7}$\orcidlink{0009-0006-9368-852X}\\
$^{1}$Manipal Centre for Natural Sciences, Centre of Excellence, Manipal Academy of Higher Education, Manipal, Karnataka 576104, India\\
$^{2}$Institute of Astronomy Space and Earth Sciences, P 177, CIT Road, Scheme 7M, Kolkata 700054, West Bengal, India\\
$^{3}$Indian Institute of Astrophysics, Koramangala II Block, Bangalore 560034, India\\
$^{4}$Institute of Theoretical Astrophysics, University of Oslo, P.O. Box 1029 Blindern, 0315 Oslo, Norway\\
$^{5}$Scuola Normale Superiore di Pisa, Piazza dei Cavalieri 7, 56126 Pisa, Italy\\
$^{6}$National Centre for Radio Astrophysics (NCRA) – Tata Institute of Fundamental Research (TIFR),\\
S. P. Pune University Campus, Ganeshkhind, Pune 411007, India\\
$^{7}$Department of Astronomy, Astrophysics and Space Engineering, Indian Institute of Technology Indore,\\
Khandwa Road, Simrol, Indore 453552, Madhya Pradesh, India
}

\date{Accepted XXX. Received YYY; in original form ZZZ}
\pubyear{\the\year}
\begin{document}
\label{firstpage}
\pagerange{\pageref{firstpage}--\pageref{LastPage}}
\maketitle

\footnotetext{E-mail: \href{mailto:sumana1981@gmail.com}{sumana1981@gmail.com}}

\begin{abstract}
Ultraluminous Infrared Galaxies (ULIRGs) are the remnants of the major mergers of gas rich galaxies. They are dusty systems and are extremely luminous in the infrared. Both starbursts and AGN activity supply power in these systems, but it is not clear how these two activities are related to the overall evolution of the merger remnant. Studies suggest that there may be an evolutionary link between ULIRGs and radio galaxies. Radio observations of ULIRGs are a useful tool for understanding this evolutionary path as well as studying the energy contributions from starbursts and AGN. We perform a systematic census of these signatures in 38 ULIRGs using uGMRT observations and archival radio data. 
In the sample, 30 sources have enough frequency coverage to obtain their radio spectrum. Of these, 17 sources show a turnover frequency between 63 and 509 MHz and a spectral age variation from 0.009 to 44 Myr. Also, 8 sources have extended structures with projected linear sizes varying from 0.7 to 11 kpc. Our results indicate that the primary energy source in these systems is the AGN rather than star formation. The majority of the ULIRGs in our sample, particularly those in the post-merger stage, exhibit Gigahertz Peaked Spectrum (GPS) or Compact Steep Spectrum (CSS)-like characteristics, suggesting that they are evolving into young radio galaxies. We note that 6 sources with high star formation activity show a low-frequency turnover and a smaller spectral age; we confirm these 6 sources host buried AGN. This study enables us to trace the evolutionary link between ULIRGs and radio galaxies.
\end{abstract}





\section{Introduction}

\begin{table*}
\centering

\caption{ULIRGs in the Sample and Their Optical Nuclear Types.}
\label{Table1}
\vspace{1ex}
\begin{tabular}{ccccc}
\hline
Galaxy & RA & Dec & z & Nuclear type \\
(1)&(2)&(3)&(4)&(5)\\
\hline
IRAS 00188$-$0856$\blacklozenge$ & 00h21m26.5s & $-08^{\circ}39^{\prime}26^{\prime\prime}$ & 0.128 & LINER \\

IRAS 00456$-$2904 & 
00h48m06.8s & 
$-28^{\circ}48^{\prime}19^{\prime\prime}$ & 
0.109 & 
H\,\textsc{ii} \\

IRAS F01004$-$2237$\dagger$ & 
01h02m50.0s& 
$-22^{\circ}21^{\prime}57^{\prime\prime}$ & 
0.118 & 
H\,\textsc{ii} \\

IRAS 01494$-$1845$\dagger$ & 
01h51m51.4s & 
$-18^{\circ}30^{\prime}46^{\prime\prime}$ & 
0.158 & 
H\,\textsc{ii} \\

IRAS 01569$-$2939 & 
01h59m13.8s & 
$-29^{\circ}24^{\prime}35^{\prime\prime}$ & 
0.139 & 
H\,\textsc{ii} \\

IRAS 01572$+$0009$\blacklozenge$ & 
01h59m50.2s & 
$+00^{\circ}23^{\prime}41^{\prime\prime}$ & 
0.163 & 
Seyfert~1 \\

       IRAS 02480$-$3745$\dagger$ & 
02h50m01.7s & 
$-37^{\circ}32^{\prime}45^{\prime\prime}$ & 
0.165 & 
Unclassified \\

IRAS 04103$-$2838$\dagger$ & 
04h12m19.4s & 
$-28^{\circ}30^{\prime}25^{\prime\prime}$ & 
0.118 & 
LINER \\

IRAS 05156$-$3024 & 
05h17m32.5s & 
$-30^{\circ}21^{\prime}13^{\prime\prime}$ & 
0.171 & 
Seyfert~2 \\

IRAS 05189$-$2524 & 
05h21m01.4s & 
$-25^{\circ}21^{\prime}45^{\prime\prime}$ & 
0.044 & 
Seyfert~2 \\

IRAS F07599+6508 & 
08h04m33.1s & 
$+64^{\circ}59^{\prime}49^{\prime\prime}$ & 
0.148 & 
Seyfert~1 \\

IRAS 08559+1053 & 
08h58m41.7s & 
$+10^{\circ}41^{\prime}22^{\prime\prime}$ & 
0.148 & 
Seyfert~2 \\

IRAS 10494+4424 & 
10h52m23.5s & 
$+44^{\circ}08^{\prime}48^{\prime\prime}$ & 
0.092 & 
LINER \\

IRAS 11095$-$0238 & 
11h12m03.4s & 
$-02^{\circ}54^{\prime}22^{\prime\prime}$ & 
0.106 & 
LINER \\

IRAS 11119+3257 & 
11h14m38.9s & 
$+32^{\circ}41^{\prime}33^{\prime\prime}$ & 
0.187 & 
Seyfert~1 \\

IRAS 12032+1707 & 
12h05m47.7s & 
$+16^{\circ}51^{\prime}08^{\prime\prime}$ & 
0.217 & 
LINER \\

IRAS 12112+0305$\blacklozenge$ & 
12h13m46.0s & 
$+02^{\circ}48^{\prime}38^{\prime\prime}$ & 
0.073 & 
LINER \\

IRAS 12127$-$1412 & 
12h15m18.9s & 
$-14^{\circ}29^{\prime}45^{\prime\prime}$ & 
0.133 & 
LINER \\

IRAS 12265+0219 & 
12h29m06.7s & 
$+02^{\circ}03^{\prime}09^{\prime\prime}$ & 
0.158 & 
Seyfert~1 \\

    IRAS 13305$-$1739$\blacklozenge$ & 
13h33m16.5s & 
$-17^{\circ}55^{\prime}11^{\prime\prime}$ & 
0.148 & 
Seyfert~2 \\

IRAS 13335$-$2612 & 
13h36m22.3s & 
$-26^{\circ}27^{\prime}34^{\prime\prime}$ & 
0.125 & 
LINER \\

IRAS 13428+5608 & 
13h44m42.1s & 
$+55^{\circ}53^{\prime}13^{\prime\prime}$ & 
0.037 & 
Seyfert~2 \\

IRAS 13443+0802 & 
13h46m51.1s & 
$+07^{\circ}47^{\prime}19^{\prime\prime}$ & 
0.135 & 
Seyfert~2 \\

IRAS 13451+1232 & 
13h47m33.3s & 
$+12^{\circ}17^{\prime}24^{\prime\prime}$ & 
0.122 & 
Seyfert~2 \\

       IRAS 13539+2920 & 
13h56m09.9s & 
$+29^{\circ}05^{\prime}35.2^{\prime\prime}$ & 
0.109 & 
H\,\textsc{ii} \\

IRAS 14060+2919 & 
14h08m19.0s & 
$+29^{\circ}04^{\prime}47^{\prime\prime}$ & 
0.117 & 
H\,\textsc{ii} \\

IRAS 14070+0525$\blacklozenge$ & 
14h09m31.2s & 
$+05^{\circ}11^{\prime}31^{\prime\prime}$ & 
0.264 & 
Seyfert~2 \\

IRAS 14394+5332$\blacklozenge$ & 
14h41m04.4s & 
$+53^{\circ}20^{\prime}09^{\prime\prime}$ & 
0.105 & 
Seyfert~2 \\

IRAS 15001+1433$\blacklozenge$ & 
15h02m31.9s & 
$+14^{\circ}21^{\prime}35^{\prime\prime}$ & 
0.162 & 
Seyfert~2 \\

IRAS 15327+2340 & 
15h34m57.2s & 
$+23^{\circ}30^{\prime}11^{\prime\prime}$ & 
0.018 & 
LINER \\

IRAS 16155+0146$\blacklozenge$ & 
16h18m09.4s & 
$+01^{\circ}39^{\prime}21^{\prime\prime}$ & 
0.132 & 
Seyfert~2 \\

IRAS 17028+5817$\blacklozenge$ & 
17h03m41.9s & 
$+58^{\circ}13^{\prime}45^{\prime\prime}$ & 
0.106 & 
LINER \\

       IRAS 17044+6720$\blacklozenge$ & 
17h04m28.4s & 
$+67^{\circ}16^{\prime}23^{\prime\prime}$ & 
0.136 & 
LINER \\

IRAS 17179+5444$\blacklozenge$ & 
17h18m54.4s& 
$+54^{\circ}41^{\prime}48^{\prime\prime}$ & 
0.147 & 
Seyfert~2 \\

IRAS 23060+0505$\blacklozenge$ & 
23h08m33.9s & 
$+05^{\circ}21^{\prime}30^{\prime\prime}$ & 
0.172 & 
Seyfert~2 \\

IRAS 23233+2817 & 
23h25m49.4s & 
$+28^{\circ}34^{\prime}21^{\prime\prime}$ & 
0.113 & 
Seyfert~2 \\

IRAS F23234+0946 & 
23h25m56.2s & 
$+10^{\circ}02^{\prime}49^{\prime\prime}$ & 
0.128 & 
LINER \\

IRAS 23389+0300$\blacklozenge$ & 
23h41m30.3s& 
$+03^{\circ}17^{\prime}26^{\prime\prime}$ & 
0.145 & 
Seyfert~2 \\

    \hline
\end{tabular}

\vspace{1ex}

\smallskip
\text The table presents properties of the ULIRG sample. The columns are as follows: (1) Source name: sources marked with ($\blacklozenge$): 13 sources were extracted from \citet{2021MNRAS.503.5746N}; sources marked with ($\dagger$) have their spectral indices estimated using two different methods: three sources were fitted with a power-law model using \texttt{gnuplot}, while one source had its two-point spectral index calculated; (2) and (3) Right Ascension and Declination were obtained from NASA/IPAC Extragalactic Database (NED); (4) Redshift ($z$); (5) Optical nuclear type classification based on \citet{Veilleux_2002}.

\end{table*}

Ultraluminous infrared galaxies (ULIRGs) are gas rich merger remnants that have high infrared luminosities of $L_\mathrm{IR}>10^{12}L_{\odot}$. They are an important population to understand the galaxy formation and evolution processes. They are distributed in both the local and high redshift universe  \citep[$z\gtrsim1-2$;][]{2006asup.book..285L, 2007ApJ...667L..17S}. The abundance of dust at sub kpc scales in ULIRGs has made it difficult to study the main power production mechanism by direct optical observations. However, X-ray, near-infrared (IR), and radio observations which can penetrate the dust, have demonstrated that a significant number of ULIRGs are powered by both starburst and AGN activity. It has also been shown that a large fraction of ULIRGs show signatures of interaction in their optical images, which confirms that they are the result of the merger of two gas-rich spiral galaxies \citep{2009ApJS..182..628V}. Initially, these merger events affect the star formation rates \citep{CalderonCastillo2024}. However, as the merger evolves, star formation is quenched and the ULIRGs transform into either un-obscured quasars or elliptical galaxies \citep{2023ApJS..265...37Y}. Therefore, the study of local universe ULIRGs can connect galaxy mergers, triggered star formation, nuclear activity and the formation of quasars or elliptical galaxies \citep{2022Univ....8..392U}. 

\begin{figure*}
\centering
\includegraphics[width=15cm,height = 8.7cm]{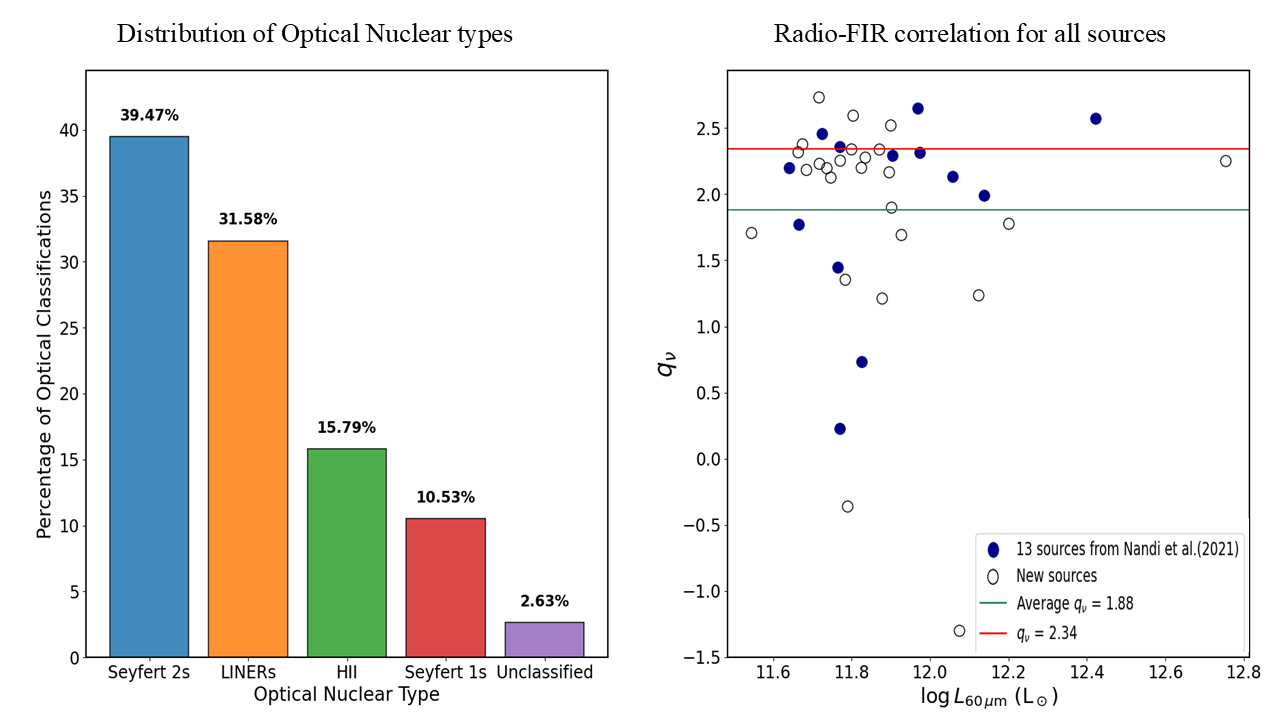}

\caption{The left panel displays a percentage bar plot illustrating the optical nuclear classification of the complete sample of 38 ULIRGs. The right panel shows a scatter plot of $q_{\nu}$ versus $\log L_{60\,\mu\mathrm{m}}$ for the 38 ULIRGs. Filled points indicate the 13 sources from the first phase of this study, taken from \citet{2021MNRAS.503.5746N}. The green line denotes the average \( q_\nu \) value of 1.88 for the 38 sources in our sample, while the red line marks the \( q_\nu = 2.34 \) threshold from \citet{2001ApJ...554..803Y}. Sources below this threshold exhibit characteristics of radio excess, whereas those above the threshold indicate infrared excess.}

\label{FIGURE1}
\end{figure*}

The radio emission in ULIRGs is mainly due to relativistic synchrotron radiation and free-free emission. Radio observations are thus important for tracing star-forming activity as well as AGN activity in ULIRGs. The well-known correlation between radio and far-IR (FIR) luminosities can be explained by star formation and starburst activity \citep{2001ApJ...554..803Y}, whereas the integrated synchrotron spectrum, detection of turnover, spectral age, identification of peak radio intensity, and brightness temperature estimation can verify the presence of AGN in ULIRGs. Several excellent works have described properties of star formation versus AGN power \citep{2009ApJS..182..628V, 2022Univ....8..392U}, interactions and mergers \citep{Veilleux_2002}, molecular outflows \citep{2018ApJ...863..143C}, dust content in ULIRGs \citep{2011ApJ...734...87R}. In some cases the AGNs are hidden behind thick dust columns, and it is extremely difficult to identify these buried AGNs. So far, it is not clear whether these buried AGNs are luminous enough to detect beneath the heavy starburst activity or whether they are highly obscured by dust. 
Mid-infrared (MIR) spectroscopy and Sub-millimeter molecular line observations are useful techniques to differentiate AGN activity from the star-forming regions in nearby ULIRGs \citep{Imanishi2008, Papadopoulos2010}. Radio observations can also play a major role in identifying ULIRGs that host deeply buried AGNs.  

Previous studies suggest that ULIRGs are the low redshift counterpart of submillimeter galaxies and may finally evolve into radio galaxies \citep{2021MNRAS.503.5746N, 2012MNRAS.422.1453N}. Powerful radio galaxies often host peculiar optical interacting systems with tidal tails,  bridges of emission, shells and dust-lanes \citep{RamosAlmeida2011}. These studies suggest that ULIRGs are the strongest candidates that can connect the evolutionary sequence of mergers and radio galaxies  \citep{2003A&A...399..511M, 1999ApJ...521L.107E, 1996ARA&A..34..749S}. The link between radio galaxy and ULIRGs has been identified in few interacting systems \citep{2004A&A...424..119M, Kukreti2022}.  However the evolution process is still not well known. 



Recently, we investigated the evolutionary sequence for the early stages of AGN development in 13 ULIRGs, with a particular focus on GHz Peaked Spectrum (GPS) sources, Compact Steep Spectrum (CSS) sources or young radio sources \citep{2021MNRAS.503.5746N}. These classes are widely interpreted as representing early evolutionary phases of radio galaxies, with GPS sources evolving into CSS sources as the radio jets expand within the host galaxy interstellar medium.  The typical projected linear size of a GPS source is $\leq$ 1 kpc  and it shows a radio spectrum turnover frequency of $\sim$ 1 GHz. Observations show that the range of the peak frequency of GPS source can vary from 400 MHz to 6 GHz \citep{1998AAS131303S}. The peak shifts towards low frequency, as the angular size of the source increases \citep{OdeaSaikia2021}. A CSS  source is $\sim$ 1 to 10 kpc in size and its radio spectrum turns over at a few hundred MHz. Generally, CSS sources peak at frequencies below 400 MHz and exhibit steep radio spectra at high frequencies \citep{OdeaSaikia2021}. Both classes have  symmetric double structure which can be resolved through high-frequency observations. We studied the distribution of the radio flux density of ULIRGs over a wide range of frequencies. We estimated the high-frequency break and  low frequency turnover from their integrated radio spectra. The turnovers at low-frequencies are due to synchrotron self-absorption or free–free absorption \citep{2010MNRAS.405..887C, hayashi2021radio}. The high-frequency steepening is interpreted as particle energy losses. 
In our previous work \citep{2021MNRAS.503.5746N}, the integrated radio spectra of 9 out of 13 ULIRGs derived using archival data and Giant Metrewave Radio Telescope (GMRT) observations, show characteristics that are similar to that of GPS/CSS/young radio sources. The variation of spectral age of these sources is in between 0.4 Myr to 20 Myr. Although this result supports the hypothesis that the ULIRGs are young radio sources, we need a larger population to understand the connection between ULIRGs and radio galaxies which is not statistically clear yet. Thus, studying ULIRGs at radio wavelengths is a way to determine  whether these system are progenitors of young radio loud elliptical galaxies. 

In this paper we investigate a large sample of ULIRGs with multifrequency radio data using new observations from the upgraded GMRT (uGMRT) as well as other ancillary data from different surveys. The paper is arranged as follows. We describe the sample selection process in the Section \ref{sec:Sample_Selection}. The main methods used to study these sources include modeling the radio spectra and estimating the radio synchrotron age to determine the source's age and characteristics. New observations, data reduction and archival data availability have been described in Section \ref{radiodataanalysis}. In Section \ref{sec:result} we describe the results of uGMRT observations, spectra modeling, radio-to-FIR correlation analysis for all sources, the Wide-field Infrared Survey Explorer (WISE) color-cut criteria and optical merger classification. In Section \ref{SECTION5} we present the discussion of our findings and in Section \ref{sec:conclusion} we present our final conclusions. We provide a detailed interaction stage classification in Appendix \ref{appendix:interaction}, highlight the ten most interesting sources in Appendix \ref{appendix:interestingsources}, and Panoramic Survey Telescope and Rapid Response System (Pan-STARRS) images of all sources in Appendix \ref{appendix:PANfigures}.
 \\ 

\label{sec:int}

\section{SAMPLE SELECTION}

We selected our sample from the IRAS 1 Jy Survey of ULIRGs \citep{Kim1998}, which consists of a large sample of 118 sources. We have selected 38 ULIRGs (see Table \ref{Table1}) based on their detection in both National Radio Astronomy Observatory (NRAO) Very Large Array (VLA) Sky Survey (NVSS) and the Tata Institute of Fundamental Research (TIFR) Giant Metrewave Radio Telescope (GMRT) Sky Survey (TGSS). In the first phase of our work, \citet{2021MNRAS.503.5746N}, we examined 13 of these 38 ULIRGs. In this paper we study the radio spectra of the remaining 25 new sources as well as examine their AGN activity, host galaxy evolution for the sample of 38 sources. The entire sample, along with their right ascension, declination, redshift, and optical nuclear type information, is summarized in Table \ref{Table1}. The sources marked with ($\blacklozenge$) represent the 13 ULIRGs analyzed in the first phase of this study by \citet{2021MNRAS.503.5746N}. Among our sample, 39.47\% exhibit Seyfert 2 characteristics, 10.53\% show Seyfert 1 characteristics, 31.58\% are classified as LINERs, and 15.79\% display H II region characteristics (see Figure \ref{FIGURE1}, left panel), which presents these optical nuclear classifications in a bar plot showing their percentage distribution. We detail each method and the morphological, physical, and emission characteristics of all sources in our sample in the following sections.

\label{sec:Sample_Selection}

\section{RADIO DATA AND ANALYSIS}

This section focuses on a new sample of 25 sources, selected from the total set of 38 sources studied in the second phase of our research. The first phase, which included 13 sources, was previously discussed in \citet{2021MNRAS.503.5746N}. In this second phase, we conducted uGMRT observations and utilized archival data to study the radio spectra of these sources. The details of the uGMRT observations , data reduction and the availability of archival data are provided below.

\label{radiodataanalysis}

\subsection{uGMRT observations }

\begin{table*}
\centering
\caption{uGMRT Band 3 and Band 4 Observational Details of Selected ULIRGs in the Sample.}
\label{band3and4table}

\vspace{1ex}

\begin{tabular}{c|c|c|c|c|c|c|c}
\multicolumn{8}{c}{\textbf{Band 3}} \\
\hline
Galaxy & Obs Date & Primary Calibrator & Phase Calibrator & Beam Size ("×") & PA ($^\circ$) & Time (min) & rms (mJy/b) \\
(1)&(2)&(3)&(4)&(5)&(6)&(7)&(8)\\
\hline
   IRAS 05156$-$3024 & 12-07-2021 & 3C147, 3C286 & 0521$-$207 & 11.23$\times$4.64 & 33.55 & 40 & 0.064 \\
   IRAS F07599$+$6508 & 12-07-2021 & 3C147, 3C286 & 0834$+$555 & 9.05$\times$4.90 & 42.54 & 40 & 0.075 \\
   IRAS 10494$+$4424 & 12-07-2021 & 3C147, 3C286 & 1021$+$219 & 9.12$\times$4.59 & 69.42 & 40 & 0.078 \\
   IRAS 11119$+$3257 & 12-07-2021 & 3C147, 3C286 & 1021$+$219 & 7.13$\times$4.62 & 61.68 & 40 & 0.105 \\
   IRAS 13428$+$5608 & 12-07-2021 & 3C147, 3C286 & 1459$+$716 & 12.89$\times$4.49 & 75.52 & 5 & 0.490 \\
   \hline

   \multicolumn{8}{c}{\textbf{Band 4}} \\
   \hline
   Galaxy & Obs Date & Primary Calibrator & Phase Calibrator & Beam Size ("×") & PA ($^\circ$) & Time (min) & rms (mJy/b) \\
   (1)&(2)&(3)&(4)&(5)&(6)&(7)&(8)\\

   \hline
   IRAS 05156$-$3024 & 10-07-2021 & 3C147, 3C286 & 0521$-$207 & 11.23$\times$4.64 & 33.55 & 37 & 0.089 \\
   IRAS F07599$+$6508 & 10-07-2021 & 3C147, 3C286 & 0834$+$555 & 9.05$\times$4.90 & 42.54 & 37 & 0.116 \\
   IRAS 10494$+$4424 & 10-07-2021 & 3C147, 3C286 & 1021$+$219 & 9.12$\times$4.59 & 69.42 & 32 & 0.142 \\
   IRAS 13428$+$5608 & 10-07-2021 & 3C147, 3C286 & 1459$+$716 & 12.89$\times$4.49 & 75.52 & 21 & 0.216 \\
  \hline
\end{tabular}

\vspace{1ex}

\smallskip
\text The table presents the observation log of ULIRGs at Band 3 and Band 4. The columns are as follows: (1) Source name; (2) Observation date; (3) Primary calibrator; (4) Phase calibrator; (5) Synthesized beam size (in arcsec); (6) Position angle (PA) in degrees; (7) On-source integration time (minutes); (8) RMS noise (mJy/beam).
\end{table*}
\sloppy

From the new sample of 25 sources, we proposed to map the low-frequency radio continuum emission of five ULIRGs: IRAS 05156-3024, IRAS F07599+6508, IRAS 10494+4424, IRAS 11119+3257, and IRAS 13428+5608, under the observation code 40\textunderscore103.

The observations were carried out using the uGMRT, which was developed to provide wider frequency coverage and substantially improved sensitivity compared to the legacy GMRT system. The upgrade introduced wideband receiver and backend systems in place of the earlier narrow-band configuration, allowing instantaneous bandwidths of up to 400~MHz and improved dynamic range. In this study, we used uGMRT Band~3 (250--500~MHz) and Band~4 (550--850~MHz). 
These improvements significantly enhance continuum sensitivity at low radio frequencies, making the uGMRT well suited for probing compact and young radio sources \citep{Gupta2017}.

To determine their integrated radio spectra, we conducted band 3 observations for all five sources. Band 4 observations were carried out for four of them, excluding IRAS 11119+3257. In addition to our uGMRT observations, we utilized various radio surveys and archival data to achieve broad frequency coverage for these five sources. For the remaining 20 sources, only archival data were used, without any new observations.

The observation dates, primary and phase calibrators, beam sizes, position angles, integration times, and rms noise values for these sources in bands 3 and 4 are provided in Table \ref{band3and4table}. We performed data reduction using available pipelines to obtain the best possible imaging quality for each source. In general, we used the Source Peeling and Atmospheric Modeling pipeline (\texttt{SPAM}; \citealt{intemma.soft08006I}), an Astronomical Image Processing System (AIPS)-based Python extension \citep{Greisen2003} which performs direction-dependent calibration and applies ionospheric corrections. We also used the Common Astronomy Software Applications package (\texttt{CASA}; \citealt{bean2022casa}). For IRAS 13428+5608 (observed in both band 3 and band 4), we adopted the CASA Pipeline-cum-Toolkit for Upgraded GMRT data Reduction (\texttt{CAPTURE}; \citealt{2021ExA....51...95K}). This is because the SPAM calibration did not recover any extended emission in this source, while CAPTURE was able to recover faint extended structure, resulting in a higher-quality image. Using flux measurements across different frequency bands, we constructed the radio spectra for our sample sources.

\subsection{Data Reduction Using SPAM}  

For the observed five sources, data reduction was performed using SPAM. The process began with the conversion of raw Long-Term Accumulation (LTA) files into UVFITS format to ensure compatibility with standard interferometric processing tools. Since SPAM cannot process wideband data in a single step, the dataset was automatically divided into multiple subbands for independent calibration.  

For our observations, five sources were processed in band 3, where the data was split into six subbands centered at 450 MHz. Four of these sources were also observed in band 4, and their data was divided into seven subbands centered at 650 MHz. This subbanding approach was necessary to improve calibration accuracy, as SPAM does not support processing large fractional bandwidths ($\Delta f / f > 0.2$) in a single run.
Each subband underwent pre-calibration, beginning with initial flagging to remove corrupted data caused by radio frequency interference (RFI), malfunctioning antennas, and instrumental errors. Residual instrumental effects were corrected, followed by direction-dependent ionospheric calibration. The final imaging was performed with (\texttt{WSClean}; \citealt{offringa2014WSCLEAN}), a multiscale, multifrequency synthesis deconvolution tool that uses the w-stacking algorithm for efficient Fourier transformation. The calibrated visibilities from all subbands were combined, converted into measurement sets with CASA, and processed to generate the final continuum image. The final wideband image, saved as an \texttt{*MFS-image.fits*} file, provided a high resolution map of the radio sources in our sample. This systematic approach ensured precise calibration and high fidelity imaging, improving the accuracy of the uGMRT data for our sources.

\subsection{Archival Radio Data}  

To determine the integrated radio spectrum of our sample, we utilized various radio surveys and archival data, ensuring broad frequency coverage for all sources. Archival data for the 25 newly selected sources were obtained from multiple surveys that covered a wide range of frequencies and resolutions. Data were obtained from the Very Large Array Low-Frequency Sky Survey Redux (VLSSr) at 73.8–74 MHz with an angular resolution of 75 arcseconds \citep{2014MNRAS.440..327L}, the Galactic and Extra-Galactic All-Sky MWA Survey (GLEAM) at 72–231 MHz with a resolution around 2 arcminutes \citep{2015PASA...32...25W}, TGSS at 150 MHz with a resolution of approximately 25 arcseconds \citep{GMRTSurvey}, the 6th Cambridge Survey of Radio Sources at 151 MHz with a resolution of 4.2×4.2 cosec(declination) arcmin$^2$ (EW×NS) \citep{1990MNRAS.246..256H}, and the Westerbork Northern Sky Survey (WENSS) at 325 MHz with a resolution of 54 arcseconds \citep{Rengelink1997}.  

Additionally, we used data from the Texas Survey of Discrete Radio Sources at 365 MHz \citep{1996AJ....111.1945D}, the Bologna Northern Cross Telescope at 408 MHz \citep{1990A&AS...82..339F}, the Rapid Australian Square Kilometre Array Pathfinder (ASKAP) Continuum Survey (RACS) at 880–887 MHz at 880–887 MHz with an angular resolution of 25 arcseconds \citep{2020PASA...37...48M}, the Faint Images of the Radio Sky at Twenty-cm (FIRST) at 1400 MHz at resolution of 5 arcseconds \citep{2015ApJ...801...26H}, NVSS at 1400 MHz with a resolution of about 45 arcseconds \citep{1998AJ....115.1693C}, and the Very Large Array Sky Survey (VLASS) at 2–4 GHz with a resolution of 2.5 arcseconds \citep{2021ApJS..255...30G}. We also included data from the Green Bank (GB6) Survey at 4850 MHz, which has a resolution of 3 arcminutes \citep{1996ApJS..103..427G}, along with archival Very Large Array (VLA) images at various frequencies and array configurations. This extensive archival data set provided broad frequency coverage, enabling a detailed analysis of the radio spectra of our sources.

\section{RESULTS}
\label{sec:result}

\subsection{The uGMRT images}

\begin{figure*}
    \centering
    \includegraphics[width=3.0in]{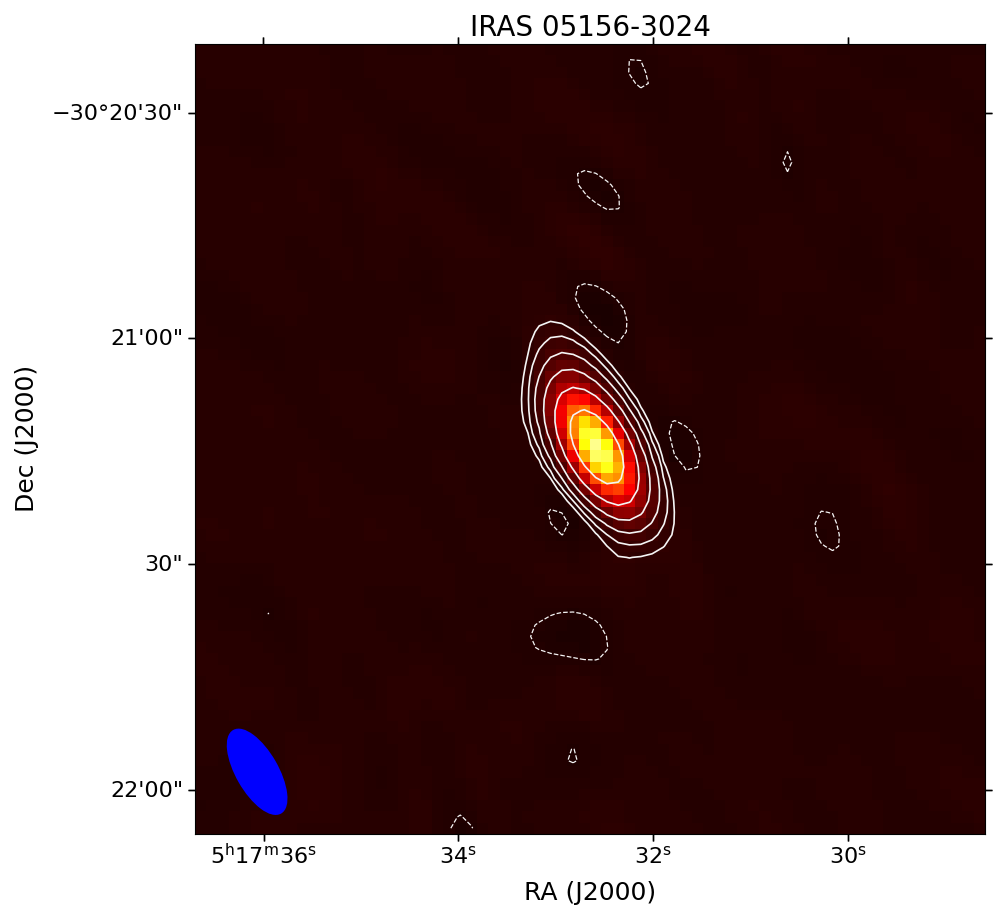} \quad
    \includegraphics[width=3.0in]{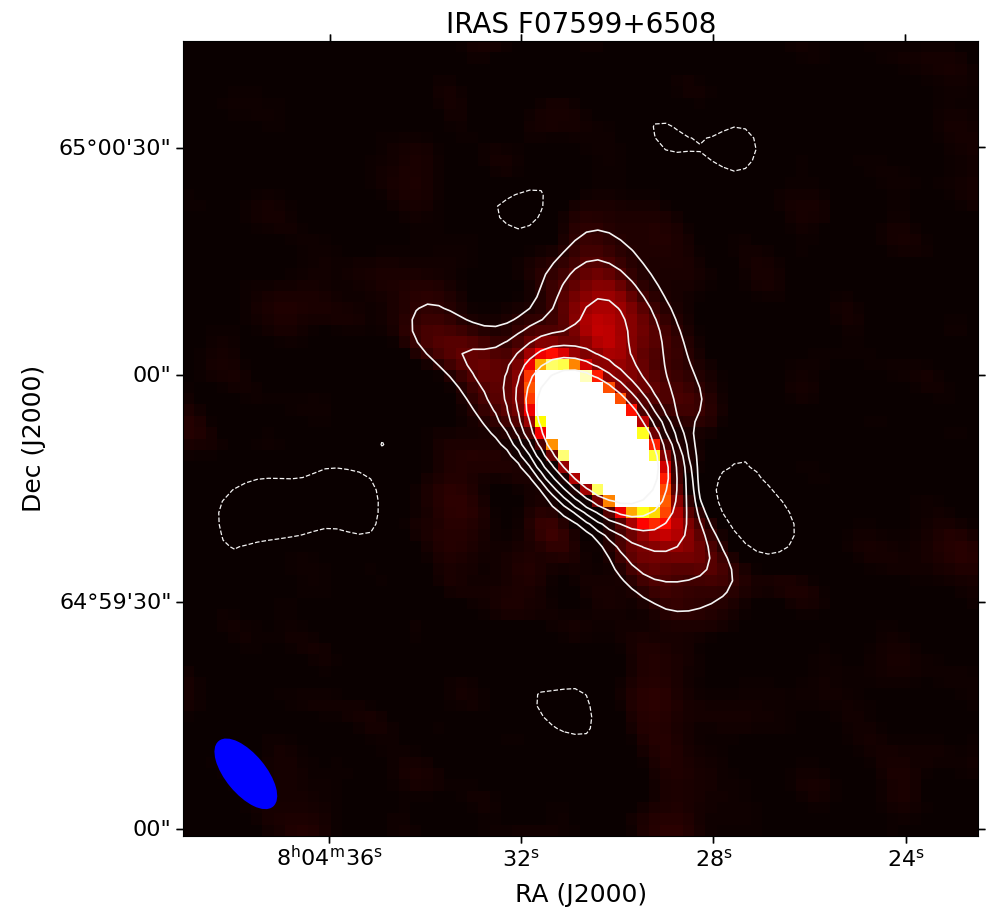} \\
    \vspace{0.1cm}
    \includegraphics[width=3.0in]{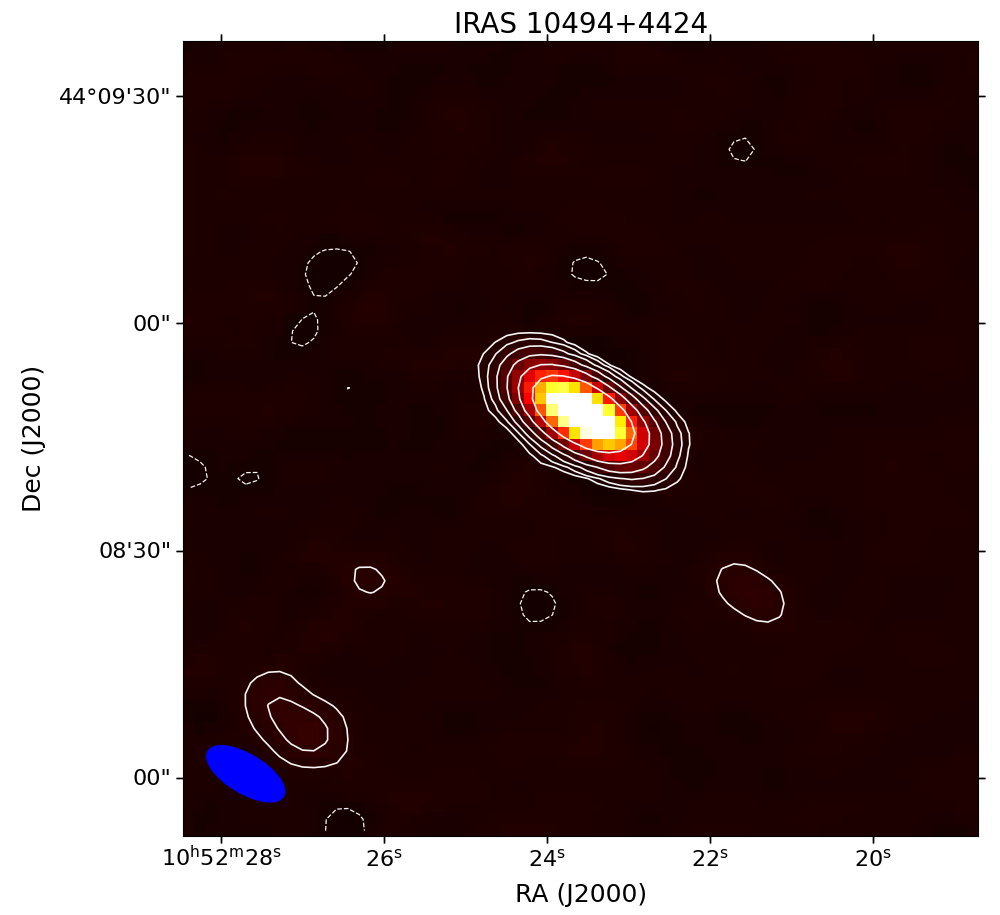} \quad
    \includegraphics[width=3.0in]{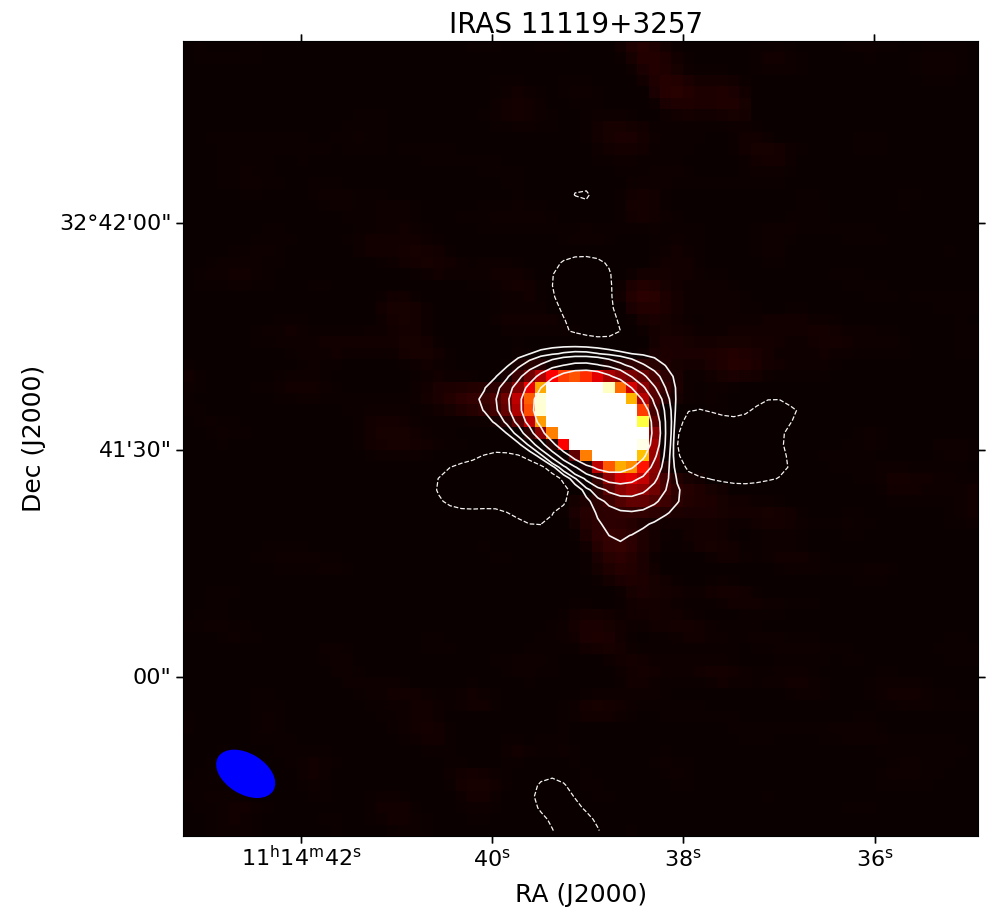} \\
    \vspace{0.1cm}
    \includegraphics[width=3.0in]{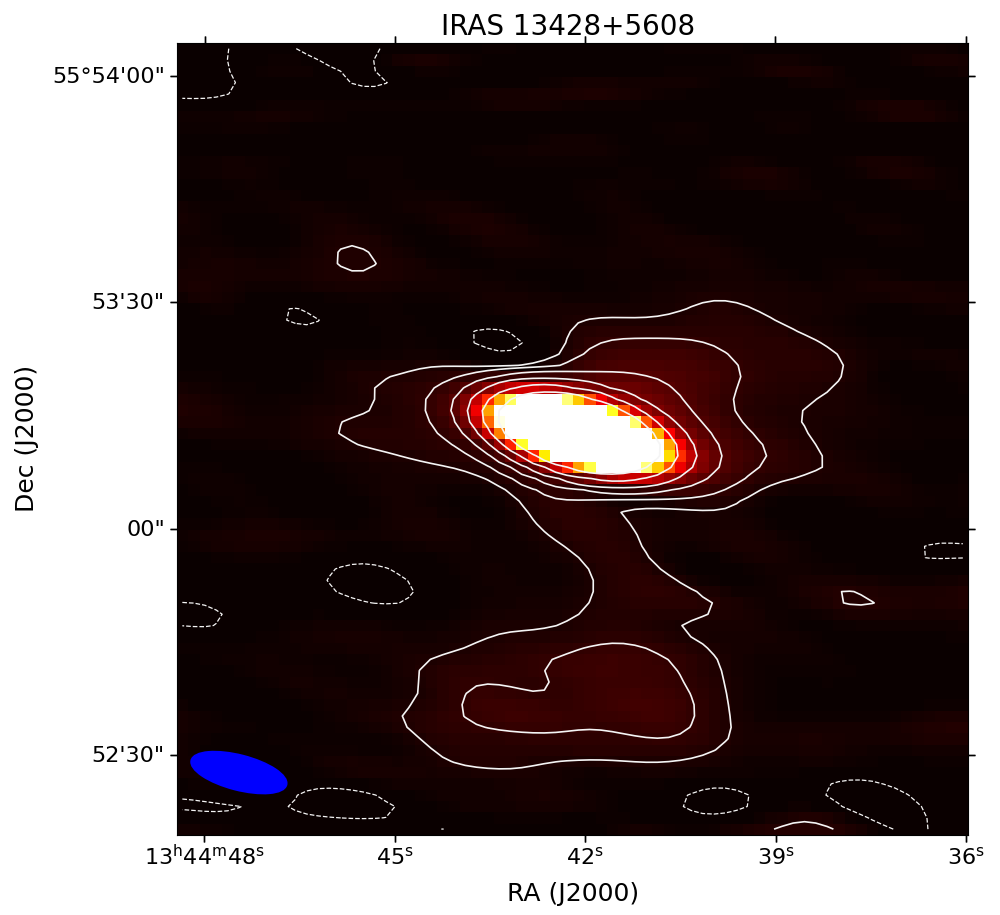} 
\caption{uGMRT Band 3 contour images of the five ULIRGs. Radio contours start at 5$\sigma$ and increase by factors of 2, where $\sigma$ represents the local rms noise level: 0.064, 0.075, 0.078, 0.105, and 0.490 mJy beam$^{-1}$ for IRAS 05156$-$3024, IRAS F07599+6508, IRAS 10494+4424, IRAS 11119+3257, and IRAS 13428+5608, respectively.}

    \label{band3}
\end{figure*}

\begin{figure*}
    \centering
    \includegraphics[width=3.0in]{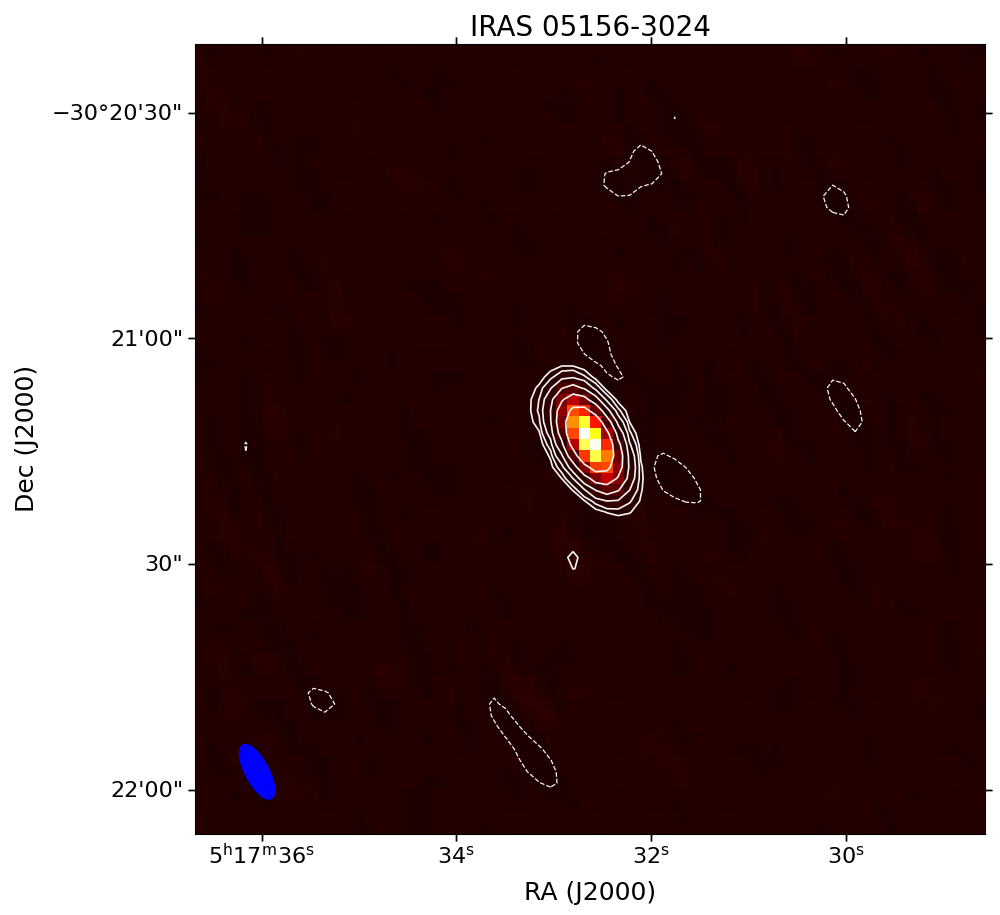} \quad
    \includegraphics[width=3.0in]{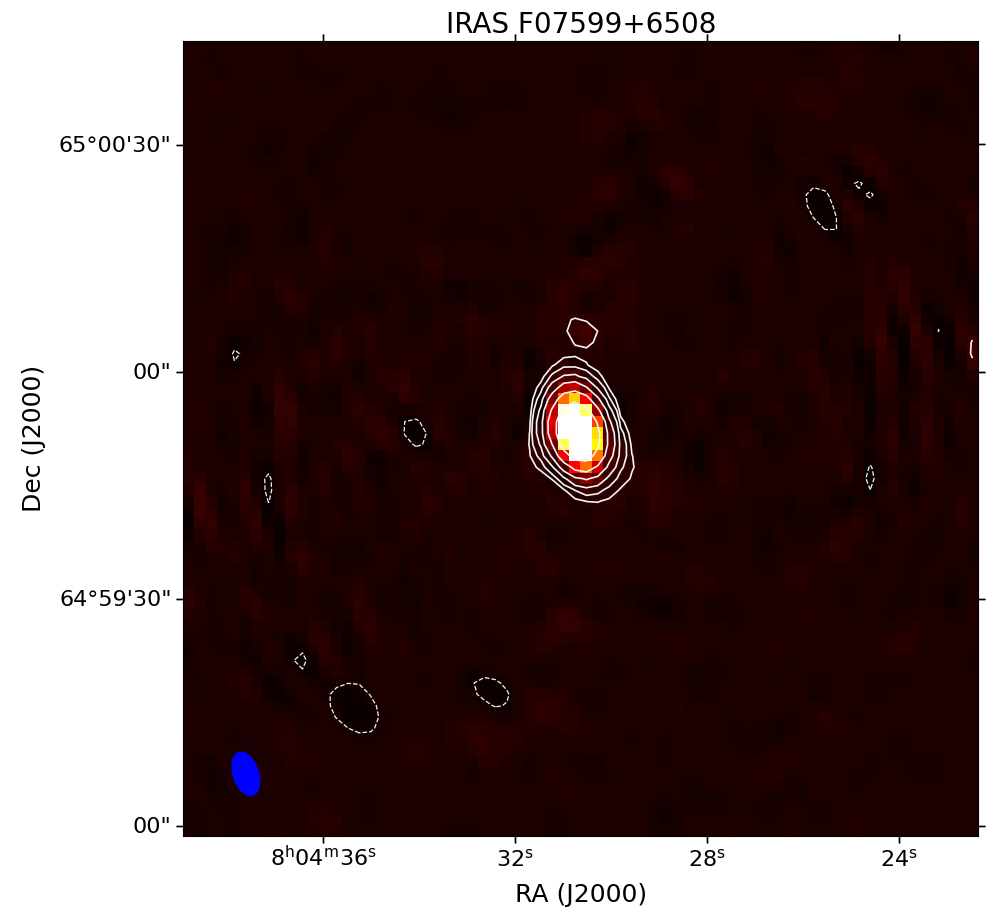} \\
    \vspace{0.1cm}
    \includegraphics[width=3.0in]{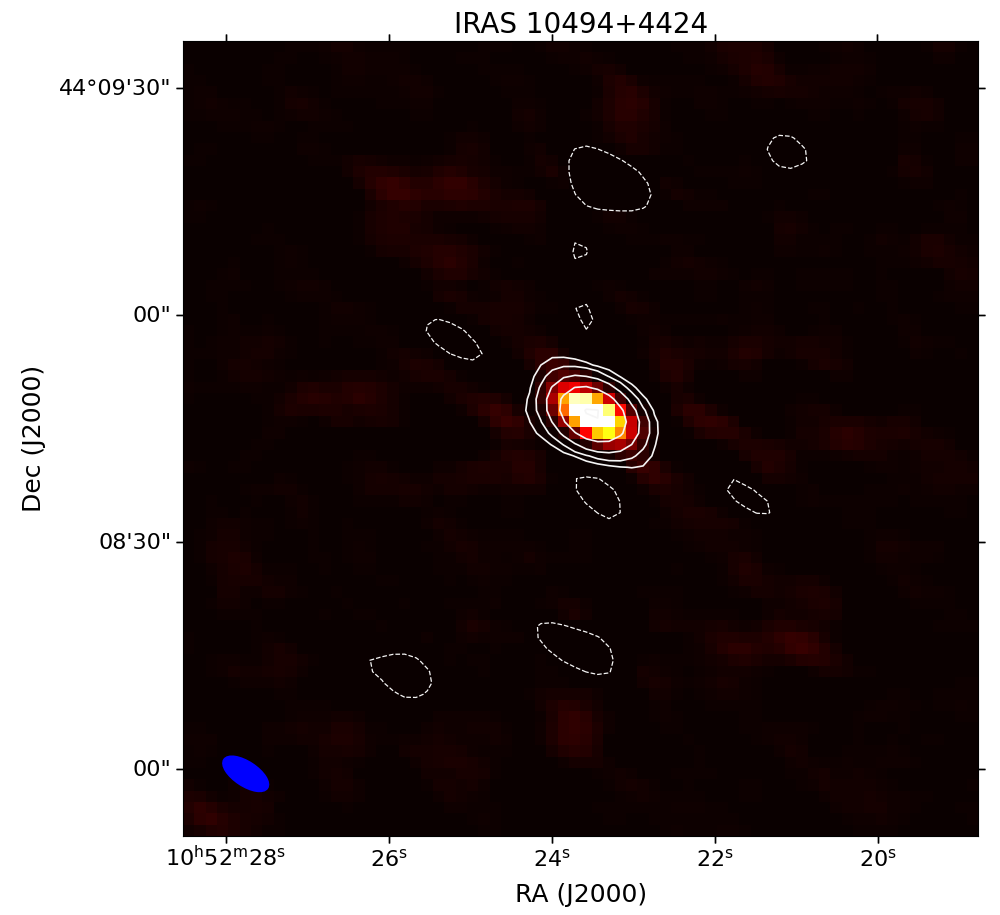} \quad
    \includegraphics[width=3.0in]{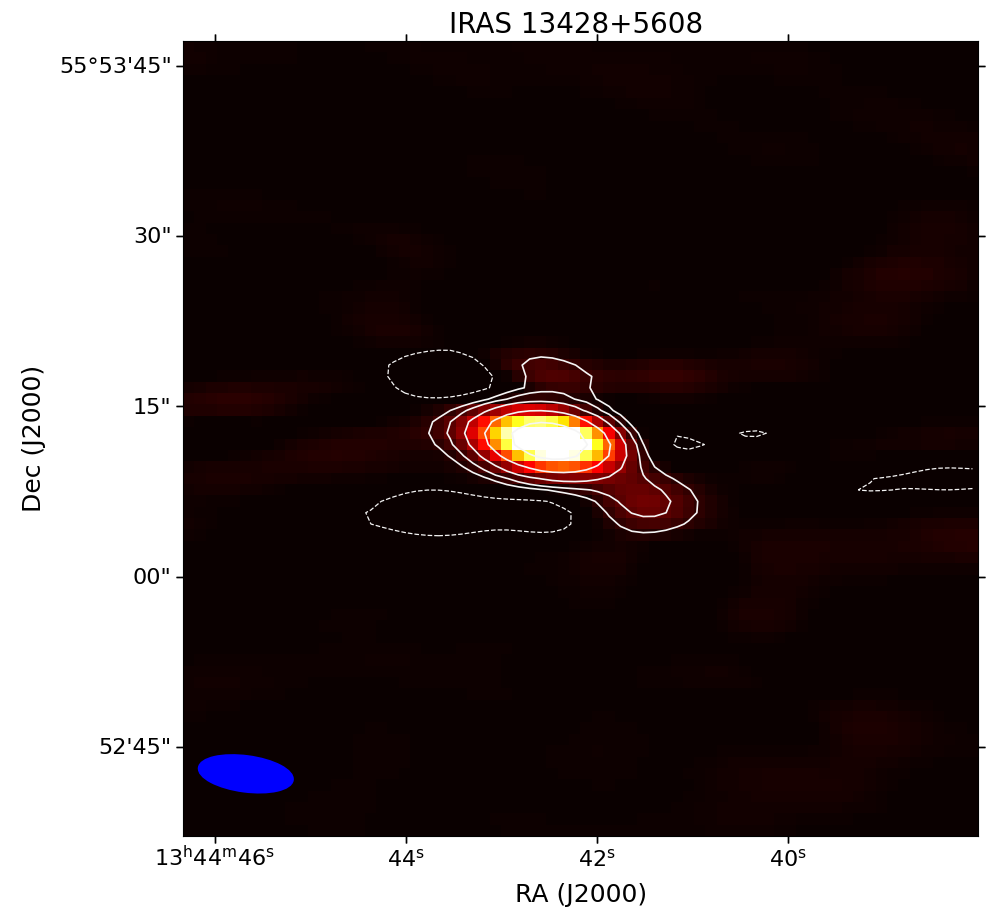} \\
   \caption{uGMRT Band 4 contour images of the four ULIRGs. Radio contours start at 5$\sigma$ and increase by factors of 2, where $\sigma$ represents the local rms noise level: 0.089, 0.116, 0.142, and 0.216 mJy beam$^{-1}$ for IRAS 05156$-$3024, IRAS F07599+6508, IRAS 10494+4424, and IRAS 13428+5608, respectively.}

    \label{band4}
\end{figure*}

We present the band 3 images of five ULIRGs and the band 4 images of four ULIRGs obtained using uGMRT (see Figures \ref{band3} and \ref{band4}). The resulting images, obtained after SPAM processing, not only provide sensitive radio images of the ULIRGs but also reveal minor extended structures, possibly due to radio jets or stellar/AGN outflows. IRAS 05156$-$3024 does not exhibit any extended structure in either band (Figures \ref{band3} \& \ref{band4}). Both IRAS F07599+6508 and IRAS 11119+3257 display minor extended radio structures in band 3 (Figure \ref{band3}). However, IRAS F07599+6508 appears as a point source in band 4 (Figure \ref{band4}). The source IRAS 10494+4424 appears very compact in both bands (Figures \ref{band3} \& \ref{band4}). The source IRAS 13428+5608, also known as Mrk 273, is a well-known source that shows extended emission in both bands. A detailed examination of its extended structure requires longer observation times. Large-scale radio emission from this source has already been reported at 144 MHz by \citet{Kukreti2022} using Low Frequency Array (LOFAR).

\subsection
{Radio Spectra of New Sources in the Second Phase}
\begin{figure*}
    \centering
    \includegraphics[width=3.0in]{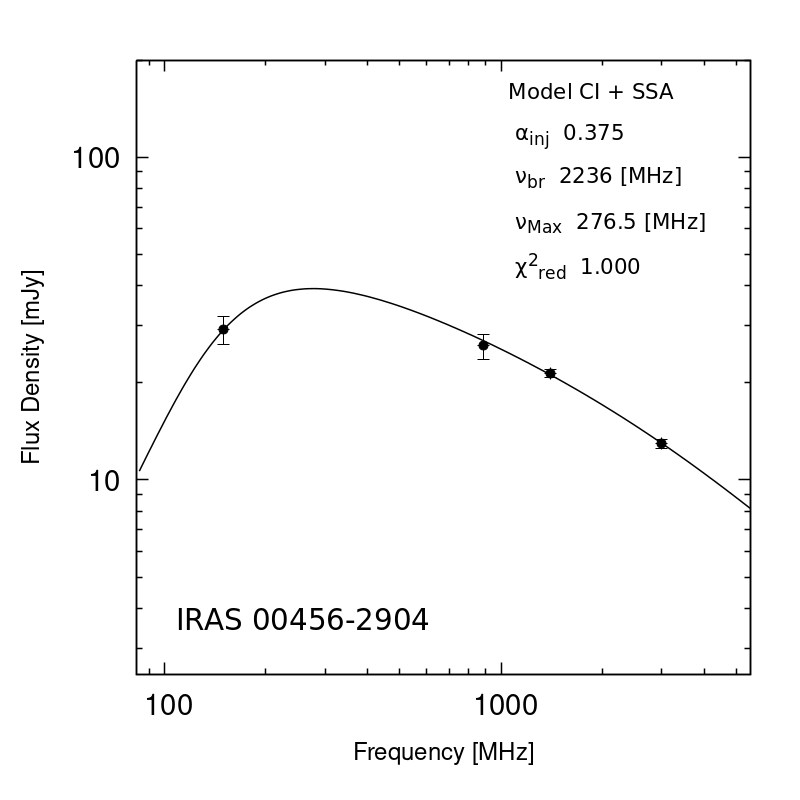} \quad
    \includegraphics[width=3.0in]{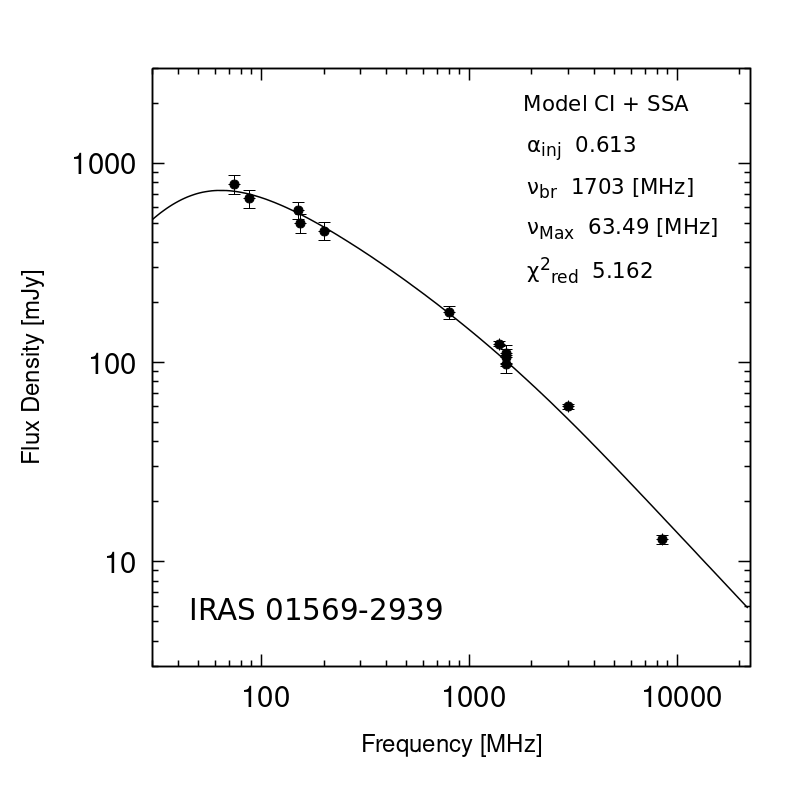} \\
    \vspace{0.1cm}
    \includegraphics[width=3.0in]{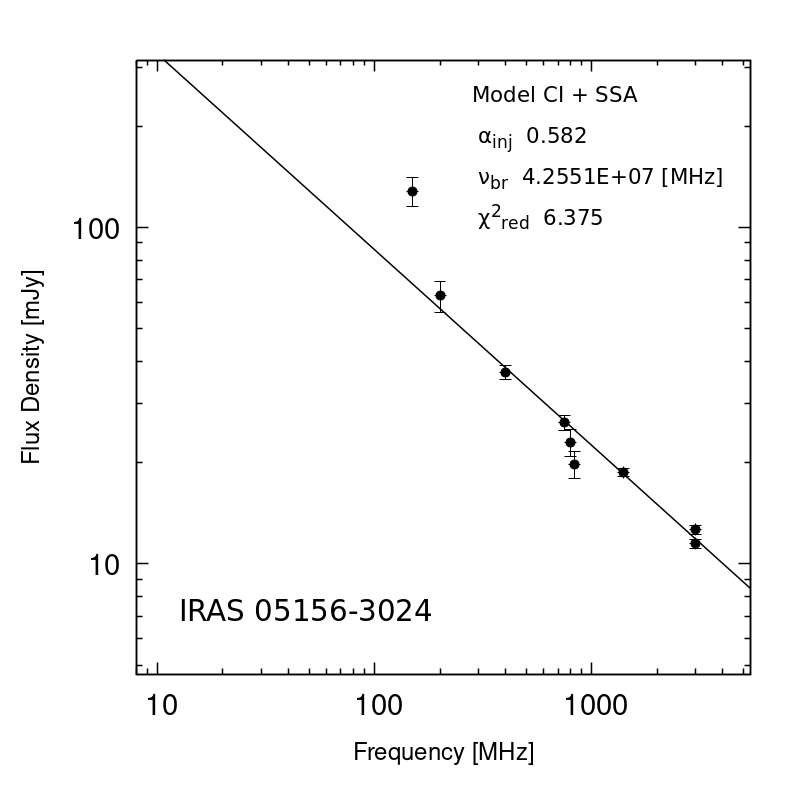} \quad
    \includegraphics[width=3.0in]{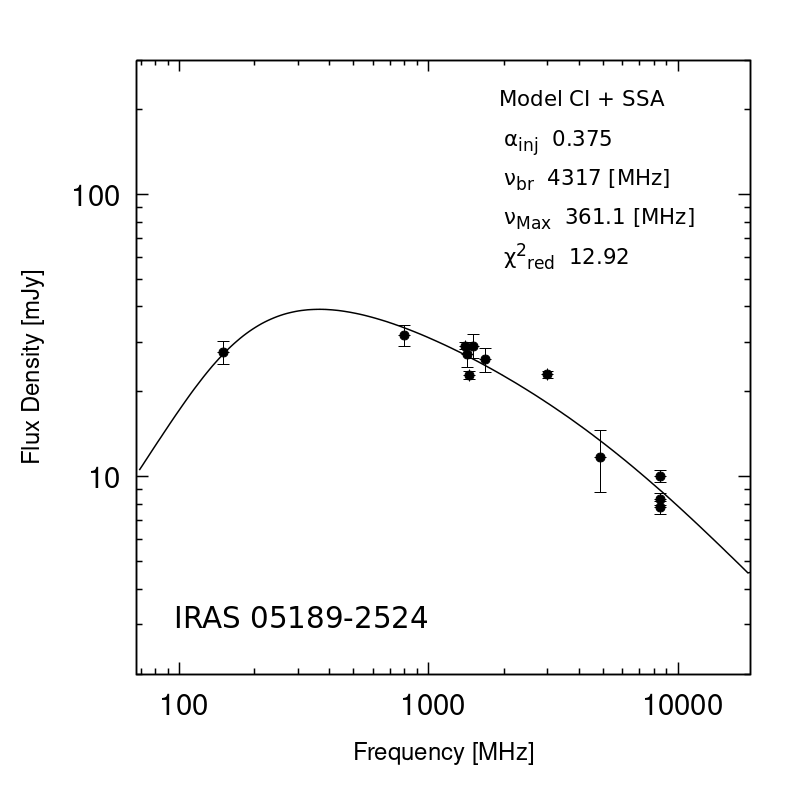} \\
    \vspace{0.1cm}
    \includegraphics[width=3.0in]{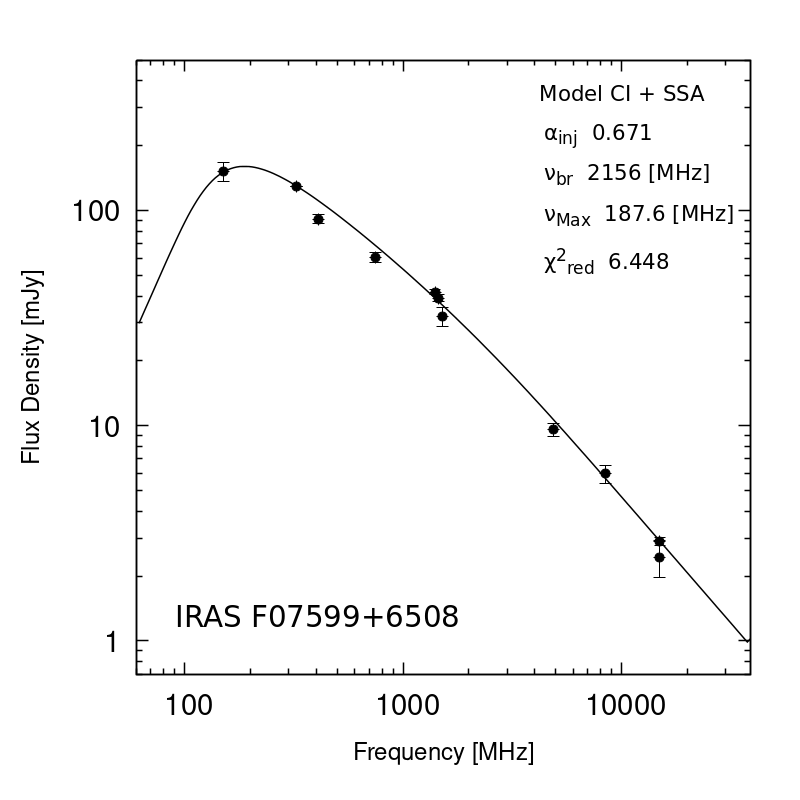} \quad
    \includegraphics[width=3.0in]{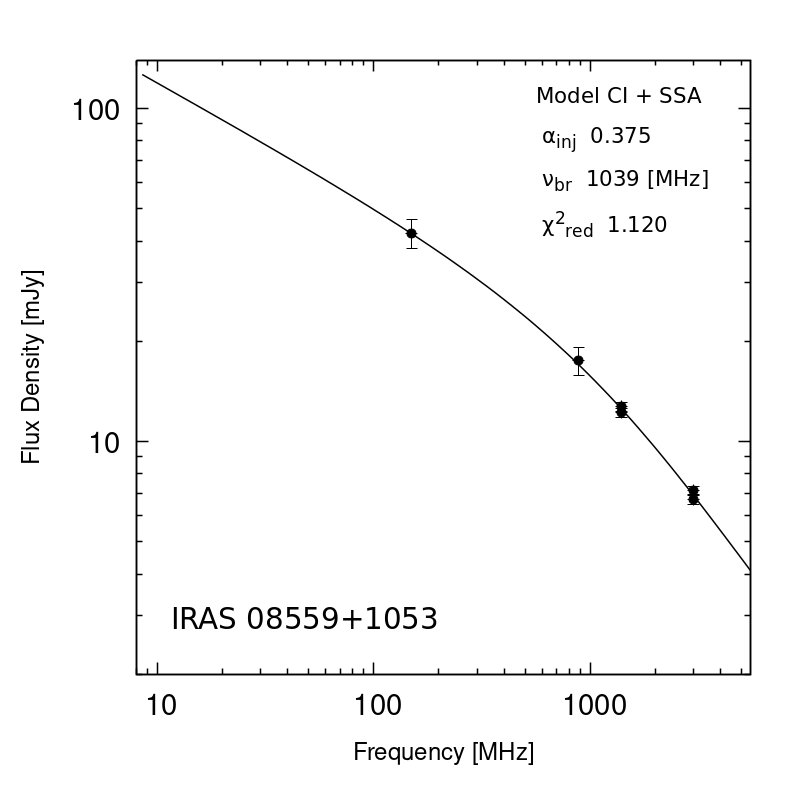} 
    
     \caption{Radio spectra modeled using SYNAGE. Panels show the SEDs for IRAS 00456$-$2904, IRAS 01569$-$2939, IRAS 05156$-$3024, IRAS 05189$-$2524, IRAS F07599$+$6508, and IRAS 08559$+$1053.}
    \label{fig:synage4}
\end{figure*}

\begin{figure*}
    \centering
    \includegraphics[width=3.0in]{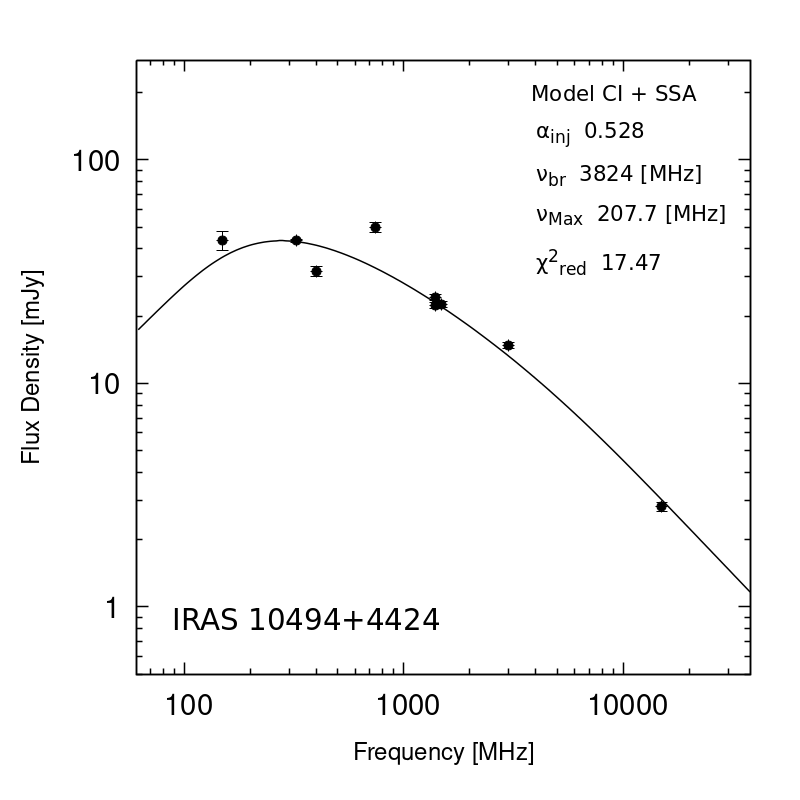} \quad
    \includegraphics[width=3.0in]{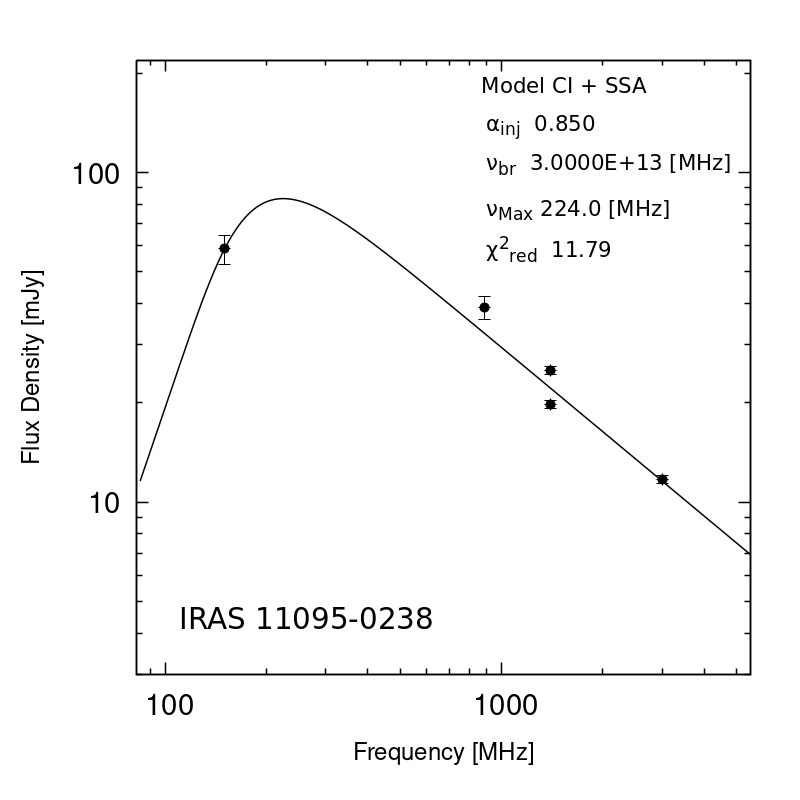} \\
    \vspace{0.1cm}
    \includegraphics[width=3.0in]{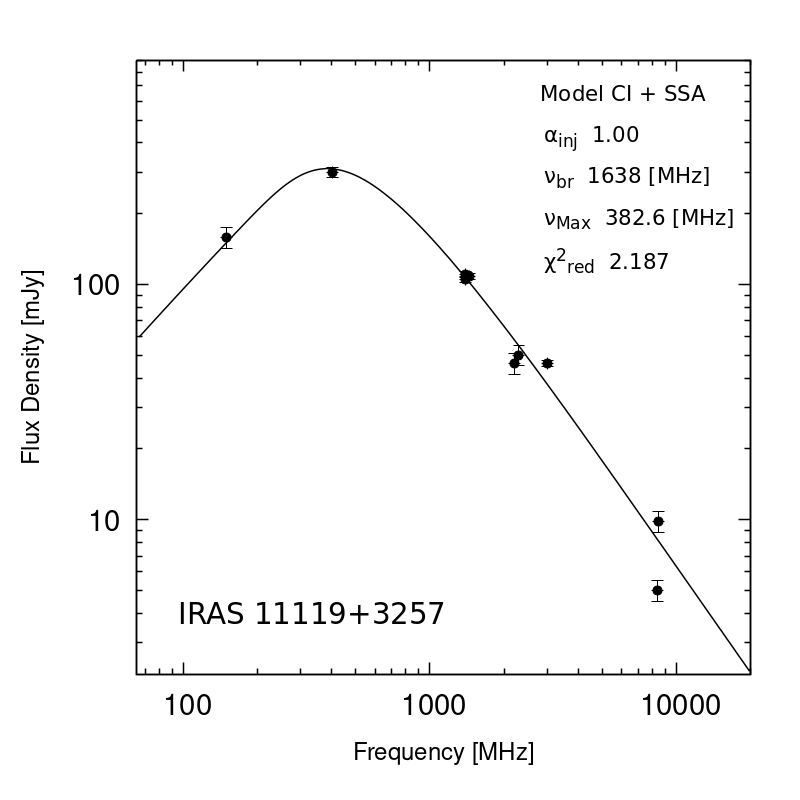} \quad
    \includegraphics[width=3.0in]{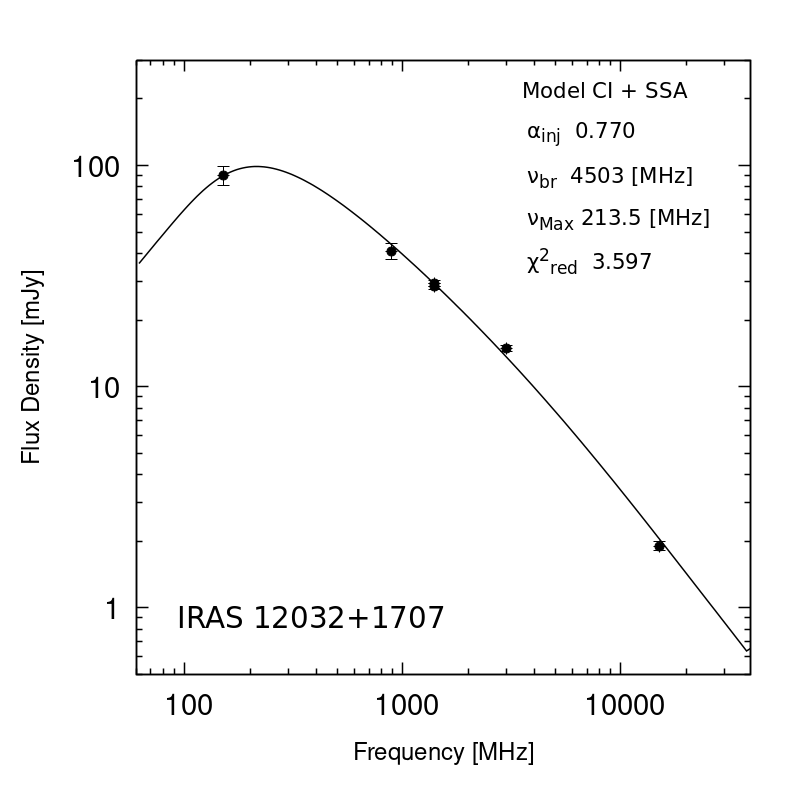} \\
    \vspace{0.1cm}
    \includegraphics[width=3.0in]{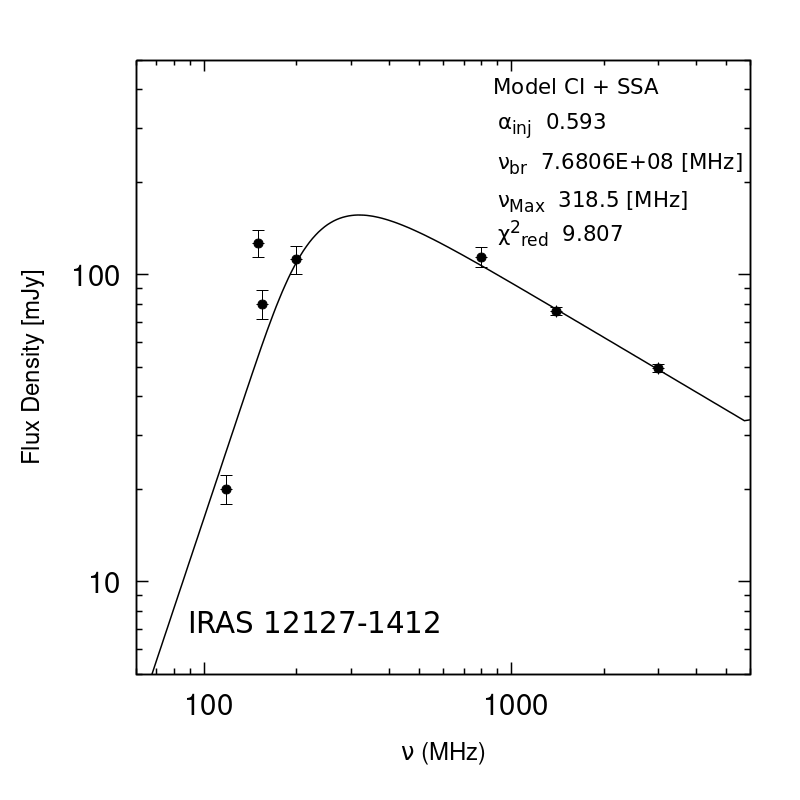} \quad
    \includegraphics[width=3.0in]{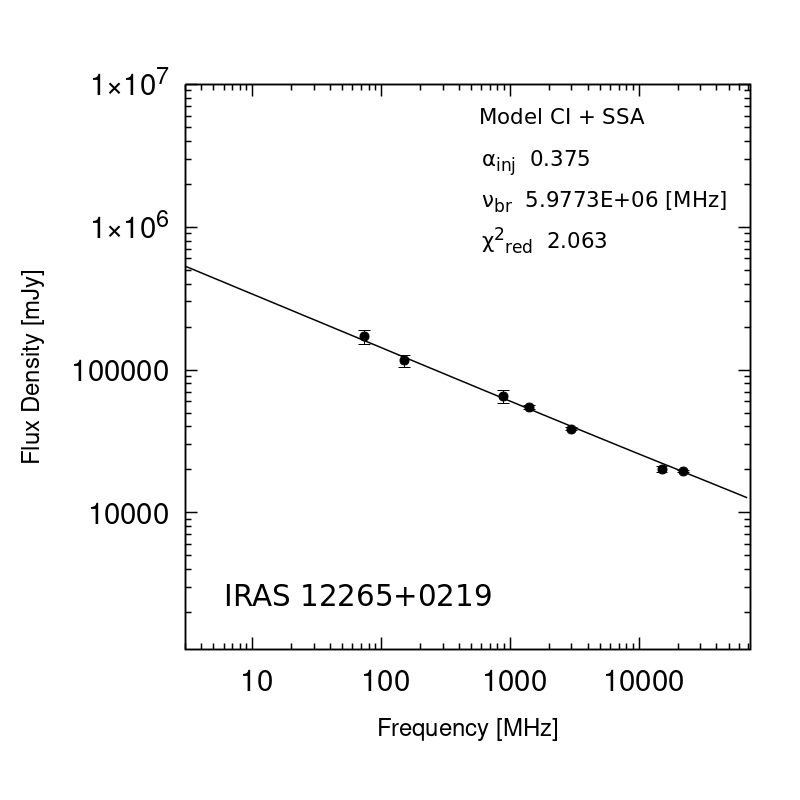} 
    \caption{Radio spectra modeled using SYNAGE. Panels show the SEDs for IRAS 10494$+$4424, IRAS 11095$-$0238, IRAS 11119$+$3257, IRAS 12032$+$1707, IRAS 12127$-$1412, and IRAS 12265$+$0219.}
        \label{fig:synage5}

\end{figure*}

\begin{figure*}
    \centering
    \includegraphics[width=3.0in]{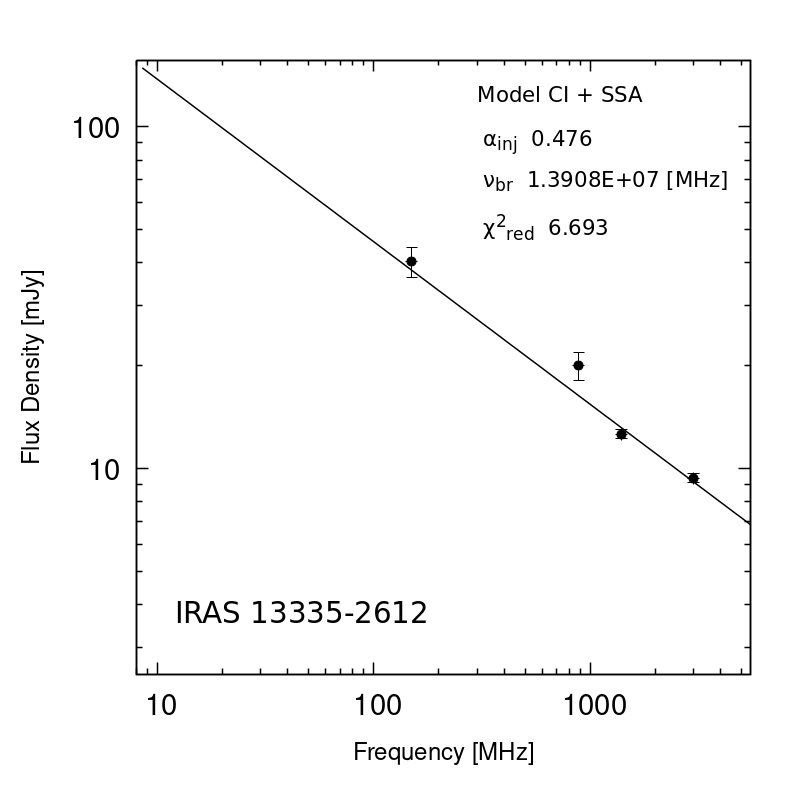} \quad
    \includegraphics[width=3.0in]{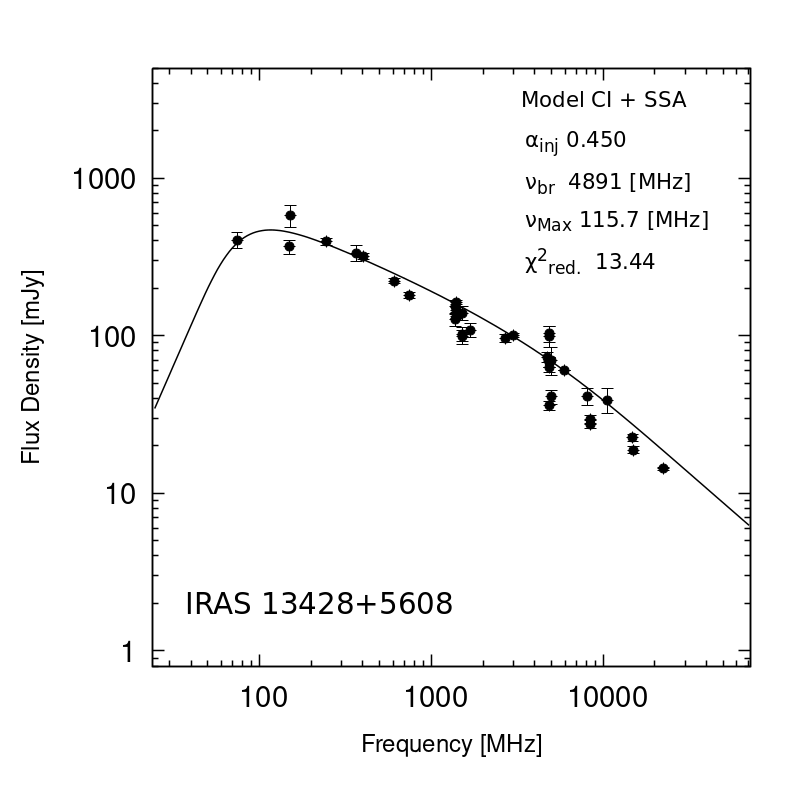} \\
    \vspace{0.1cm}
    \includegraphics[width=3.0in]{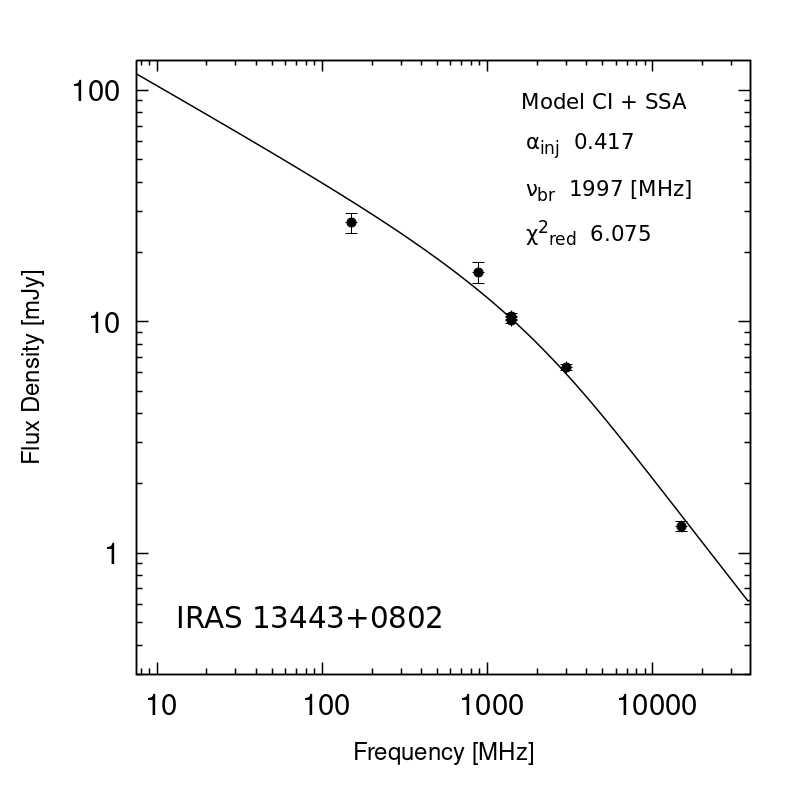} \quad
    \includegraphics[width=3.0in]{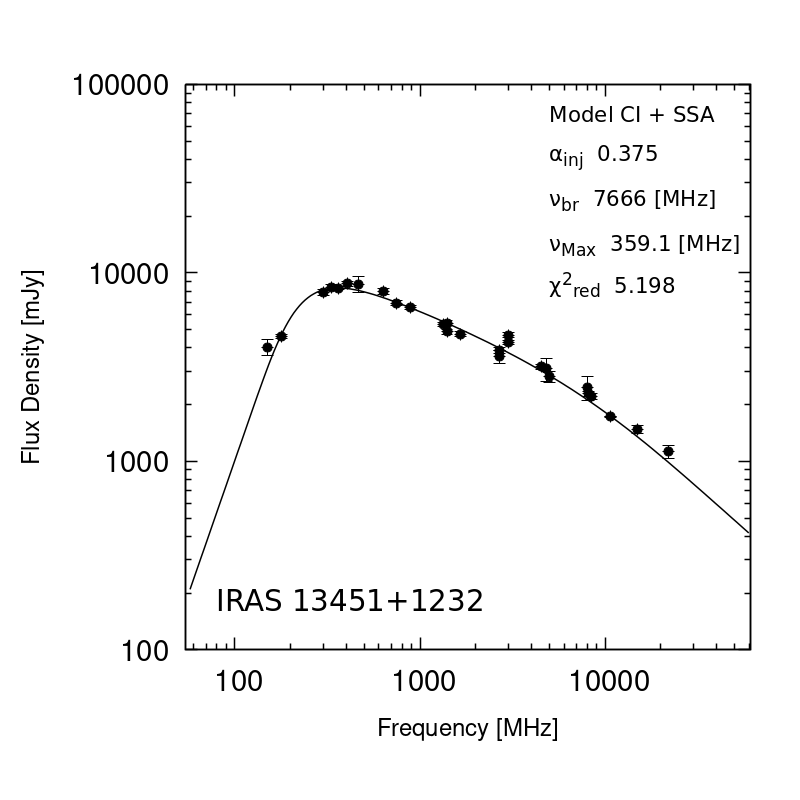} \\
    \vspace{0.1cm}
    \includegraphics[width=3.0in]{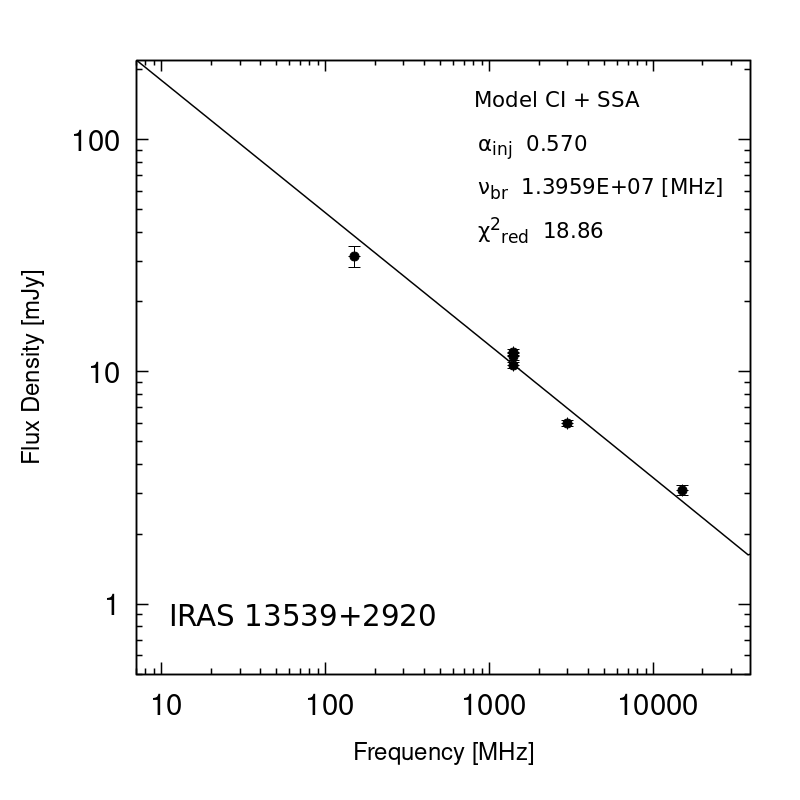} \quad
    \includegraphics[width=3.0in]{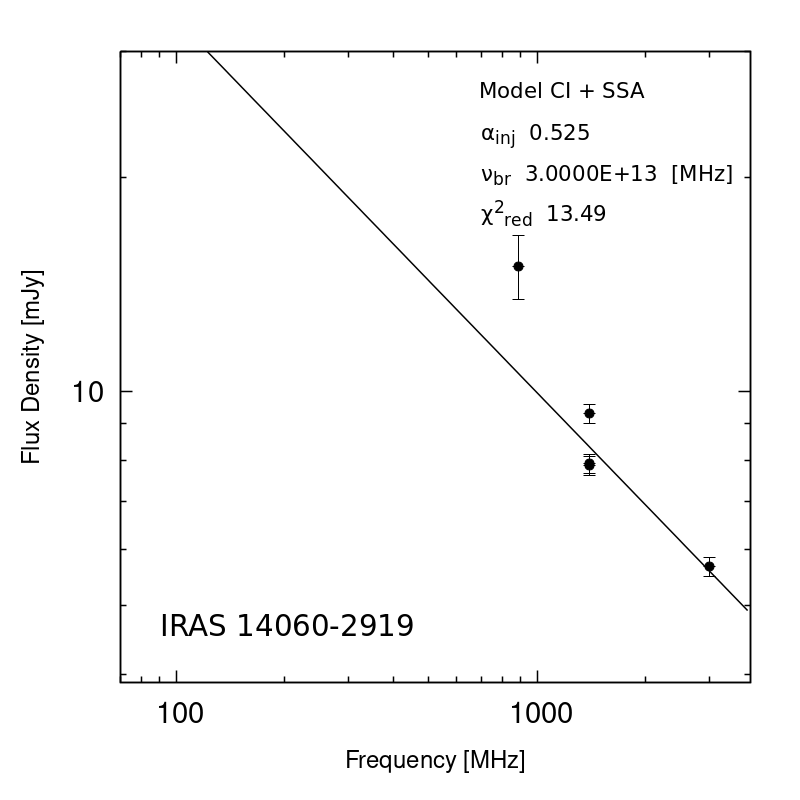} 
    
    \caption{Radio spectra modeled using SYNAGE. Panels show the SEDs for IRAS 13335$-$2612, IRAS 13428$+$5608, IRAS 13443$+$0802, IRAS 13451$+$1232, IRAS 13539$+$2920, and IRAS 14060$-$2919.}
        \label{fig:synage6}

\end{figure*}

\begin{figure*}
    \centering
    \includegraphics[width=3.0in]{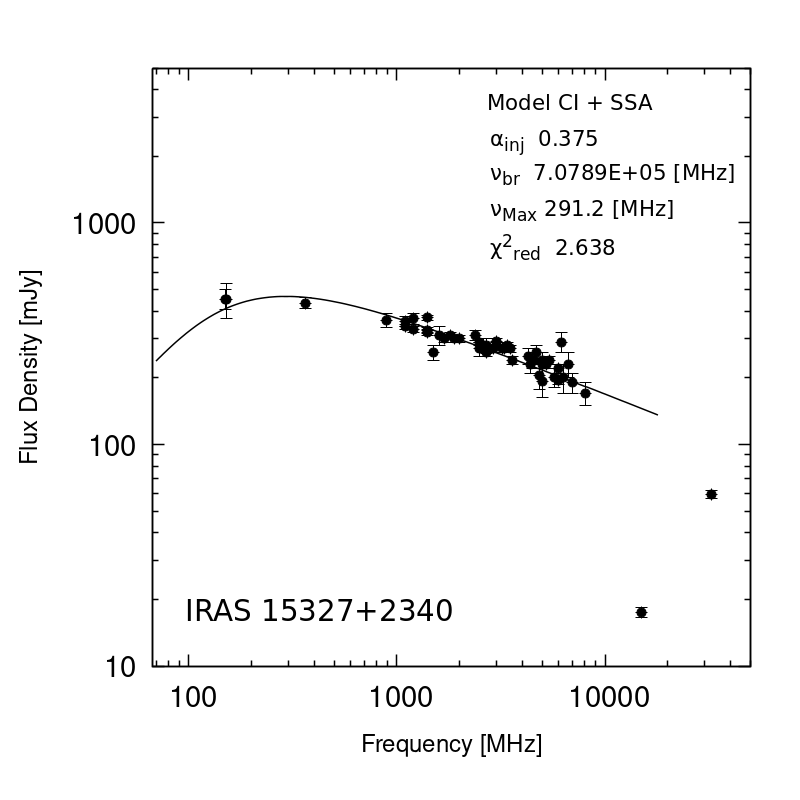} \quad
    \includegraphics[width=3.0in]{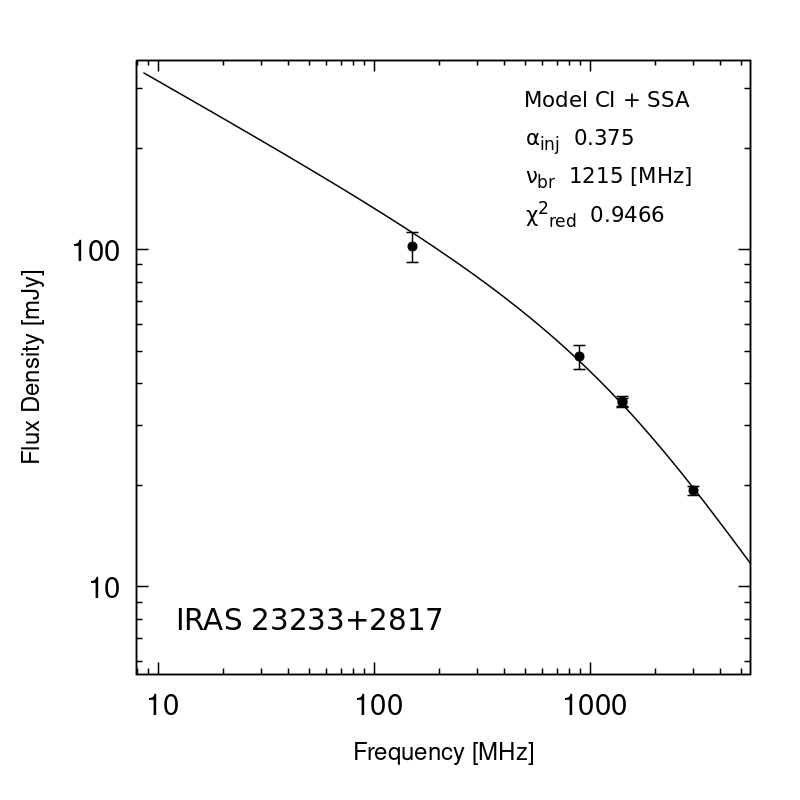} \\
    \vspace{0.1cm}
    \includegraphics[width=3.0in]{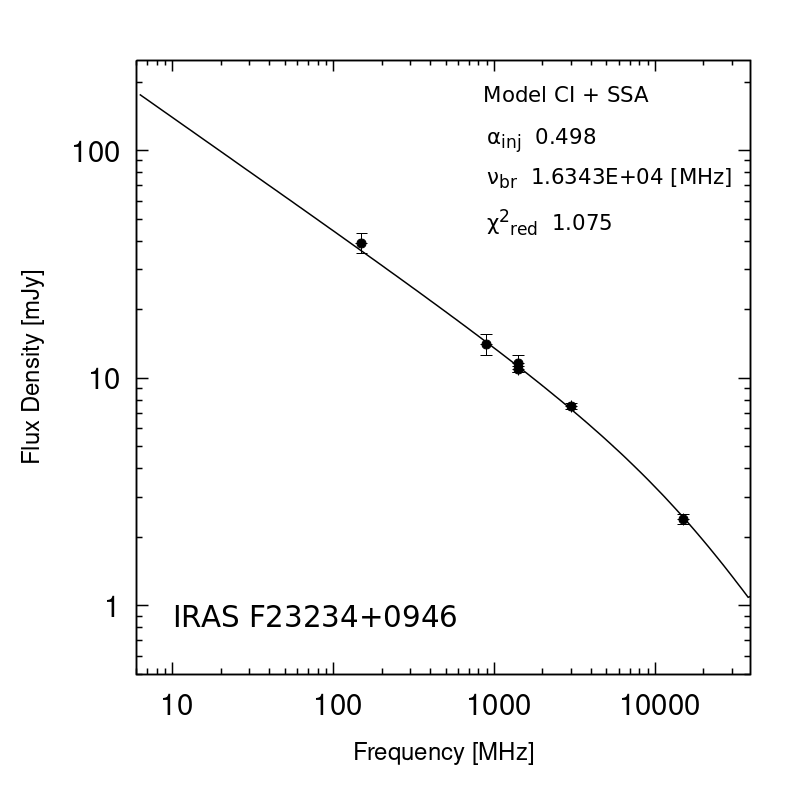} \quad
    \includegraphics[width=0.41\textwidth]{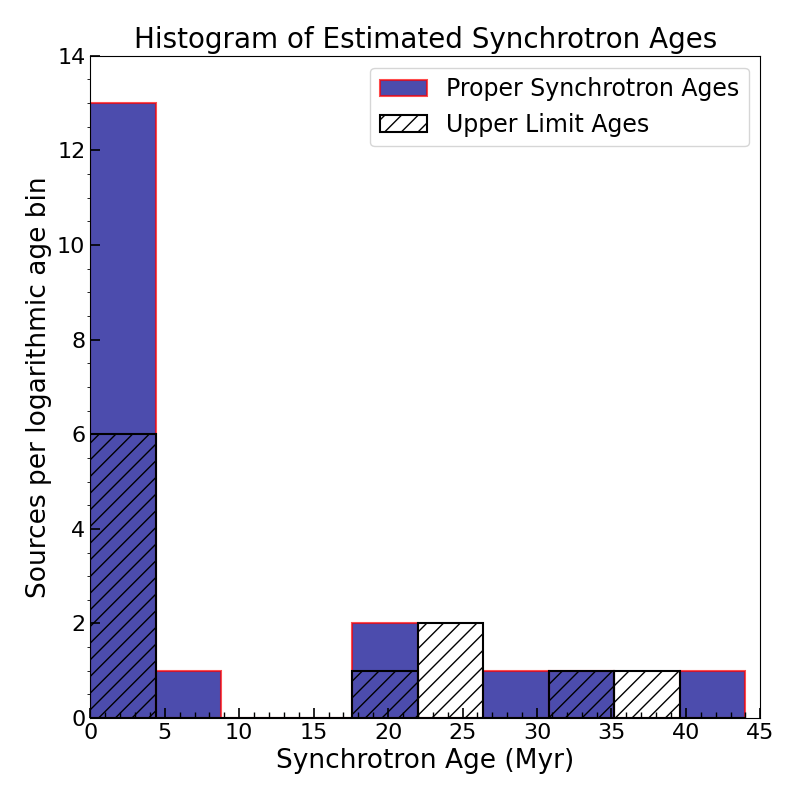}
   
    \caption{Radio spectra modeled using SYNAGE. Panels show the SEDs for IRAS 15327+2340, IRAS 23233+2817, IRAS F23234+0946 and Distribution of synchrotron ages for the radio sources. Filled bars show proper age estimates; hatched bars indicate upper limits.}
            \label{fig:synage7}

\end{figure*}

Studying the radio spectra of ULIRGs can help answer questions about the physical processes driving their radio emission, such as synchrotron radiation from magnetic fields or emission from ionized gas and dust. Spectral characteristics such as turnover frequencies, spectral breaks and the slope of the spectrum allow us to distinguish between AGN-driven activity and star formation processes. These features can also provide clues about the evolutionary stage of the ULIRGs, indicating whether the radio emission arises from a young, compact source, an aging electron population, or a system undergoing merger-driven activity. Together, this information helps determine their potential as progenitors of radio galaxies. In this second phase of our study of ULIRG radio emission, we focus on the new sample of 25 ULIRGs. Using uGMRT data along with archival observations covering a wide range of frequencies, we determined the integrated radio spectra for 21 out of the 25 sources. Here we note that the GMRT beam size and some other low frequency survey beam size is relatively large, compared with that of the Very Long Baseline Interferometry (VLBI). So, in a few cases measured radio emission may include contributions from components other than the central source. The flux density measurements and corresponding frequencies used in the spectral fitting are provided in Table~\ref{fluxtable}, along with their associated uncertainties and references. The remaining four sources lacked sufficient data points to perform reliable spectral fitting as indicated by the $\dagger$ in Table \ref{agetable}. We performed radio SED modeling for these 21 sources using the SYNAGE software \citep{1999A&A...345..769M}, as shown in Figures \ref{fig:synage4}, \ref{fig:synage5}, \ref{fig:synage6} and \ref{fig:synage7}. The SYNAGE output parameters, including the injection spectral index ($\alpha_{\rm inj}$), spectral break frequencies, the reduced values $\chi^2$ for the goodness of fit, the magnetic field strengths and the synchrotron ages, are shown in Table~\ref{agetable}. Most of the sources in the second phase exhibit a low turnover frequency, ranging from 63 to 383 MHz. To estimate spectral ages and magnetic field strengths, we use the method described in \citet{2021MNRAS.503.5746N}. The spectral ages, estimated using the break frequency, range from 0.0093 to 44 Myr, with eight sources showing very young spectral ages. Furthermore, we observed a very high break frequency for 11 sources. We have calculated upper age limits for these sources. Figure \ref{fig:synage7} shows a histogram of synchrotron age estimates for the full sample, including both estimated ages and upper age limits. We computed the spectral index for 3 of these sources by fitting power law model using gnuplot. For the source IRAS 02480$-$3745, which had only two data points, we performed a two-point spectral index calculation. Using the spectral index information, we have estimated flux density and luminosity for each source at 6 GHz (Table~\ref{agetable}). In our sample $\sim$82\% sources are radio quiet with $\log$ $L_{6 GHz}$ $\lesssim$ 23.2 {WHz}$^{-1}$ \citep{2016ApJ...831..168K} while only 7 sources can be categorised as radio-loud AGNs.

\begin{table*}
\begin{center}

\caption{Observed Flux Densities at Various Radio Frequencies with Associated Errors for ULIRGs in the Sample.}
\label{fluxtable}

\vspace{1ex}

    \begin{tabular}{|c|c|c|c|c|l|}
    \hline
      Galaxy & Frequency & Flux density & Flux density error & Reference \\
   ~ &(MHz) &(milliJy)& (milliJy)& \\
   (1)&(2)&(3)&(4)&(5)\\

 \hline
        IRAS 00456$-$2904  & 150 & 29.252 &2.925 &\citet{GMRTSurvey} \\
        ~ & 887 & 26.000 &2.320 & \cite{Hale2021} \\
        ~ & 1400 & 21.400&0.642  &\cite{1998AJ....115.1693C} \\ 
    
  ~ & 3000 &12.943 &0.388  & \cite{2021ApJS..255...30G}
      \\ 
       \hline

     IRAS F01004$-$2237$\dagger$ & 887 & 20.000& 1.900 & \cite{2020PASA...37...48M} \\ 
            ~ & 1400 & 13.700& 0.600& \cite{1998AJ....115.1693C} \\ 
       ~ & 1400& 13.400& 0.402&\cite{2015ApJ...801...26H}\\
          ~ & 3000 & 4.367& 0.131& \cite{2021ApJS..255...30G}\\ 
                 \hline

                  IRAS 01494$-$1845$\dagger$  & 887 & 11.900&  1.190 & \cite{2020PASA...37...48M} \\ ~ & 1400 & 8.000& 0.240  & \cite{1998AJ....115.1693C} \\ 
            ~ & 3000 &  4.902&0.147 & \cite{2021ApJS..255...30G}\\ \hline

IRAS 01569$-$2939&74&782.800&84.960 &\citet{2014MNRAS.440..327L}\\
~ & 88 & 661.780& 70.140& \citet{2015PASA...32...25W} \\
~ & 150& 578.860 & 57.880& \cite{GMRTSurvey} \\
~ & 155 &  498.670 & 52.850 &\citet{2015PASA...32...25W}\\
~ & 200& 456.610 & 48.400 &\citet{2015PASA...32...25W} \\
~ &  800 &  177.600 & 12.930  &\cite{Hale2021}\\
~ & 1400& 123.900 & 3.710 & \cite{1998AJ....115.1693C} \\
~ & 3000& 59.870 & 1.790 & \cite{2021ApJS..255...30G}\\
~ & 1510 & 110.410 & 11.040 &VLA archival image\\ 
~ & 1510 & 105.790 & 10.580  &VLA archival image\\
~ & 1510 &  98.230 & 9.820 & VLA archival image\\
~ & 8460 & 12.880 &  0.610 &VLA archival image\\
 \hline

 IRAS 02480$-$3745$\dagger$& 150 &98.880& 9.888  & \cite{GMRTSurvey} \\
 ~ & 1400 & 13.100 & 0.393 & \cite{1998AJ....115.1693C} \\ 
 \hline

     IRAS 04103$-$2838$\dagger$& 150 & 35.020& 3.502& \cite{GMRTSurvey} \\
    
     ~ & 887 & 19.240 & 1.846 & \cite{2020PASA...37...48M} \\
       
       ~ & 1400 & 13.700 &  0.240 & \cite{1998AJ....115.1693C} \\ 
    
      ~ & 3000 & 8.173 & 0.245 & \cite{2021ApJS..255...30G}\\  
    
       \hline

IRAS 05156$-$3024&150& 128.290& 12.890&\cite{GMRTSurvey}\\
~& 200 &62.650 & 6.640&\citet{2015PASA...32...25W} \\
~& 402 &37.100& 1.850&Present work\\
~& 745& 26.310& 1.310 &Present work\\
~& 800& 23.000& 2.110&\cite{2020PASA...37...48M}\\
~& 834& 19.800& 1.850&\cite{2013yCat.8081....0M}\\
~& 1400& 18.700& 0.560&\cite{1998AJ....115.1693C}\\
~& 3000& 11.490& 0.340&\cite{2021ApJS..255...30G} \\
~& 3000&  12.640& 0.370&\cite{2021ApJS..255...30G}\\
 \hline

IRAS 05189$-$2524&150& 27.560& 2.560&\cite{GMRTSurvey}\\
~& 800& 31.600& 2.710& \cite{2020PASA...37...48M}\\
~& 1400& 29.100&  0.870&\cite{1998AJ....115.1693C}\\
~& 1430&  27.110&  2.710&VLA archival image\\
~& 1450& 22.880& 0.810&VLA archival image\\
~& 1510& 29.080& 2.910&VLA archival image\\
~& 1690  &  26.000 &2.600 &VLA archival image\\
 ~& 3000& 22.980& 0.680&\cite{2021ApJS..255...30G}\\
~& 4890&11.700&2.900&VLA archival image\\      
~& 8490& 7.780& 0.390&VLA archival image\\
~& 8490& 8.320& 0.420&VLA archival image\\
~& 8490& 10.040& 0.500&VLA archival image\\

\hline

    IRAS F07599+6580& 150& 151.970& 15.190&\cite{GMRTSurvey}\\    
         ~ & 325& 129.740& 0.130& \cite{Rengelink1997}\\   
        ~ & 408& 91.200& 4.560&Present work\\    
       ~ & 745& 60.700& 3.320&Present work\\   
     ~ & 1400& 41.760& 1.250&\cite{1998AJ....115.1693C}\\    
         ~ & 1450& 39.150& 1.390&VLA archival image\\    
        ~ & 1510& 32.250& 3.220&VLA archival image\\   
         ~ & 4890& 9.590& 0.690&VLA archival image\\   
         ~ & 8460& 5.970& 0.590&VLA archival image\\    
       ~ & 14900& 2.450& 0.480&VLA archival image\\   
      ~ & 15000& 2.900& 0.130&\cite{nagar2003agn}\\ 
         \hline 

 \end{tabular}
\end{center}
\end{table*}

\begin{table*}

\begin{center}
   \begin{tabular}{|c|c|c|c|c|l|}
     \hline
       Galaxy & Frequency & Flux density & Flux density error & Reference \\
   ~ &(MHz) &(milliJy)& (milliJy)& \\
         (1)&(2)&(3)&(4)&(5)\\
  \hline

\hline

IRAS 08559+1053  & 150 & 42.127&4.212& \cite{GMRTSurvey} \\
~ & 887 & 17.500 & 1.725 & \cite{2020PASA...37...48M} \\
       ~ & 1400 & 12.200 & 0.366 & \cite{1998AJ....115.1693C} \\ 
      ~ & 1400 & 12.720 &  0.381 & \cite{2018AJ....155..188L} \\ 
     ~ & 3000 &6.685 & 0.200 & \cite{2021ApJS..255...30G}\\
    ~& 3000& 7.126& 0.213&\cite{2021ApJS..255...30G}\\
       \hline
IRAS 10494+4424 & 150& 43.480&  4.340&\cite{GMRTSurvey}\\
               ~& 325& 43.420& 0.550&\cite{Rengelink1997} \\
             ~& 402& 31.680& 1.590&Present work\\
             ~& 745& 50.000&  2.500&Present work\\
           ~& 1400&  22.400&  0.670&\cite{1998AJ....115.1693C}\\
              ~& 1400& 24.200&  0.720&\cite{2015ApJ...801...26H}\\
       ~& 1490& 22.520& 0.670&VLA archival image\\
~& 3000& 14.820& 0.440&\cite{2021ApJS..255...30G}\\
~& 15000& 2.800& 0.140&\cite{nagar2003agn}\\
\hline
        
        IRAS 11095$-$0238  & 150 &  58.710& 5.871 &\cite{GMRTSurvey} \\
        ~& 887 &39.000& 3.230 & \cite{2020PASA...37...48M} \\
        ~ & 1400 &  25.100&  0.753 & \cite{1998AJ....115.1693C} \\ 
        ~ & 1400 & 19.780& 0.593 & \cite{2015ApJ...801...26H} \\
        ~ & 3000 & 11.724& 0.351 & \cite{2021ApJS..255...30G} \\ \hline

IRAS 11119+3257& 150& 158.000 &15.800&\cite{GMRTSurvey}\\
~& 402& 298.800& 14.940&Present work\\
~& 1400& 110.400 & 3.310&\cite{1998AJ....115.1693C}\\
~& 1400& 105.350 & 3.160&\citet{2015ApJ...801...26H}\\
~& 1400& 108.240 & 3.240&\cite{2015ApJ...801...26H}\\
~& 2200& 46.000& 4.600&\citet{2013AJ....146....5P}\\
~& 2300& 50.000& 5.000&\citet{2010113B}\\
~& 3000& 45.970& 1.370&\cite{2021ApJS..255...30G}\\
~& 8400& 5.000& 0.500&\citet{2010113B}\\
~& 8460& 9.820& 0.980& \citet{2010113B}\\
\hline

        IRAS 12032+1707 & 150 & 90.228& 9.022& \cite{GMRTSurvey} \\
        ~ & 887 & 41.000& 3.370 & \cite{2020PASA...37...48M} \\
        ~ & 1400 & 29.310&  0.879 & \cite{2015ApJ...801...26H} \\ 
        ~ & 1400 & 28.400& 0.852& \cite{1998AJ....115.1693C} \\ 
        ~ & 3000 & 14.902& 0.447 & \cite{2021ApJS..255...30G}\\
        ~ & 15000 &1.900 &0.095 & \citet{nagar2003agn} \\
        \hline

IRAS 12127$-$1412&118& 20.000&  2.120&\citet{2015PASA...32...25W}\\
~& 150& 126.690 & 12.660&\cite{GMRTSurvey}\\
~& 155& 80.000& 8.480&\citet{2015PASA...32...25W}\\
~& 200& 112.000  & 11.870&\citet{2015PASA...32...25W}\\
~& 800& 114.000& 8.480& \cite{2020PASA...37...48M}\\
~& 1400& 76.000& 2.280&\cite{1998AJ....115.1693C}\\
~& 3000& 49.420& 1.480&\cite{2021ApJS..255...30G}\\
\hline

 IRAS 12265+0219 (3C273)& 73.8 & 170340.000&  19044.000&\citet{2014MNRAS.440..327L}\\
       ~& 150 &  119198.000& 11919.800  & \cite{GMRTSurvey} \\
      ~& 887& 65020.000& 4551.900&\cite{2020PASA...37...48M} \\
     ~ & 1400 &5499.2 &110.950& \cite{2015ApJ...801...26H} \\
    ~&  3000& 38611.500& 1158.340&\cite{2021ApJS..255...30G}\\
       ~ & 15000 & 21614.000&  1028.830& \citet{nagar2003agn} \\
      ~& 22000& 41000.000&  28700.000&\cite{1994MNRAS.267..167G}\\
      \hline

       IRAS 13335$-$2612 & 150 & 40.273& 4.027 & \cite{GMRTSurvey} \\
        ~ & 887 & 20.000&  1.900 & \cite{2020PASA...37...48M} \\
        ~ & 1400 & 12.600& 0.378& \cite{1998AJ....115.1693C} \\ 
        ~ & 3000 & 9.364& 0.280& \cite{2021ApJS..255...30G}\\ \hline

       IRAS 13428+5608& 74& 405.000& 45.270 &\citet{2014MNRAS.440..327L}\\
 ~& 150& 366.600&  36.660&\citet{2017104S}\\
 ~& 151& 580.000 &90.000&\citet{1990MNRAS.246..256H}\\
 ~& 244& 394.000&  20.000&\citet{2010MNRAS.405..887C}\\
 ~& 365& 335.000&  37.000&\cite{1996AJ....111.1945D}\\
 ~& 402& 318.810 & 15.940&Present work\\
 ~& 610& 222.000& 8.000&\citet{2010MNRAS.405..887C} \\
 ~& 745& 180.190 & 9.000&Present work\\

 \hline
       
 \end{tabular}
\end{center}
\end{table*}

\begin{table*}
\begin{center}
    \begin{tabular}{|c|c|c|c|c|l|}
    \hline
       Galaxy & Frequency & Flux density & Flux density error & Reference \\
   ~ &(MHz) &(milliJy)& (milliJy)& \\
      (1)&(2)&(3)&(4)&(5)\\

        \hline 

 ~& 1380& 126.630& 12.660&VLA archival image\\

 ~& 1380& 152.760 & 15.280&VLA archival image\\
~& 1400& 145.400 & 5.200&\cite{1998AJ....115.1693C}\\
 ~& 1400& 132.000& 3.960&\cite{Maccagni2017}\\
 ~& 1400& 163.000& 4.890&\cite{2017ApJ...847..136B}\\
 ~& 1430& 138.460& 13.850&VLA archival image\\
 ~& 1450& 139.340 & 4.950&VLA archival image\\
 ~& 1510& 98.510& 9.850&VLA archival image\\
 ~& 1510& 102.580 &0.260&VLA archival image\\
 ~& 1510& 139.390& 13.940&VLA archival image\\
 ~& 1690& 108.560& 10.860&VLA archival image\\
 ~& 2700& 96.000& 6.000&\citet{1976ApJ...207..725S}\\
 ~& 3000& 100.000& 3.000&\cite{2021ApJS..255...30G}\\
 ~& 4755& 73.000& 5.000&\citet{1995ApJS...98..369B} \\
 ~& 4850& 99.000& 14.850&\citet{1991ApJS...75....1B}\\
  ~& 4850& 103.000& 12.000&\citet{1991ApJS...75.1011G}\\

~& 4890& 35.870& 2.590&VLA archival image\\
 ~& 4890& 67.500& 4.880&VLA archival image\\
 ~& 4890& 63.100& 4.560&VLA archival image\\
 ~& 4890& 68.280& 4.970&VLA archival image\\
 ~& 5000& 70.000& 14.000& \citet{1980AJ.....85.1462K}\\
 ~& 5010& 40.890& 4.090&VLA archival image\\
 ~& 5950& 60.300& 0.080&\citet{leroy2011complex}\\
 ~& 8100& 41.000& 5.000&\citet{1976ApJ...207..725S}\\
   ~& 8460& 29.390& 1.850&VLA archival image\\
 ~& 8460& 27.410& 1.730&VLA archival image\\
 ~& 10630& 39.000& 7.000&\citet{1995ApJS...98..369B}\\
 ~& 14900& 22.460& 1.100&VLA archival image\\
 ~& 15000& 18.800& 0.940&\citet{nagar2003agn}\\
 ~& 22500& 14.300& 0.300&\citet{2008AAA...477...95C}\\
 
  \hline

 IRAS 13443+0802 & 150 & 26.780&  2.678 &\cite{GMRTSurvey} \\ 
        ~ & 887 &  16.300 & 1.641 &\cite{2020PASA...37...48M} \\
        ~ & 1400 & 10.500& 0.315 & \cite{1998AJ....115.1693C} \\ 
        ~ & 1400 & 10.090& 0.302& \cite{2015ApJ...801...26H} \\ 
    
        ~ & 3000 &6.313& 0.189 & \cite{2021ApJS..255...30G}\\ 
        ~ & 15000 & 1.300 & 0.062& \citet{nagar2003agn}\\
        \hline

        IRAS 13451+1232 (PKS 1345+12) &150& 4036.980& 403.698&\cite{GMRTSurvey} \\
        ~ & 178& 4600.000& 125.000 &\cite{1967MmRAS..71...49G}  \\
        ~ & 302& 7860.000& 300.000& \cite{1998AAS131303S}\\
        ~ & 333& 8390.000& 300.000&\cite{1998AAS131303S}\\
        ~ & 365 &8309.000& 169.000& \cite{1996AJ....111.1945D}\\ 
        ~ & 408& 8780.000& 270.000&\cite{1998AAS131303S}\\
        ~ & 468 & 8720.000 & 850.000 & \cite{Kuehr1981}\\
        ~ & 635& 7970.000& 330.000&\cite{Kuehr1981}\\
        ~ & 750& 6900.000& 200.000&\cite{Kuehr1981}\\
        ~ & 887 &6551.000& 161.916 &\cite{2020PASA...37...48M}  \\
        ~ & 1335& 5330.000& 150.000&\cite{1998AAS131303S}\\
        ~ & 1400 & 5397.200& 161.900& \cite{1998AJ....115.1693C} \\ 
        ~ & 1400 & 4859.880& 145.796 & \cite{2015ApJ...801...26H} \\
        ~ & 1660& 4734.000& 150.000&\cite{Stanghellini2005}\\
        ~ & 2695& 3600.000& 290.000&\cite{Adgie1972}\\
        ~ & 2700& 3880.000& 128.103&\cite{Wright1990}\\
        ~ & 3000& 4660.200& 139.805&\cite{2021ApJS..255...30G}\\
        ~ & 3000 &  4270.110& 111.450& \cite{2021ApJS..255...30G}\\ 
        ~ & 4535& 3180.000& 100.000&\cite{1998AAS131303S}\\
        ~ & 4850& 3092.000& 427.000&\cite{1991ApJS...75.1011G}\\
        ~ & 5000& 2810.000&  200.000&\cite{Kuehr1981}\\
        ~ & 8000& 2460.000& 369.000&\citet{Stull1971}\\
        ~ & 8085& 2280.000& 70.000&\cite{1998AAS131303S}\\
        ~ & 8400& 2210.000& 70.000&\cite{1998AAS131303S}\\
        ~ & 10695& 1720.000& 10.000&\cite{Kuehr1981}\\
        ~ & 15000 & 1475.000 &73.750 & \cite{nagar2003agn}\\
        ~ & 22000& 1130.000&  90.000&\cite{1998AAS131303S}\\
        \hline

    \end{tabular}
\end{center}
\end{table*}

\begin{table*}
\begin{center}
    \begin{tabular}{|c|c|c|c|c|l|}  
    \hline
        Galaxy & Frequency & Flux density & Flux density error & Reference \\
   ~ &(MHz) &(milliJy)& (milliJy)& \\
      (1)&(2)&(3)&(4)&(5)\\

        \hline
   IRAS 13539+2920 & 150 & 31.415& 3.141& \cite{GMRTSurvey} \\~ & 1400 & 12.100& 0.360 & \cite{1998AJ....115.1693C} \\ 
 ~ & 1400 & 11.600& 0.348&\cite{2018AJ....155..188L} \\ 
 ~ & 1400 & 10.700& 0.321& \cite{1995ApJ...450..559B}\\
 ~ & 3000 & 6.020& 0.180 & \cite{2021ApJS..255...30G}\\ 
~ & 15000 & 3.100 &0.155 & \cite{nagar2003agn} \\ \hline

     IRAS 14060+2919  & 887 &15.000& 1.550 &\cite{2020PASA...37...48M}\\
     ~ & 1400 &  9.300& 0.279 & \cite{2018AJ....155..188L}\\ 
    ~ & 1400 & 7.860& 0.235 & \cite{1995ApJ...450..559B}\\
    ~ & 1400 & 7.910& 0.237& \cite{1998AJ....115.1693C}\\
 ~ & 3000 &5.666& 0.169 &\cite{2021ApJS..255...30G} \\ \hline

        IRAS 15327+2340 (Arp 220) & 150 & 440.3 & 44.7 & \cite{GMRTSurvey} \\
         ~ & 151& 450.000& 81.000&\cite{waldram19967c}\\
         ~ & 365 & 435.000& 26.000& \cite{1996AJ....111.1945D} \\
         ~ & 887 & 364.000& 25.980& \cite{2020PASA...37...48M}\\
         ~ & 1100& 360.000& 20.000&\cite{williams2010evaluating}\\
         ~ & 1100& 340.000& 10.000&\cite{williams2010evaluating}\\
         ~ & 1200& 330.000& 10.000&\cite{williams2010evaluating}\\
         ~ & 1200& 370.000& 20.000&\cite{williams2010evaluating}\\
        ~ & 1400& 373.000& 11.190&\cite{1998AJ....115.1693C}\\
         ~ & 1400& 326.300& 9.789&\cite{allison2014search}\\
         ~ & 1400& 320.000& 10.000&\cite{williams2010evaluating}\\
         ~ & 1500& 260.000& 20.000&\cite{williams2010evaluating}\\
         ~ & 1600& 310.000& 30.000&\cite{williams2010evaluating}\\
         ~ & 1700& 300.000& 10.000&\cite{williams2010evaluating}\\
         ~ & 1800& 310.000& 10.000&\cite{williams2010evaluating}\\  
         ~ & 1900& 300.000& 10.000&\cite{williams2010evaluating}\\   
         ~ & 2000& 300.000& 10.000&\cite{williams2010evaluating}\\   
         ~ & 2380& 312.000& 16.000&\cite{dressel1978arecibo}\\ 
         ~ & 2500& 290.000& 10.000&\cite{williams2010evaluating} \\  
         ~ & 2500& 270.000& 20.000&\cite{williams2010evaluating}\\
         ~ & 2600& 280.000& 20.000&\cite{williams2010evaluating}\\    
         ~ & 2695& 279.000& 22.000&\cite{sulentic1976radio}\\ 
         ~ & 2695& 260.000& 10.000& \cite{condon1983optical}\\
         ~ & 2900& 270.000& 10.000&\cite{williams2010evaluating}\\    
         ~ & 3000& 279.062& 8.371&\cite{2021ApJS..255...30G}\\  
         ~ & 3000& 291.954& 8.758&\cite{2021ApJS..255...30G}\\    
         ~ & 3200& 270.000& 10.000&\cite{williams2010evaluating}\\  
         ~ & 3300& 270.000& 10.000&\cite{williams2010evaluating}\\   
     ~ & 3400& 280.000& 10.000&\cite{williams2010evaluating}\\  
        ~ & 3500& 270.000& 10.000&\cite{williams2010evaluating}\\ 
         ~ & 3600& 240.000& 10.000&\cite{williams2010evaluating}\\  
         ~ & 4300& 250.000& 20.000&\cite{williams2010evaluating}\\   
         ~ & 4400& 230.000& 20.000&\cite{williams2010evaluating}\\   
         ~ & 4500& 240.000& 20.000&\cite{williams2010evaluating}\\   
         ~ & 4600& 240.000& 20.000&\cite{williams2010evaluating}\\    
         ~ & 4700& 260.000& 20.000&\cite{williams2010evaluating}\\    
        ~ & 4850& 204.000& 27.000&\cite{1991ApJS...75.1011G}\\   
         ~ & 5000& 192.000& 29.000&\cite{sulentic1976radio}\\  
         ~ & 5000& 230.000& 10.000&\cite{williams2010evaluating}\\   
         ~ & 5000& 240.000& 20.000&\cite{williams2010evaluating}\\
        ~ & 5200& 230.000& 10.000&\cite{williams2010evaluating}\\   
        ~ & 5400& 240.000& 10.000&\cite{williams2010evaluating}\\    
         ~ & 5700& 200.000& 20.000&\cite{williams2010evaluating}\\   
         ~ & 5950& 194.500& 8.000&\cite{leroy2011complex}\\    
     ~ & 6000& 220.000& 10.000&\cite{williams2010evaluating}\\    
         ~ & 6200& 290.000& 30.000&\cite{williams2010evaluating}\\    
         ~ & 6300& 200.000& 30.000&\cite{williams2010evaluating}\\  
         ~ & 6700& 230.000& 30.000&\cite{williams2010evaluating}\\  
          ~ & 7000& 190.000& 20.000&\cite{williams2010evaluating}\\    
         ~ & 8085& 170.000& 20.000&\cite{condon1983optical}\\
\hline
    \end{tabular}
      
    \end{center}
\end{table*}

\begin{table*}
\centering
    \begin{tabular}{|c|c|c|c|c|l|}
      \hline
        Galaxy & Frequency & Flux density & Flux density error & Reference \\
   ~ &(MHz) &(milliJy)& (milliJy)& \\

      (1)&(2)&(3)&(4)&(5)\\
      \hline
 
        \hline
IRAS 23233+2817 & 150& 101.970& 10.197&\cite{GMRTSurvey} \\
~ & 887& 48.000& 3.860&\cite{2020PASA...37...48M} \\
~ & 1400& 35.500& 1.200& \cite{1998AJ....115.1693C} \\
~ & 1400& 35.100& 1.053&\cite{2015ApJ...801...26H}\\
~ & 3000& 19.221& 0.576& \cite{2021ApJS..255...30G}\\ 

\hline

 IRAS F23234+0946& 150 & 39.140& 3.914 & \cite{GMRTSurvey} \\ 
 ~ & 887 & 14.000& 1.480& \cite{2020PASA...37...48M} \\ 
 ~ & 1400 & 11.600& 1.000& \cite{1998AJ....115.1693C}\\ 
 ~ & 1400& 10.900& 0.327&\cite{2015ApJ...801...26H}\\
 ~ & 3000 & 7.527& 0.225 & \cite{2021ApJS..255...30G}\\ 
 ~ & 15000 & 2.400& 0.120& \cite{nagar2003agn} \\  
 \hline
\end{tabular}

\vspace{1ex}

\smallskip
\text The table presents flux density measurements for the ULIRGs in our sample at multiple radio frequencies. The columns are as follows: (1) Source name; (2) Observing frequency in MHz; (3) Measured flux density in milliJanskys (mJy); (4) Associated uncertainty in flux density (mJy); and (5) Reference for the measurement, including published surveys, VLA archival images, and the present work.

\end{table*}

\subsection{Radio-FIR correlation}

An analysis of radio-FIR correlation was conducted to identify whether the sources show characteristics of starbursts or AGN \citep{2001ApJ...554..803Y}. This correlation reflects the connection between radio and FIR emissions due to star formation in the galaxy. Relativistic electrons in supernova remnants produce non-thermal synchrotron emission, while short-lived massive stars generate these remnants. These stars also emit in the FIR due to re-radiation by dust heated by ultraviolet (UV) emission \citep{2012ApJ...756..141B}. Non-linearity in the correlation may arise if the radio luminosity is not proportional to the star formation activity. The degree of dispersion or non-linearity can be quantified using the ‘\( q_\nu \)’ parameter (logarithmic FIR to radio flux density ratio) as described by \citet{1990ApJS...73..359C,1991ApJ...376...95C}. In the IRAS 2 Jy Sample \citep{2001ApJ...554..803Y}, the mean ‘\( q_\nu \)’ value is approximately 2.34. 
The points below \( q_\nu \) = 2.34 signify radio-dominated emission, and those above indicate IR-dominated emission. We carried out this analysis at the 1.4 GHz frequency band, as the flux density at this frequency is available for all the sources. The 1.4 GHz luminosity was calculated using the following relation:

\begin{equation}
{\log L_{1.4 \text{ GHz}} (\text{W Hz}^{-1}) = 20.08 - 2 \log D + \log S_{1.4 \text{ GHz}}}
\end{equation} 
where \( D \) is the distance in Mpc and \( S_{1.4 \text{ GHz}} \) is the flux density in units of Jy. Similarly, we estimated the luminosity at 60 \(\mu m\) using the following relation:

\begin{equation}
{\log L_{60 \mu m} (\text{L}_\odot) = 6.014 - 2 \log D + \log S_{60\mu m}}
\end{equation} 
According to \citet{Helou1985, hayashi2021radio}, the correlation between the radio and FIR luminosity, denoted as \( q_\nu \), is given by the following equation:

\begin{equation}
q_\nu = \log\left[\left( \frac{f_{\text{FIR}}}{3.75 \times 10^{12} \text{ Hz}} \right) / f_\nu\right]
\end{equation}

 where \( f_{\text{FIR}} \) is the FIR flux density in units of  \( \text{W m}^{-2} \text{Hz}^{-1} \) commonly defined as,

\begin{equation}
f_{\text{FIR}} = 1.26 \times 10^{14} \left( 2.58 \times f_{60} + f_{100} \right)
\end{equation}
The radio flux densities at 1.4\,GHz ($f_\nu$) and their associated errors, estimated from radio images, along with the flux densities and errors at 60\,$\mu$m and 100\,$\mu$m, were collected from \citet{Moshir1990}. The FIR-to-radio ratios at \(1.4 \, \text{GHz}\) and the corresponding values of \(q_\nu\) for the 38 sources are tabulated in Table~\ref{qtable}. IRAS 12265+0219 stands out as the most luminous source in our sample, with \(\log L_{1.4\text{GHz}} = 29.397\,\text{WHz}^{-1}\) followed by IRAS 13451+1232, which has \(\log L_{1.4\text{GHz}} = 28.311 \, \text{WHz}^{-1}\). The estimated values of \(q_\nu\) for these two sources are \(-1.22\) and \(-0.361\), respectively.
The right panel of Figure \ref{FIGURE1} displays a scatter plot of $q_\nu$ values versus $\log (L_{60\,\mu\mathrm{m}})$ for all 38 ULIRGs, with filled circles indicating the 13 sources from the first phase of this study \citep{2021MNRAS.503.5746N}. The green horizontal line represents the average $q_\nu$ value of 1.88, while the red line marks the \citet{2001ApJ...554..803Y} threshold of $q_\nu = 2.34$, below which sources are classified as radio-excess and above which they exhibit infrared-excess characteristics. Approximately 80\% exhibit radio-excess characteristics with \( q_{\nu} < 2.34 \).

\begin{figure*} 
\centering

\includegraphics[width=15.5cm]{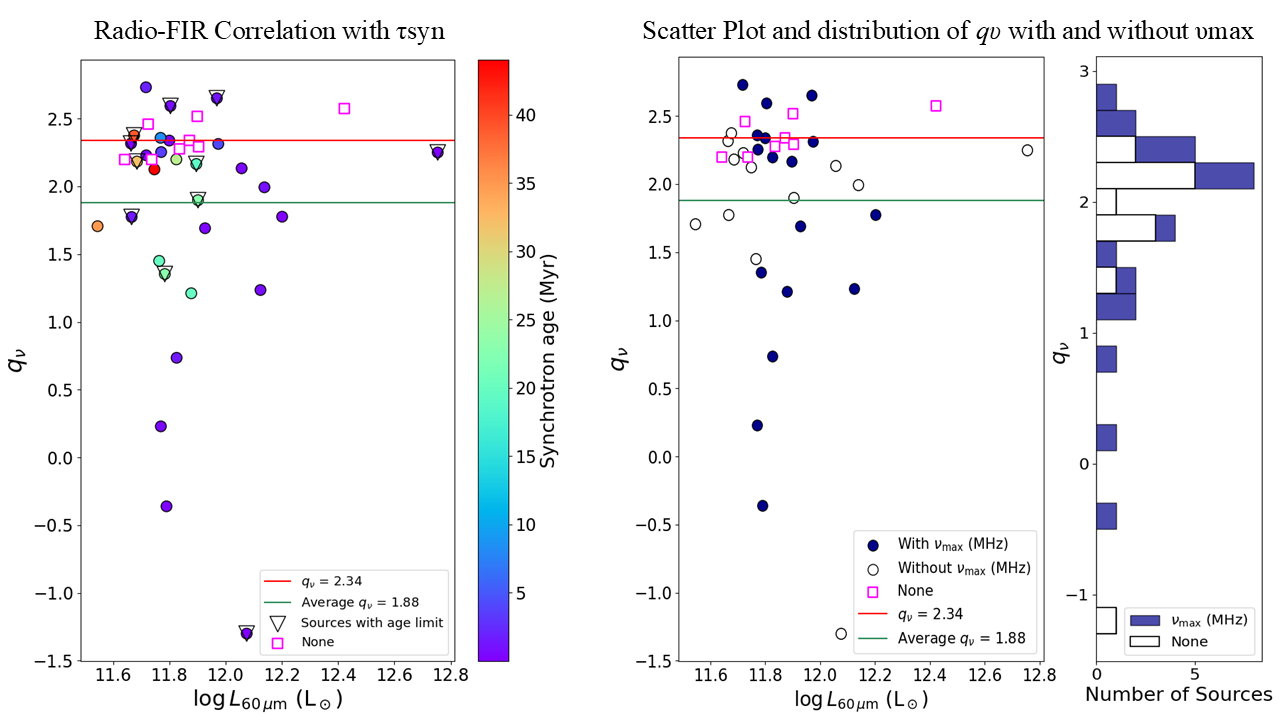}
\caption{The figure illustrates the relationship between $q_{\nu}$ and $\log L_{60\,\mu\mathrm{m}}$ in the first panel based on synchrotron age, where scatter points are color-coded by radio synchrotron age, and in the second panel based on turnover frequency, where scatter points are classified according to turnover frequency. The third panel shows a histogram of turnover classifications based on the second panel. The green line in first and second panels represents the average \(q_\nu\) value for the entire sample, while the red line marks the threshold at \(q_\nu = 2.34\). According to \citet{2001ApJ...554..803Y}, sources below this threshold are classified as radio-excess, while those above are IR-excess. In the second panel, filled circles indicate sources with turnover frequency, open circles represent sources without a turnover frequency. Squares in the first and second panels denote sources for which spectral age could not be determined due to insufficient data.}

\label{AGETURNOVERHISTOGRAM}

\end{figure*}

In Figure~\ref{AGETURNOVERHISTOGRAM}, the layout of the first and second panels illustrates the relationship between the radio-infrared flux ratio ($q_\nu$) and the luminosity $\log(60\,\mu\mathrm{m})$ for the 38 sources in our sample. The first panel presents scatter points color-mapped by radio synchrotron age, while the second panel displays scatter points categorized by turnover frequency. The third panel is a histogram showing the number of sources in the sample that exhibit turnover characteristics. Out of the 38 sources, 30 sources with estimated radio spectral ages are represented by filled circles, and color-mapped according to their ages. The remaining 8 sources, for which age estimates could not be obtained, are shown as hollow squares. Among the 30 sources with age estimates, 25 (83.3\%) exhibit \( q_{\nu} < 2.34 \). Of these 25 sources, most sources (64.0\%) appear blue in the age color map, indicating younger spectral ages. This suggests that a significant fraction of the radio-excess sources with spectral age estimates, are in an early evolutionary stage.

The second panel presents a scatter plot of turnover frequencies for all 38 sources. Most of the radio-excess AGNs identified in the first panel also show a low-frequency turnover, indicating their young age. There are 8 sources that exhibit infrared-excess characteristics, defined by \( q_{\nu} > 2.34 \). Among these, 6 sources appear blue in the first panel, suggesting young spectral ages. Interestingly, these same 6 sources display very low turnover frequencies in the second panel, which was unexpected.

The third panel shows a histogram of the turnover frequencies, providing a clearer view of the overall distribution corresponding to the scatter plot in the second panel. It reveals that 17 sources in our sample exhibit turnover characteristics. 

\begin{table*}
\centering

\caption{Synage Output Parameters, Estimated Magnetic Fields, Synchrotron Ages, and Logarithmic Luminosities at 6 GHz for the ULIRGs in the Sample.}
\label{agetable}

\vspace{1ex}

\begin{tabular}{cccccccc}
\hline
Galaxy & $\alpha_{inj}$ & $\nu_{\rm br}$ & ${\chi}^2_{red}$ & B & $\tau_{\rm syn}$ & $\nu_{\rm max}$ & $\log L_{6\,\mathrm{GHz}}$ \\
       &                & (GHz)          &                   & (nT) & (Myr)         & (MHz)           & (W Hz$^{-1}$) \\
(1) & (2) & (3) & (4) & (5) & (6) & (7) & (8) \\
\hline

IRAS 00188$-$0856$\blacklozenge$ & -0.533 & 5.38    &1.03 & 2.8        &4.3   &263 & 22.51 \\  

IRAS 00456$-$2904 & -0.375 &  2.236 &1.000& 1.004   &  27 &276.5&22.64\\

IRAS F01004$-$2237$\dagger$&-0.980&   & &    &  &  &22.34\\

IRAS 04194$-$1845$\dagger$&-0.661&   & &    &  &  &22.19\\

IRAS 01569$-$2939& -0.613 & 1.703   &5.162 & 1.39  &  20  & 63.49&  23.57 \\

IRAS 01572+0009$\blacklozenge$ & -0.757 &24.60    &16.44 & 4.8        &0.87  &&22.81 \\

IRAS 02480$-$3745$\dagger$&-0.891&   & &    &  & & 22.42\\

IRAS 04103$-$2838$\dagger$& -0.2304 &  & &  &  &  & 22.53\\

IRAS 05156$-$3024& -0.582 & $>100$  & 6.375 & 0.97  &  <23   & &    22.81  \\

IRAS 05189$-$2524& -0.375 & 4.317 &12.92 & 5.2  &  1.9  &361.1  & 21.87 \\

IRAS F07599+6508& -0.671 &2.156   & 6.448&  28.5 &  0.21 & 187.6&22.96\\

IRAS 08559+1053&-0.375  &  1.039 &1.120&  0.8  & 44 & &22.61\\

IRAS 10494+4424 & -0.528 & 3.824  & 17.47  &24.8   &  0.19  &207.7  &22.34\\

IRAS 11095$-$0238& -0.850 & $>100$ &11.79&  1.16  & <19 & 224.0& 22.31\\

IRAS 11119+3257& -1.000 & 1.638   &2.187 & 13.4  & 0.73  & 382.6&  23.40   \\

IRAS 12032+1707& -0.770 &  4.503 & 3.597&   26 & 0.16 &213.5 &23.10\\

IRAS 12112+0305$\blacklozenge$ & -0.546 & $>100$   &1.929&4.34   &$<$1.12  &290  &22.14\\

IRAS 12127$-$1412& -0.593 & $>100$ & 9.807&  1.25 &   $<$23 &318.5 &  23.17 \\

IRAS 12265+0219&-0.375 & $>100$  &2.063 & 105   & <0.0093 && 26.33\\


IRAS 13305$-$1739$\blacklozenge$ & -0.591 & 1760   &0.95& 1.4       &20    & & 23.07\\

IRAS 13335$-$2612&-0.476& $>100$& 6.693&  0.74  & <32 &  &22.40\\

IRAS 13428+5608& -0.450& 4.891 &13.44 &  3.4 &  3.5  & 115.7&22.38\\

IRAS 13443+0802&-0.417& 1.997 & 6.075 &  18  &  0.43& & 22.43\\

IRAS 13451+1232& -0.375&7.666 & 5.198&   18.6 & 0.21 &359.1 & 25.07\\


IRAS 13539+2920&-0.570&  $>100$&   18.86&23 & <0.107   & & 22.19\\

IRAS 14060+2919&-0.525& $>100$& 13.49&  0.63  & <39 &  &22.17\\

IRAS 14070+0525$\blacklozenge$& -0.775&&&&&&22.54\\


IRAS 14394+5332$\blacklozenge$ & -0.768 & $>100$ &24.03&  4.07      &$<$1.5   && 22.58  \\ 

IRAS 15001+1433$\blacklozenge$& -0.513 & 12.400  &0.28  &10.39    &0.40  &  &22.76\\

IRAS 15327+2340&-0.375 & $>100$& 2.638&  80  & <0.76 & 291.2 &22.13\\

IRAS 16156+0146$\blacklozenge$& -0.297&&&&&&22.39\\
IRAS 17028+5817$\blacklozenge$ & -0.665 & 2300 &2.01&    2.43    &   8.10   &299& 22.21  \\            
IRAS 17044+6720$\blacklozenge$& -0.254&&&&&&22.22 \\  

IRAS 17179+5444$\blacklozenge$& -0.734 &  5.290 &2.29& 5.9    &1.4  &509& 23.82\\

IRAS 23060+0505$\blacklozenge$& -0.451&&&&&&22.44\\

IRAS 23233+2817&-0.375&  1.215 & 0.9466& 1.0   &  35& &22.82\\

IRAS F23234+0946 & -0.498 & 16.343 & 1.075 & 19 & <0.14 & &23.38\\

IRAS 23389+0300$\blacklozenge$& -0.740 &2.500   &3.630&12.4    &0.66  &177& 24.21\\
\hline
\end{tabular}

\vspace{1ex}

\smallskip
\

\text The table presents derived physical properties of the ULIRG sample. The columns are as follows: (1) Source name: sources marked with  ($\blacklozenge$): 13 sources were extracted from the first phase study by \citet{2021MNRAS.503.5746N}; sources marked with ($\dagger$) have their spectral indices estimated using two different methods: three sources were fitted with a power law model using \texttt{gnuplot}, while one source had its two-point spectral index calculated; (2) Injection spectral index; (3) Spectral break frequencies in MHz; (4) Reduced \(\chi^2\) values for the goodness of fit; (5) Magnetic fields in nT; (6) Synchrotron age in Myr; (7) Low-frequency turnover in MHz; (8) Logarithmic radio luminosity at $\mathrm{6\,GHz}$ in units of W Hz$^{-1}$.

\end{table*}

\begin{table*}
\centering

\caption{Radio and infrared flux densities, logarithmic luminosities, and FIR-to-radio ratios (\(q_\nu\)) for the ULIRGs in the Sample.}

 \label{qtable}

\vspace{1ex}

   \begin{tabular}{cccccccccc}
  \hline
     Galaxy & $S_{1.4\mathrm{GHz}}$ & $\log L_{1.4\mathrm{GHz}}$ & $S_{60\mu m}$ & err$_{S_{60\mu m}}$ & $\log L_{60\mu m}$ & $S_{100\mu m}$ & err$_{S_{100\mu m}}$ & FIR & $q_\nu$ \\ 
~ & ($10^{-28}\ \mathrm{W\,m^{-2}\,Hz^{-1}}$) & (W Hz$^{-1}$) & (Jy) & (Jy) & $(L_\odot)$ & (Jy) & (Jy) & ~ & \\

        (1)&(2)&(3)&(4)&(5)&(6)&(7)&(8)&(9)&(10)\\

         \hline
        
      IRAS 00188$-$0856$\blacklozenge$& $1.65\pm0.05$ & 23.844&2.592E+00 & 2.33E-01 &11.974 & 3.403E+00 & 3.40E-01 & 1.27E-13    & 2.313  \\

        IRAS 00456$-$2904 & $2.14\pm0.06$  & 23.806 & 2.598E+00 & 1.82E-01 & 11.825&3.377E+00 & 3.71E-01 & 1.27E-13          & 2.199  \\

        IRAS F01004$-$2237$\dagger$ & $1.37\pm0.06$  & 23.678 & 2.287E+00 & 1.60E-01 & 11.835&1.790E+00 & 1.79E-01 & 9.69E-14 & 2.276  \\

        IRAS 01494$-$1845$\dagger$ & $0.80\pm0.02$  &23.729 & 1.288E+00 & 9.02E-02 &11.871&1.846E+00 & 2.58E-01 & 6.51E-14   & 2.337  \\

        IRAS 01569$-$2939 & $12.39\pm0.37$  &24.797 & 1.734E+00 & 1.39E-01 &11.878& 1.514E+00 & 1.82E-01 & 7.54E-14           & 1.211 \\ 
        
       IRAS 01572+0009$\blacklozenge$& $2.70\pm0.09$  & 24.236& 2.224E+00 & 1.78E-01 &12.138& 2.164E+00 & 2.81E-01 & 9.96E-14   & 1.993 \\
        
        IRAS 02480$-$3745$\dagger$ & $1.31\pm0.03$ &23.985 & 1.252E+00 & 6.26E-02 & 11.900&1.487E+00 & 1.34E-01 & 5.94E-14   & 2.519  \\ 
        
        IRAS 04103$-$2838$\dagger$ & $1.37\pm0.02 $&23.678 &1.823E+00 & 9.12E-02 & 11.737&1.707E+00 & 1.71E-01 & 8.08E-14    & 2.196  \\ 
        
        IRAS 05156$-$3024 & $1.87\pm0.05$  &24.174& 1.162E+00 & 6.97E-02 &11.902 &1.402E+00 & 1.68E-01 & 5.54E-14            & 1.898  \\ 
        
        IRAS 05189$-$2524 & $2.91\pm0.08$  &23.111& 1.367E+01 & 5.47E-01 &11.717 &1.136E+01 & 5.68E-01 & 5.88E-13             & 2.731  \\ 

         IRAS F07599+6508 &$4.17\pm0.12$~& 24.384& 1.692E+00 & 8.46E-02 & 11.927&1.730E+00 & 1.90E-01 & 7.68E-14           & 1.691 \\ 
        
        IRAS 08559+1053 & $1.22\pm0.03$  &23.850& 1.119E+00 & 6.71E-02 & 11.747&1.952E+00 & 2.54E-01 & 6.10E-14            & 2.125  \\

        IRAS 10494+4424 & $2.24\pm0.08$ &23.668 &3.527E+00 & 2.12E-01 & 11.800&  5.412E+00 & 3.25E-01 & 1.83E-13           & 2.338  \\ 
        
        IRAS 11095$-$0238 & $2.51\pm0.07$  &23.849& 3.249E+00 & 1.62E-01 & 11.896&2.531E+00 & 2.53E-01 & 1.37E-13     & 2.165  \\
        
        IRAS 11119+3257 & $11.00\pm0.33$ &25.030 &1.588E+00 & 1.75E-01 & 12.124&1.523E+00 & 1.68E-01 & 7.08E-14              & 1.235  \\
        
        IRAS 12032+1707 & $2.84\pm0.08$ &24.587& 1.358E+00 & 9.51E-02 & 12.201&1.543E+00 & 1.85E-01 & 6.36E-14             & 1.776  \\
        
        IRAS 12112+0305$\blacklozenge$ &$ 2.40\pm0.12$  &23.537& 8.503E+00 & 5.10E-01 & 11.969&9.976E+00 & 7.98E-01 & 4.02E-13   & 2.650  \\ 
        
        IRAS 12127$-$1412 & $7.60\pm0.22$ &24.543 & 1.538E+00 & 9.23E-02 &11.784 &1.134E+00 & 1.81E-01 & 6.43E-14            & 1.353  \\
        
        IRAS 12265+0219 &$5499.2\pm110.95$ &27.699& 2.060E+00 & 1.44E-01 & 12.075&2.891E+00 & 2.02E-01 & 1.03E-13     & -1.300 \\

     IRAS 13305$-$1739$\blacklozenge$& $4.83\pm0.15 $ &24.445& 1.164E+00 & 9.31E-02 &11.764 &1.044E+00 & 2.51E-01 & 5.09E-14     & 1.450 \\ 
        
        IRAS 13335$-$2612 &$ 1.26\pm0.03$  &23.704 &1.402E+00 & 1.12E-01 & 11.685&2.101E+00 & 2.10E-01 & 7.20E-14            & 2.183  \\
        
        IRAS 13428+5608 & $14.5\pm0.52 $ &23.662& 2.174E+01 & 8.70E-01 &11.771& 2.138E+01 & 8.55E-01 & 9.76E-13           & 2.253  \\

        IRAS 13443+0802 & $1.05\pm0.03$&23.691 & 1.300E+00 & 1.43E-01 & 11.718&1.940E+00 & 1.74E-01 & 6.67E-14           & 2.229 \\
        
        IRAS 13451+1232 & $539.70\pm16.19$  &26.305& 1.916E+00 & 2.11E-01 & 11.790&2.060E+00 & 1.85E-01 & 8.82E-14            & -0.361  \\

        IRAS 13539+2920 & $1.21\pm0.03$ &23.550 & 1.830E+00 & 1.28E-01 &11.664 &2.730E+00 & 1.91E-01 & 9.39E-14           & 2.316  \\ 
        
        IRAS 14060+2919 & $0.93\pm0.02$  &23.502& 1.611E+00 & 1.61E-01 &11.675 &2.417E+00 & 1.69E-01 & 8.28E-14 & 2.376  \\ 
        
        IRAS 14070+0525$\blacklozenge$&$ 0.50\pm0.02$  & 24.023 & 1.447E+00 & 8.68E-02 &12.422 &1.821E+00 & 1.64E-01 & 7.00E-14 & 2.572  \\

      IRAS 14394+5332$\blacklozenge$ & $4.20\pm0.21$  &  24.064&1.954E+00 & 7.82E-02 &11.666 &2.395E+00 & 1.20E-01 & 9.37E-14 & 1.774\\ 
        
        IRAS 15001+1433$\blacklozenge$ & $1.70\pm0.01$ &24.080& 1.871E+00 & 9.36E-02 & 12.057&2.043E+00 & 1.84E-01 & 8.66E-14 & 2.133  \\
        
        IRAS 15327+2340 & $32.60\pm0.97$  &23.367& 1.038E+02 & 4.15E+00 & 11.804&1.124E+02 & 3.37E+00 & 4.79E-12 & 2.593  \\ 
        
       IRAS 16156+0146$\blacklozenge$& $0.83\pm0.04 $ & 23.574&1.126E+00 & 6.76E-02 & 11.641&1.003E+00 & 2.21E-01 & 4.92E-14 & 2.199  \\ 
        
        IRAS 17028+5817$\blacklozenge$& $1.50\pm0.01$ & 23.625& 2.430E+00 & 1.46E-01 & 11.769&3.910E+00 & 1.96E-01 & 1.28E-13 & 2.358  \\ 
        
      IRAS 17044+6720$\blacklozenge$& $0.50\pm0.02$  &23.382& 1.281E+00 & 6.41E-02 & 11.725&0.978E+00& 1.57E-01 & 5.40E-14 & 2.459  \\ 
        
        IRAS 17179+5444$\blacklozenge$& $33.60\pm1.70$  &25.284& 1.362E+00 & 6.81E-02 & 11.826&1.915E+00 & 1.53E-01 & 6.84E-14 & 0.735  \\
        
       IRAS 23060+0505$\blacklozenge$& $0.65\pm0.03$  &23.720& 1.152E+00 & 8.06E-02 &11.904& 8.329E-01 & 1.75E-01 & 4.79E-14 & 2.294  \\
        
        IRAS 23233+2817 & $3.55\pm0.12$ & 24.059&1.262E+00 & 1.26E-01 & 11.545&2.107E+00 & 3.37E-01 & 6.76E-14 & 1.706 \\ 
        
        IRAS F23234+0946 & $1.16\pm0.01$  &23.691 & 1.561E+00 & 9.37E-02 & 12.754&2.108E+00 & 2.95E-01 & 7.73E-14 & 2.250 \\
        
        IRAS 23389+0300$\blacklozenge$ & $86.30\pm2.60$  &25.680 &1.233E+00 & 1.48E-01 & 11.770&1.171E+00 & 2.58E-01 & 5.48E-14 & 0.229  \\

         \hline 
         
\end{tabular}

\vspace{1ex}

\smallskip
\text{The table presents derived radio and infrared flux densities, logarithmic luminosities, and FIR-to-radio ratios ($q_{\nu}$) for the ULIRG sample. The columns are as follows:} (1) Source name: sources names marked with ($\blacklozenge$): 13 sources were extracted from \citet{2021MNRAS.503.5746N}; sources marked with ($\dagger$) have their spectral indices estimated using two different methods: three sources were fitted with a power-law model using \texttt{gnuplot}, while one source had its two-point spectral index calculated; (2) Radio flux of the object at $\mathrm{1.4\,GHz}$ in units of $10^{-28}\ \mathrm{W\,m^{-2}\,Hz^{-1}}$ \citep{1998AJ....115.1693C}; (3) Logarithmic radio luminosity at $\mathrm{1.4\,GHz}$ in units of W Hz$^{-1}$; (4) Flux at $60\,\mu m$ \citep{Moshir1990}; (5) Error in flux at $60\,\mu m$ \citep{Moshir1990}; (6) Logarithmic luminosity at $60\,\mu m$ in units of $L_\odot$; (7) Flux at $100\,\mu m$ \citep{Moshir1990}; (8) Error in flux at $100\,\mu m$ \citep{Moshir1990}; (9) FIR-to-radio ratio \citep{2001ApJ...554..803Y}; (10) '\(q_\nu\)' factor at $\mathrm{1.4\,GHz}$ calculated as suggested in \citet{2001ApJ...554..803Y}.

\end{table*}

\subsection{Optical Interaction Class and Analysis of WISE Data}
The interaction stage of each ULIRG in our sample is classified as pre-merger, merger, or post-merger based on the optical morphological criteria of \citet{Veilleux_2002}. A detailed description of the classification scheme is provided in  Appendix~\ref{appendix:interaction}. The interaction class assigned to each source is listed in Table~\ref{merger}, and the Pan-STARRS optical images illustrating these stages are shown in Appendix Figures~\ref{panstar1} and \ref{panstar2}. Following the classification of sources based on their optical interaction stages, we investigated their mid-infrared properties using data from the WISE. We cross-matched the source positions with the WISE catalog and found that 37 out of 38 sources in our sample are detected. The only exception is IRAS F07599+6508, which is not detected in WISE. WISE employs a straightforward method to identify AGN based on the W1$-$W2 color cut criterion. According to \citet{Blecha2018}, galaxies with a W1$-$W2 color below 0.5 are classified as blue, suggesting potential starburst activity. A color between 0.51 and 0.8 suggests a possible presence of AGN (pre-merger phase), while a color greater than 0.8 confirms AGN activity (merger or post-merger phase). We applied this color-cut criterion to our sample and tabulated the magnitudes W1 and W2 for all sources in Table~\ref{merger}. To analyze the infrared color properties of our sample, we examined the W1$-$W2 color index. We found that four sources IRAS 13443+0802, IRAS 13451+1232, IRAS 17044+6720, and IRAS F23234+0946 have binary detections in WISE. For these sources, WISE provides two separate detections at the same RA and Dec coordinates, each with its own W1 and W2 magnitudes. As a result, we obtained two different W1$-$W2 values for each of these sources. Among binary-detected sources, we found that two have at least one W1$-$W2 value $\leq$ 0.5. Based on the classification by \citet{Blecha2018} and \citet{Assef2013Mid}, most of the sources in our sample exhibit characteristics indicative of mid-infrared AGNs (see Figure~\ref{Wise,age,classplot}).

\begin{figure*}
\centering
\includegraphics[width=16.5cm,height =8cm]{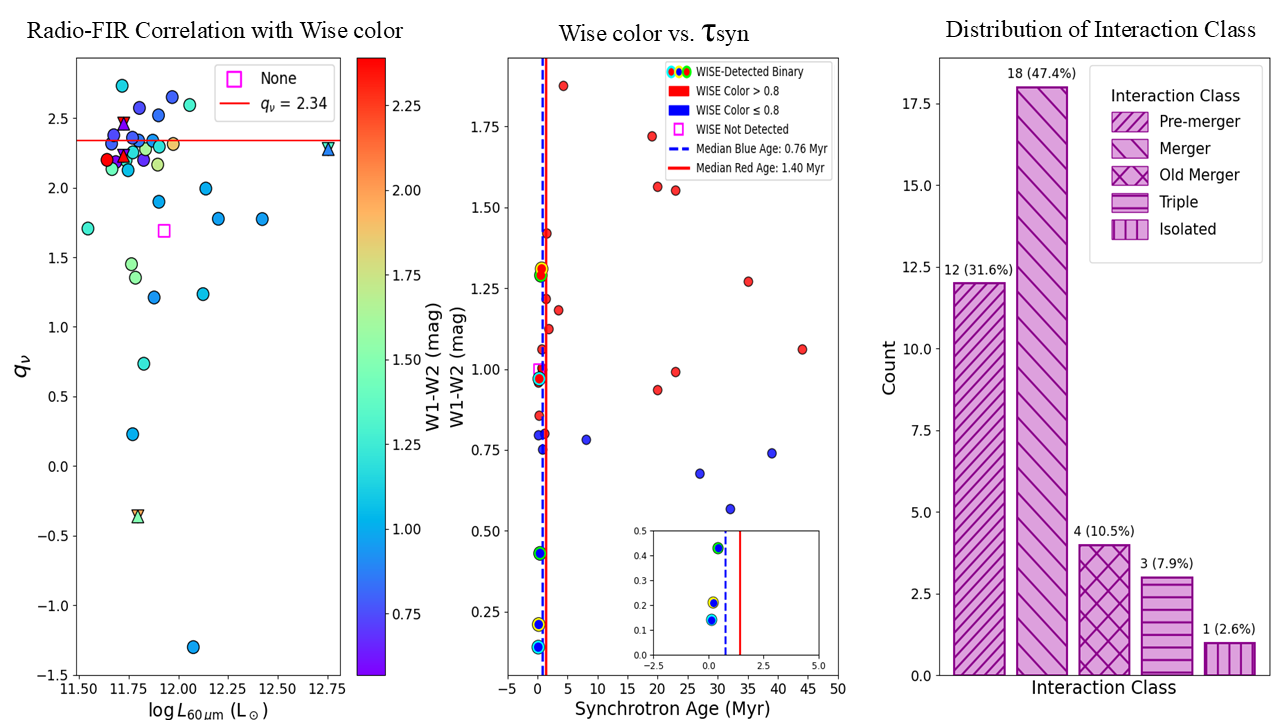}

\caption{
The first panel shows a scatter plot of $q_{\nu}$ vs. $\log L_{60\mu\mathrm{m}}$, with points classified based on the WISE color cut criteria (\( W1 - W2\)) for AGN detection, as defined by \citet{Blecha2018}. Upward and downward triangles indicate sources detected as binaries in the WISE catalog, located at the same RA and Dec coordinates but exhibiting different WISE magnitudes. A square indicates the source not detected in the WISE catalog. The red horizontal line marks the average $q_{\nu}$ value of 2.34 \citep{2001ApJ...554..803Y}. The second panel presents a scatter plot of WISE color cut (\( W1 - W2\)) values against synchrotron age. The WISE color cut categorizes sources as follows: values $\leq 0.8$ correspond to star formation or a possible AGN presence, while values $> 0.8$ confirm AGN activity, typically associated with the merger or post-merger phase. Binary-detected sources with two different WISE color values are represented by overlaid concentric circles, each outlined in a distinct color. The blue and red vertical lines mark the median synchrotron ages of sources with \( W1 - W2 \leq 0.8 \) (0.76 Myr) and \( W1 - W2 > 0.8 \) (1.40 Myr) respectively. An inset within this panel shows a zoomed view of the synchrotron age range $-2.5$ to $5$~Myr, highlighting the separation between the median age distributions. The third panel displays a bar plot showing the distribution of interaction classes, based on optical classification \citep{Veilleux_2002}, expressed as a percentage of the total sample.}

\label{Wise,age,classplot}
\end{figure*}

 In Figure~\ref{Wise,age,classplot}, we present a multi-panel illustration of the WISE color analysis in relation to various properties of our sample. The first panel displays a scatter plot illustrates the distribution of \(q_\nu\) values plotted against the logarithm of luminosity at ${60\mu m}$ for our sample. The scatter data points are color-coded based on mid-infrared (mid-IR) WISE color selection criteria. Upward and downward triangles denote sources detected as binaries in the WISE catalog, located at identical RA and Dec coordinates. The source IRAS F07599+6508, which is not detected in WISE, is marked by a hollow square. The red line in this plot is the average $q_{\nu}$ value of 2.34.

The second panel presents scatter plot of synchrotron spectral age versus WISE \( W1 - W2 \) color. Points are color-coded such that blue indicates \( W1 - W2 \leq 0.8 \), which typically suggests either star-forming galaxies or obscured AGN activity, while red denotes \( W1 - W2 > 0.8 \), generally associated with AGN-dominated systems. The blue vertical line represents the median synchrotron age (0.76 Myr) for sources with \( W1 - W2 \leq 0.8 \), and the red vertical line marks the median age (1.40 Myr) for sources with \( W1 - W2 > 0.8 \). Binary detected sources for which spectral ages have been estimated are further highlighted using concentric circles: IRAS 13443+0802 (lime green), IRAS 13451+1232 (yellow), and IRAS F23234+0946 (cyan). According to \citet{Veilleux_2002} and \citet{Rodr2009}, our sample includes 12 pre-merger sources, 18 merger-stage sources, and 4 post-merger sources, classified based on optical wavelength observations. The interaction classification for each source is provided in Table \ref{merger}. In this second panel, sources with \( W1 - W2 \leq 0.8 \) suggest a possible AGN presence, typically associated with the pre-merger phase. Whereas sources with \( W1 - W2 > 0.8 \) indicate confirmed AGN activity, corresponding to the merger and post-merger phases. Most of the sources in our sample are in the merger stage and have \( W1 - W2 > 0.8 \). The majority of sources in this plot with \( W1 - W2 > 0.8 \) have lower synchrotron ages, suggesting a connection between the WISE color, the merger stage identified in optical observations, and radio synchrotron aging.

The third panel displays a bar plot showing the distribution of optical interaction classes, based on the classification scheme from \citet{Veilleux_2002}, expressed as a percentage of the total sample. Most of the sources in our sample are in the merger stage and exhibit \( W1 - W2 > 0.8 \). The majority of these sources also exhibit relatively lower synchrotron ages, suggesting a possible connection between the WISE color, the merger stage (as identified through optical observations), and radio synchrotron aging. Specifically, the median synchrotron age for sources with \( W1 - W2 > 0.8 \) (red points), which tend to be in the merger or post-merger stages, is 1.46~Myr, indicating older radio-emitting plasma. In comparison, sources with \( W1 - W2 \leq 0.8 \) (blue points), typically associated with the pre-merger stage, have a median age of 0.73~Myr, reflecting younger synchrotron plasma. This trend suggests a potential relationship between mid-infrared color, galaxy interaction stage, and the synchrotron age.

Although our sample size is small, the observed trend suggests a possible link between synchrotron age, WISE mid-infrared color, and the galaxy's merger stage. We expect that with a larger sample, this relationship will become clearer and more statistically significant.

\begin{table*}
\centering


\caption{Nuclear structures, separations, interaction classes, and WISE magnitudes of ULIRGs in the sample.}
\label{merger}

\vspace{1ex}

\begin{tabular}{ccccccccc}
 
\hline

Galaxy & Nuclear structure & NS (Opt) & NS (IR) & NS (RF) & Interaction class & $W1$ & $W2$ & $W1 - W2$ \\
& & arcseconds & arcseconds & arcseconds & & (mag) & (mag) & (mag) \\
(1)&(2)&(3)&(4)&(5)&(6)&(7)&(8)&(9)\\
\hline

\hline
IRAS 00188$-$0856$\blacklozenge$& S(Opt), S(IR), S(RF) & & & & Post merger&    12.318 & 10.442& 1.876     \\
IRAS 00456$-$2904 &D(Opt), S(IR), S(RF)& 8.327&& & Pre-merger&  12.725&12.048&     0.677      \\
IRAS F01004$-$2237 $\dagger$&S(Opt), S(IR), S(RF)&  && &Post merger& 12.887&11.256& 1.631     \\
IRAS 01494$-$1845$\dagger$ &S(Opt), S(IR), S(RF)&  && &Merger&   13.401&12.470&   0.931      \\
IRAS 01569$-$2939&S(Opt), S(IR), S(RF)&  & & &Merger& 13.877&12.942&0.935                        \\
IRAS 01572+0009$\blacklozenge$&S(Opt), S(IR), T(RF)& &&0.340&Merger& 10.790&9.792&   0.998    \\
IRAS 02480$-$3745$\dagger$&S(Opt), S(IR), S(RF)&  && & Merger &14.580&13.736&     0.844                   \\
IRAS 04103$-$2838$\dagger$&S(Opt), S(IR), S(RF)&  & &&  Merger& 12.321&11.035&       1.286              \\
IRAS 05156$-$3024&S(Opt), S(IR), S(RF)&  && & Merger&    13.497&12.506&        0.991                   \\
IRAS 05189$-$2524&S(Opt), S(IR), D(RF)  &  &&1.042&Merger  &  8.649&7.524&  1.125         \\
IRAS F07599+6508&  S(Opt), S(RF)&  && & Merger &                        \\
IRAS 08559+1053&S(Opt), S(IR), S(RF) &  &&  & Merger& 11.551&10.490&        1.061          \\
IRAS 10494+4424&S(Opt), S(IR), S(RF) && & &Merger &  12.766&11.909&0.857                 \\
IRAS 11095$-$0238&S(Opt), S(IR), S(RF)&&&& Merger &  13.679&11.959&  1.720                     \\
IRAS 11119+3257&S(Opt), S(IR), S(RF)&  &  &&Merger&  9.906&8.845&       1.061                    \\
IRAS 12032+1707&D(Opt), S(IR), S(RF)&2.278 & & &Pre-merger  &  14.065&13.105&   0.960   \\
IRAS 12112+0305$\blacklozenge$&D(Opt), S(IR), D(RF)& 2.484 && 2.721& Pre-merger&   12.257&11.455&       0.802     \\
IRAS 12127$-$1412& D(Opt), S(IR), S(RF)&7.170&&&Pre-merger&  10.891&9.338&    1.553             \\
IRAS 12265+0219&S(Opt), S(IR), S(RF)&&&&Merger&  8.395&7.425&           0.970                         \\

IRAS 13305$-$1739$\blacklozenge$&S(Opt), S(IR), E(RF)&& &&Post merger& 12.626&11.061&         1.565                \\
IRAS 13335$-$2612&D(Opt), S(IR), S(RF)&1.120&&&Pre-merger&   12.942&12.374&     0.568    \\
IRAS 13428+5608&D(Opt), S(IR), E(RF)&0.880&&1.243&Merger& 10.416&9.234& 1.182  \\
IRAS 13443+0802&T(Opt), D(IR), D(RF)&5.741&5.544&7.552&Triple&  12.73, 13.85&12.16, 12.56&    0.57, 1.29  \\
IRAS 13451+1232&D(Opt), D(IR), D(RF)&1.509&4.896&2.000&Pre-merger&  11.52, 13.42& 10.21, 12.72&    1.31, 0.7               \\

IRAS 13539+2920& D(Opt), S(IR), D(RF)&3.033&&&Pre-merger& 12.861&12.064&  0.797       \\
IRAS 14060+2919&S(Opt), S(IR), S(RF)&&& &Merger&  12.744&12.003&     0.741                  \\
IRAS 14070+0525$\blacklozenge$&S(Opt), S(IR), S(RF)&&&&Post merger&                      14.244&13.287&    0.957          \\
                       
IRAS 14394+5332$\blacklozenge$&T(Opt), S(IR), S(RF)& &&&Triple&  12.384&10.963&            1.421                 \\
IRAS 15001+1433$\blacklozenge$&T(Opt), S(IR), E(RF)&&&&Triple& 12.337&11.049&  1.288                          \\
IRAS 15327+2340 &D(Opt), S(IR), D(RF)&1.049&&2.462&Pre-merger& 10.079&9.327&0.752                \\
IRAS 16155+0146$\blacklozenge$&D(Opt), S(IR), S(RF)&2.475&&&Pre-merger&  13.105&10.715&  2.39         \\
IRAS 17028+5817$\blacklozenge$&D(Opt), S(IR), S(RF)&10.288&&&Pre-merger& 13.401&12.321&      0.783           \\
IRAS 17044+6720$\blacklozenge$&S(Opt), D(IR), S(RF)&&11.268&&Merger&  11.97, 14.96& 10.44, 14.86&         1.53, 0.1          \\
IRAS 17179+5444$\blacklozenge$&S(Opt), S(IR), S(RF)&&&&Merger&12.724&11.506&    1.218                         \\
IRAS 23060+0505$\blacklozenge$&S(Opt), S(IR), S(RF)&&&&Merger&                 9.809&8.614&  1.195     \\
IRAS 23233+2817& S(Opt), S(IR), S(RF)&&&&Iso&  12.512&11.241&    1.271                    \\
IRAS F23234+0946& D(Opt), D(IR), S(RF)&2.661&5.004&&Pre-merger& 13.66, 14.73& 12.69, 14.43&  0.97, 0.3                 \\
IRAS 23389+0300$\blacklozenge$& D(Opt), S(IR), D(RF)&1.460& & 0.208  &Pre-merger&  13.698&12.694& 1.004  \\
\hline
\end{tabular}
\vspace{1ex}

\smallskip
\text The table presents the sources in our sample along with their nuclear structures, nuclear separations in arcseconds, interaction classifications and WISE magnitudes. The columns are as follows: (1) Source name: sources names marked with ($\blacklozenge$): thirteen sources were extracted from \citet{2021MNRAS.503.5746N}; sources marked with ($\dagger$) have their spectral indices estimated using two different methods: three sources were fitted with a power-law model using \texttt{gnuplot}; (2) Nuclear structure in the optical (Opt), infrared (IR), and radio frequency (RF) bands, where S = single, D = double, T = triple and E = extended; (3), (4), and (5) Nuclear separation (NS) between two nuclei measured in the optical, IR, and radio bands, respectively, in arcseconds; (6) Interaction classes are based on optical morphological features, as classified by \citet{Veilleux_2002}; (7) \& (8) WISE magnitudes and (9) Calculated WISE color $(W1 - W2)$ in magnitudes \citet{Blecha2018}.
\end{table*}

\section{DISCUSSION}
\label{SECTION5}
 \subsection{\bf GPS/CSS Characteristics}
GPS and CSS sources are typically associated with young radio galaxies, characterized by a peaked radio spectrum \citep{OdeaSaikia2021,1995A&A...302..317F,1998PASP..110..493O, collier2018high}. Above the peak, the spectrum steepens in the optically thin region. In this study, we analyze the integrated radio spectrum of a large sample of ULIRGs. Table \ref{agetable} presents SYNAGE output parameters, including magnetic field and spectral age, for 38 sources (13 sources from previous work highlighted with ($\blacklozenge$), and 25 sources from the present work). According to our findings, the integrated radio spectra of 17 ULIRGs from our total sample show low-frequency turnover ranging from 63 to 509 MHz, with spectral age spanning from 0.0093 to 44 Myr. 
 The variation of $q_{\nu}$ values spectral age and turnover frequency for the whole sample is shown in Figure \ref{AGETURNOVERHISTOGRAM}. In this sample 8 sources indicate a very young spectral age, with estimated ages \( < 1\,\mathrm{Myr} \). We note very high break frequency for 11 sources. We consider their break frequency must be higher than the highest available frequency used for SYNAGE fitting. For these 11 sources, we estimate only an upper age limit. In our sample, the majority of sources appear compact, with unresolved radio structures. From available highfrequency data, we find that 8 sources have resolved radio structure. For most ULIRGs, the highest available angular resolution data are provided by VLASS at 3~GHz, and the redshift range of the sample is from 0.034 to 0.264. Since the VLASS resolution is 2.5", we cannot resolve the nearest  source of the sample with linear size $\lesssim$1 kpc. Similarly the most distant source with linear size $\lesssim$4 kpc can not be resolved by VLASS. Unavailability of resolved images limits reliable measurements of projected linear sizes. As a result, GPS/CSS classification is not possible for many sources from their linear size criteria. However, depending on the turn over frequency range below 400 MHz (except the source IRAS 17179$+$5444 which shows turnover at 509 MHz) and spectral age, our sample shows young CSS characteristics. Few sources may also appear as double nucleus system. The optical-radio contour overlays in Figure \ref{kntr} highlight the most 5 interesting sources within our sample. High-frequency radio data is necessary to confirm their GPS/CSS nature. The presence of several young compact GPS/CSS radio sources in a relatively large sample of ULIRG strongly suggests that they represent radio galaxies in early stages of evolution.

\subsection{Buried AGN}

Buried AGNs are active galactic nuclei whose central regions are heavily obscured by surrounding dust and gas \citep{imanishi2010spitzer}. Although the black hole is actively accreting and emitting energy, the radiation output, particularly in optical and X-ray wavelengths, is often obscured, making direct detection challenging.
In the previous section, we noted that six sources IRAS 00188$-$0856, IRAS 05189$-$2524, IRAS 10494$+$4424, IRAS 12112$+$0305, IRAS 15327$+$2340, and IRAS 17028+5817 exhibit \( q_{\nu} \) values greater than 2.34, suggesting that their far-infrared emission is likely dominated by starburst activity. Interestingly, when we modeled their radio spectral energy distributions (Figures~\ref{fig:synage4}, \ref{fig:synage5}, \ref{fig:synage6} and \ref{fig:synage7}), we found that the radio emission in these sources is primarily non-thermal, with clear low-frequency turnovers and young spectral ages. These are key features of compact, young radio AGNs. The combination of starburst dominated far-infrared emission and signs of AGN activity in the radio suggests that these galaxies may host young AGNs still embedded within dusty, star forming environments.
Figure~\ref{AGETURNOVERHISTOGRAM} highlights these findings. In the left panel, these six sources lie above the \( q_{\nu} = 2.34 \) threshold (left panel), suggesting starburst dominance, while appearing in blue in the age color map, indicating their youth. In the right panel (Figure \ref{AGETURNOVERHISTOGRAM}), all six sources show detectable turnover frequencies, supporting their classification as young radio AGNs. The corresponding spectral ages and turnover frequencies are listed in Table~\ref{agetable}, while FIR fluxes and \( q_{\nu} \) values are provided in Table~\ref{qtable}.

The presence of excess radio emission, even in systems with high \( q_{\nu} \) values, highlights the obscured nature of these AGNs. While they are not easily identified in the infrared alone, the shape of their radio spectra reveals the activity of the central engine.

Additional evidence from other wavelengths supports the interpretation of these sources as buried AGNs. According to the optical nuclear classifications from \citet{Veilleux_2002}, IRAS 05189$-$2524 is classified as a Seyfert~2 galaxy, while the other five sources are classified as LINERs. Both types are typically associated with obscured AGN activity. All six sources also have WISE \( W1 - W2 \) color values greater than 0.5: IRAS 00188$-$0856 (1.876), IRAS 05189$-$2524 (1.125), IRAS 10494$+$4424 (0.857), IRAS 12112$+$0305 (0.802), IRAS 15327$+$2340 (0.752) and IRAS 17028+5817 (0.783) satisfying common mid-infrared AGN selection criteria \citep{stern2012mid, Assef2013Mid, Blecha2018}. This suggests that, although some AGN signatures can be detected in the optical or mid-infrared, the central engines in these galaxies are largely hidden and are best revealed through their compact radio emission. In a recent study \citet{Hayashi2025} confirmed IRAS 00188$-$0856 as a buried AGN using  Very Long Baseline Array (VLBA) observations.

\subsection{Interaction Class Characteristics and WISE–Radio Trends}
In our sample, 31.6\% of ULIRGs are classified as pre-merger systems, 47.4\% as mergers, and 10.5\% as post-mergers. We compared the interaction classes defined by \citet{Veilleux_2002} with the estimated WISE color \( W1 - W2 \) to constrain different phases of galaxy evolution. As shown in Table \ref{merger}, we observe a correlation between the interaction class and WISE color \( W1 - W2 \) values in our sample. There is a significant trend between radio excess and \( W1 - W2 \) color, as illustrated in Figure \ref{Wise,age,classplot} (first panel), where redder objects tend to exhibit higher radio excess though a few exceptions exist. We also found a tendency for higher synchrotron age for color $>$ 0.8. As the WISE color goes blue to red, both the radio excess, and spectral age vary from low to high. A general trend is post merger interaction has higher WISE color. We also note the nuclear structure and its component separations in optical, IR and radio wavelengths for these interaction systems. This multi-wavelength comparison indicates that the systems are about to merge or have already merged (see in Table \ref{merger}).

In Table \ref{merger}, we provide details on the interaction classification, including nuclear structure and nuclear separation across optical, infrared, radio wavelengths and WISE magnitudes. The source IRAS 13443+0802, appears as a triple system in optical (Figure \ref{kntr}), but is observed as a double system in both radio and infrared. Its estimated \( W1 - W2 \) values are 0.57 and 1.29 (mag). Based on the WISE color, one component of this system is classified as an AGN, while the other exhibits characteristics of starburst activity. Similarly, IRAS 13451+1232 is identified as a double system across optical, radio, and infrared wavelengths. The estimated W1$-$W2 values for its components are 1.31 and 0.7, both consistent with AGN activity. The source IRAS 17044+6720, classified as a single system in optical according to \citet{Veilleux_2002}, is detected as a double system in the infrared. Its estimated \( W1 - W2 \) values are 1.53 and 0.1, indicating starburst and AGN activity, respectively. The \( W1 - W2 \) values for all sources are depend on their interaction class. Most of the sources in our sample are merger systems.

\section{Conclusions}
\label{sec:conclusion}

In this study, we have investigated different statistical properties of a large sample of ULIRGs. Multi-frequency radio data of ULIRGs provides a better quantitative understanding of their radio spectrum, spectral indices, spectral age, and morphology. We performed a nonlinearity test of radio-FIR correlation, estimated the WISE color, compared the interaction class, and checked the evolution process from ULIRG to young radio source.    

The spectral shape mostly follows a power-law with a high-frequency spectral break and low-frequency concave turnover between 63 to 509 MHz. It was possible to estimate the spectral age for 30 sources using archival data and new observations. The spectral age of 19 sources is $<$5 Myr and 17 sources show a low frequency turnover. There are 3 extended structures, 8 resolved double nuclei, and 1 triple component in radio. 14 sources show the double nuclear structure and 3 triple nuclear structures in the optical wave band \citep[based on the information from][]{Veilleux_2002}. 

Four sources show double components in the WISE catalog. The extended or resolved radio structures that are associated with a single optical or IR nuclear component, may evolve into a CSS sources. Whereas a single component in radio hosting more than a single nuclear component in optical/IR may have binary AGN. Double components in all three bands strongly indicate the possibility of binary AGN. We plan to perform high-frequency radio observations of these ULIRGs in the future. 

Most of the ULIRGs in our sample show a dispersion in the radio–FIR correlation and exhibit a radio excess of AGN origin. Radio excess and spectral age have a strong correlation with WISE color. ULIRGs with higher spectral ages and higher redder colors may follow the `frustration scenario' while the major population of our sample follow the `youth hypothesis' and they may evolve into GPS/CSS sources. We also note that 6 sources with a starburst origin have low-frequency turnovers. These ULIRGs may host buried AGNs.

\section*{Acknowledgements}

The authors gratefully acknowledge the anonymous referee for the thoughtful and constructive feedback, which has greatly contributed to improving this manuscript. S.N. acknowledges support by the Anusandhan National Research Foundation (ANRF) as well as the Science and Engineering Research Board (SERB), a statutory body of Department of Science and Technology (DST), Government of
India (Scheme: State University Research Excellence (SERB
SURE); File Number: SUR/2022/003864).
We thank the GMRT staff for technical support during the observations. GMRT is run by the National Centre for Radio Astrophysics of the Tata Institute of Fundamental Research.
This research has also made use of the NASA/IPAC Extragalactic Database (NED), which is operated by the Jet Propulsion Laboratory, California Institute of Technology, under contract with the National Aeronautics and Space Administration (NASA).
MD acknowledges the support of the Science and Engineering Research Board (SERB) Core Research Grant (CRG/2022/004531) and the Department of Science and Technology (DST) grant (DST/WIDUSHIA/PM/2023/25(G)) for this research.

\section*{Data Availability}
The data underlying this article can be shared on reasonable request
to the corresponding author. The GMRT raw data can be obtained
from the following link: https://naps.ncra.tifr.res.in/goa/data/search.

\bibliographystyle{mnras}
\bibliography{example}   

\appendix

 \section{Interaction Stage Classification}
\label{appendix:interaction}

Interaction class refers to the classification of galaxies based on their level of interaction or merger activity with other galaxies. According to \citet{Veilleux_2002} there are three common phases (Pre-merger, merger and post-merger) within the interaction class on the basis of optical wavelength. Pre-merger phase is the early stages of galaxy interaction, where  signs of gravitational attraction and tidal distortion due to their mutual gravitational influence can be seen. Here, the galaxies have begun to approach each other, but have not yet merged into a single entity. The detection possibility of two nuclei or dual AGN at optical, IR and radio wavelengths is high in this phase. During the merger phase, galaxies are in the process of colliding and merging with each other. This time galaxies undergo significant morphological changes, such as the formation of tidal tails, bridges, and distorted shapes. Eventually, they merge into a single, larger galaxy. Close binary nuclei or single nucleus can be detected at optical, IR and radio wavelengths. In Post-merger phase the galaxies have already undergone a significant merger event and have coalesced into a single system. While obvious tidal features such as tails may no longer be visible, often falling below the detection threshold, many post-merger systems still exhibit disturbed central morphologies, including knots of star formation, shells, or ring-like structures. The remnant typically appears as a relaxed, high surface brightness core. Over time, the galaxy may continue to evolve through processes such as ongoing star formation and dynamical relaxation. A new AGN jet can form during this stage, which may appear as a GPS or CSS source at radio wavelengths. The PANSTARRS optical images of each ULIRGs are shown in Figures~\ref{panstar1} and \ref{panstar2}, where we can see all these interaction features.

\section{Notes on interesting sources}
\label{appendix:interestingsources}
This appendix highlights ten particularly interesting ULIRGs from our sample that exhibit distinctive radio morphologies, clear merger signatures, or evidence of enhanced AGN activity. These sources are discussed in more detail in this section.
\begin{figure*}
    \centering
    \includegraphics[width=2.27in]{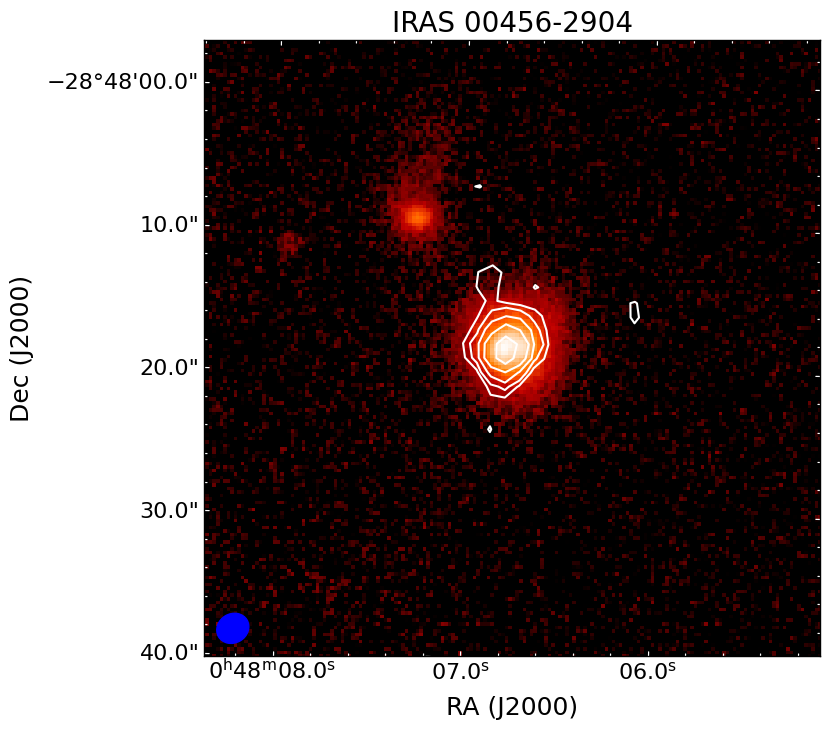} 
    \hfill
    \includegraphics[width=2.27in]{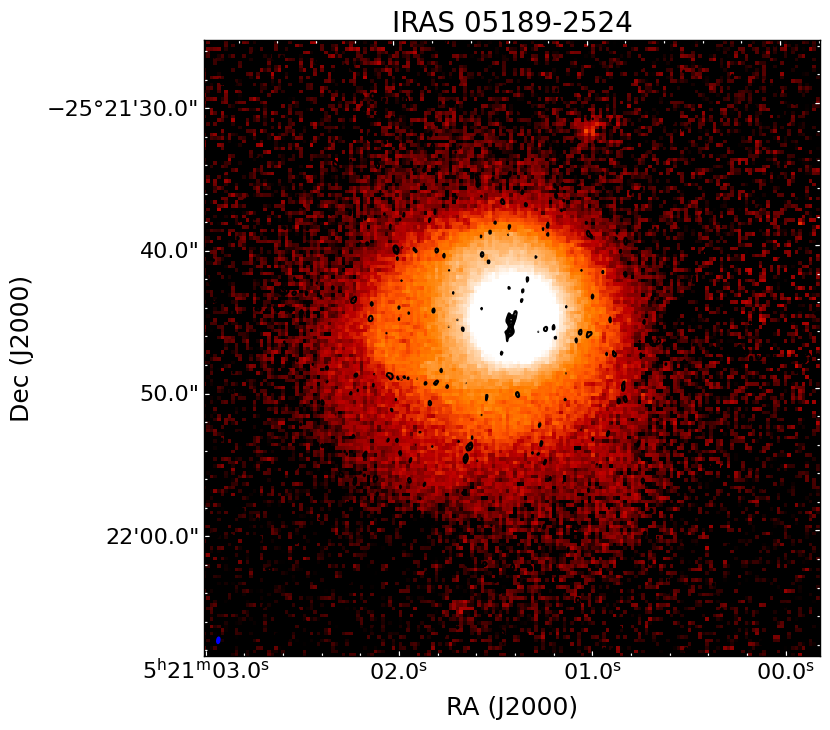} \hfill
    \includegraphics[width=2.27in]{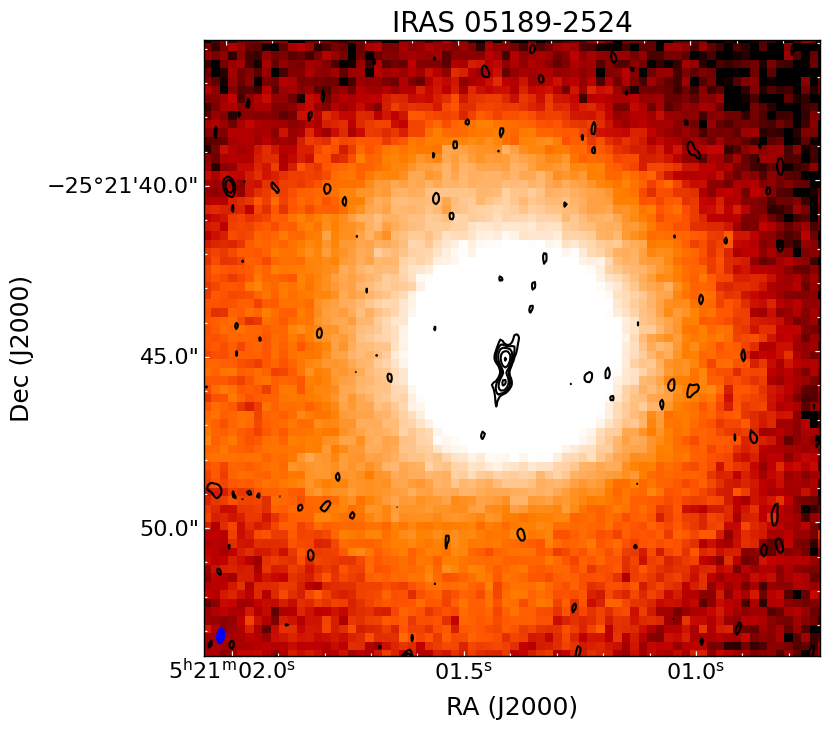} \\
    \vspace{0.1cm}
    \includegraphics[width=2.27in]{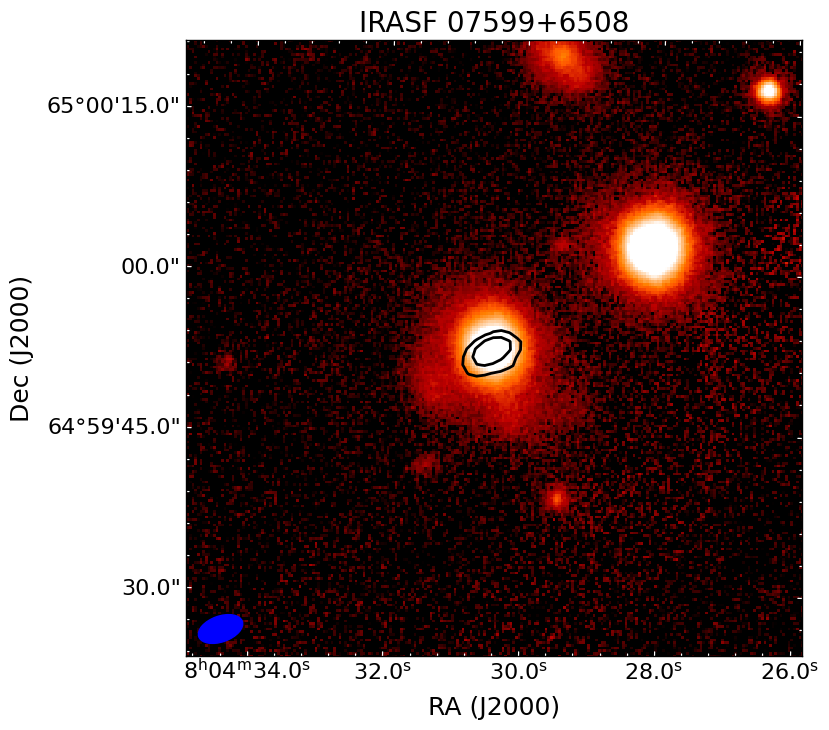} \hfill
    \includegraphics[width=2.25in]{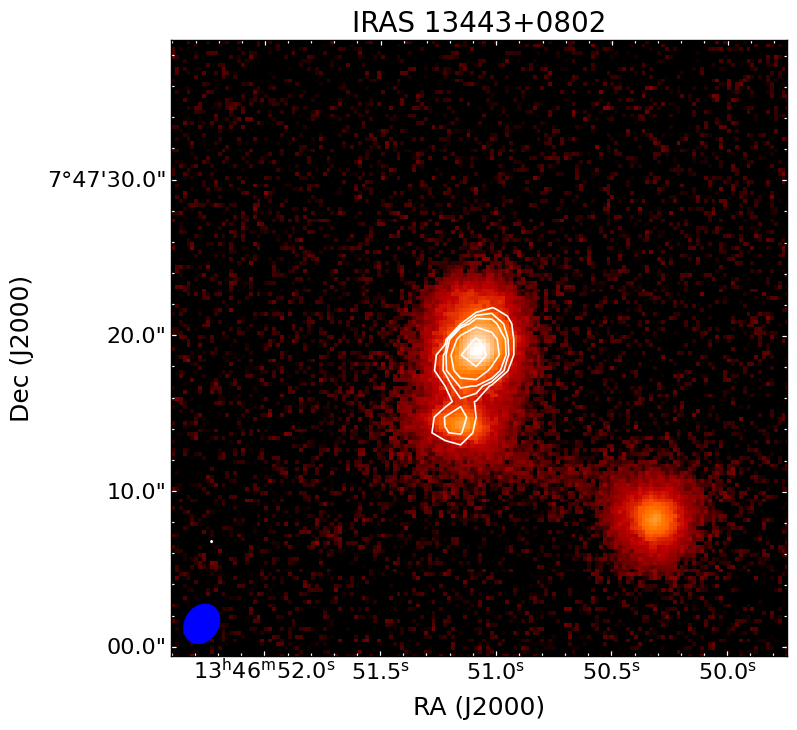} \hfill
    \includegraphics[width=2.27in]{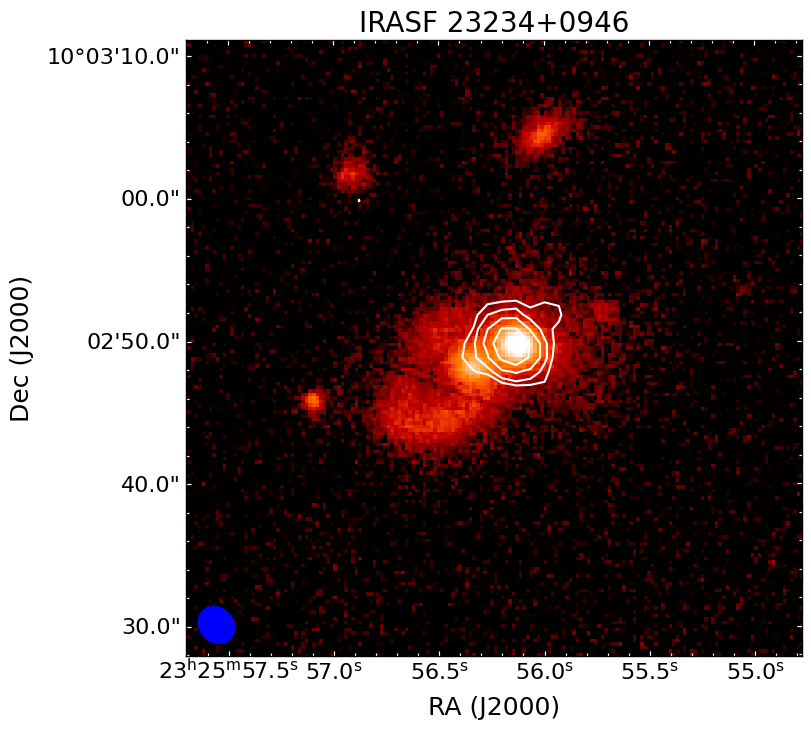}
\caption{Contours from the highest available radio–frequency maps overlaid on PANSTARRS optical images of the five ULIRGs. Radio contours start at $3\sigma$ and increase by factors of 2, where $\sigma$ represents the local rms noise level: $3.594\times10^{-4}$, $1.5\times10^{-4}$, $9.759\times10^{-4}$, $3.0\times10^{-4}$, and $3.639\times10^{-4}$ Jy beam$^{-1}$ for IRAS~00456$-$2904, IRAS~05189$-$2524, IRAS~F07599$+$6508, IRAS~13443$+$0802, and IRAS~F23234$+$0946, respectively.}

    \label{kntr}
\end{figure*}

\subsection*{IRAS 00456$-$2904}
According to \citet{Clements1996_1}, this galaxy's morphology suggests that it is an interacting system dominated by starburst activity, with a broad tidal tail associated with the northeast galaxy and a shorter one extending northwest from the main galaxy (see Figure \ref{panstar1}). The imaging analysis of \citet{veilleux2006deep} shows that the bright south-west nucleus is a late-type galaxy while the north-east source is of ambiguous type. These findings indicate that IRAS 00456$-$2904 is probably in an early-interaction phase and should be classified as a binary system \citep{dasyra2006dynamical} with a projected nuclear separation of 22.2 kpc. The WISE color \( W1 - W2 =0.677 \) also suggests it is a pre-merger system. The highest available 3 GHz VLASS image shows radio emission associated with the southern nucleus only (see Figure \ref{kntr}). It has a small amount of extended emission towards the north-west direction at this frequency. 
The integrated radio spectrum shows a low-frequency turnover at 276.5 MHz and a break frequency at 2236 MHz. The maximum spectral age limit of this source is estimated to be 27 Myrs. Radio excess fraction, spectrum, and spectral age confirm that the AGN contribution is significant for this source.

\subsection*{IRAS 05189$-$2524}
The Palomar optical telescope images of the system exhibit intersecting tails and shells. \citet{Sanders1988} indicates that it is an evolved merger of two spiral galaxies (see in Figure \ref{panstar1}). In contrast, Hubble images of the system reveal off-center nuclei, distorted outer sections, and a dim, arcing tail around the galaxies \citep{Charlene1994}. The system is classified as a Seyfert 2 galaxy \citep{Surace_1998}. 
Although \citet{veilleux2006deep} classified it as a merger/interaction class, the WISE color \( W1 - W2 =1.125 \) indicates it is in a post/evolved merger stage. It exhibits a low turnover frequency of 361.1 MHz and a break frequency at 4317 MHz, as determined through the spectral fitting. The system is resolved at a radio frequency of 8.49 GHz, as seen in Figure \ref{kntr}. The 10.9 kpc double radio structure is just at the center of the optical host galaxy.
The spectral age of the source is 1.9 Myrs. The estimated \(q_\nu\) value is 2.731, which shows high star formation as well. Our study indicates that IRAS 05189$-$2524 is hosting a buried AGN and CSS-like new young radio source. 

\subsection*{IRAS F07599+6508}

The galaxy has broad line absorption and is categorized as a radio-quiet quasi-stellar object (QSO). Hubble images reveal clumps of emissions surrounding the QSO, which could be associated with emissions from massive stars and a recent star-forming event \citep{Tacconi2002}. R and K band images show the presence of tidal tails \cite{Veilleux_2002}, and the host galaxy can be seen in Figure \ref{panstar1}. In Figure~\ref{kntr}, the radio contours are overlaid on the optical image. 
There is no WISE data available for this source. The system has a low-frequency turnover at 187.6 MHz and a break frequency of 2156 MHz. The spectral age of the ULIRG is 0.21 Myr. Both GMRT images do not show any extended radio emission beyond the optical host (see Figure \ref{band3} and Figure \ref{band4}). The source is not resolved in the highest available 15 GHz data \citep{nagar2003agn}. This is one of the examples which show the transformation of gas-rich mergers into radio quasars.

\subsection*{IRAS 11095$-$0238}
\begin{figure*}

    \centering
    \includegraphics[width=2.295in,height=1.95in]{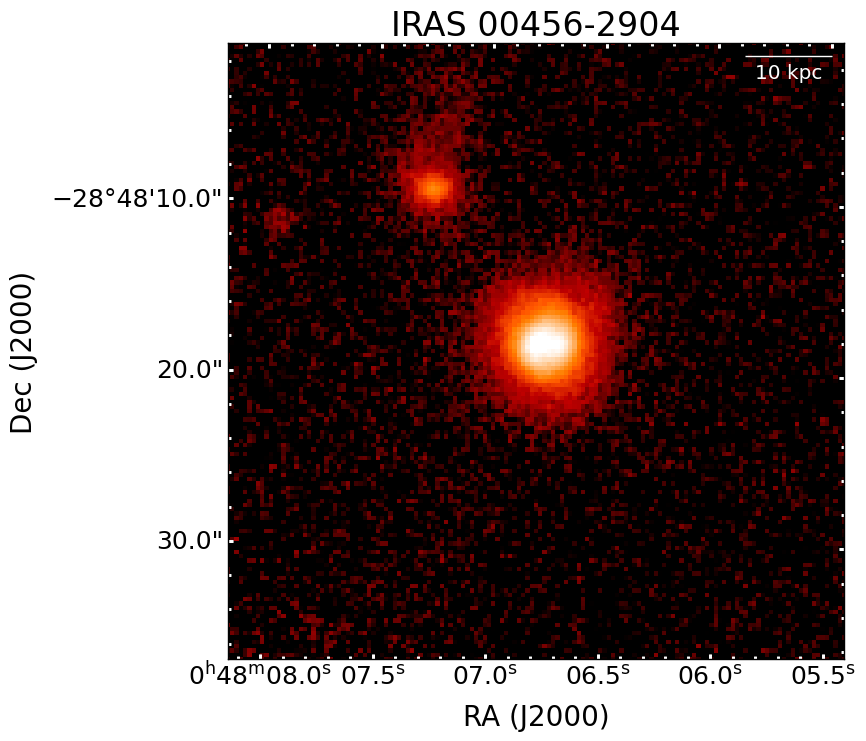} 
    \hfill
    \includegraphics[width=2.295in,height=1.95in]{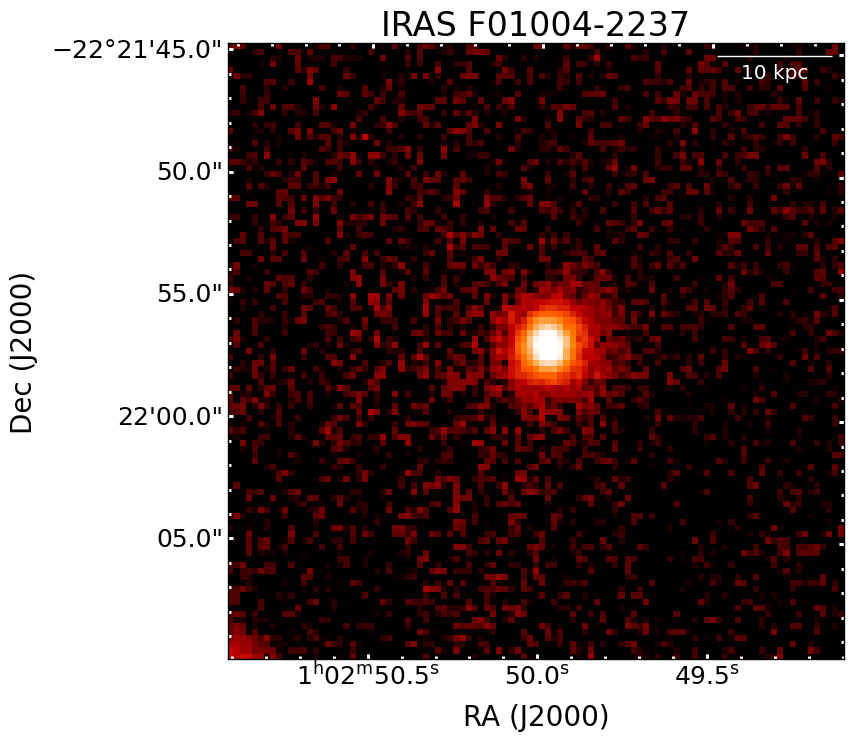} \hfill
    \includegraphics[width=2.2in,height=1.95in]{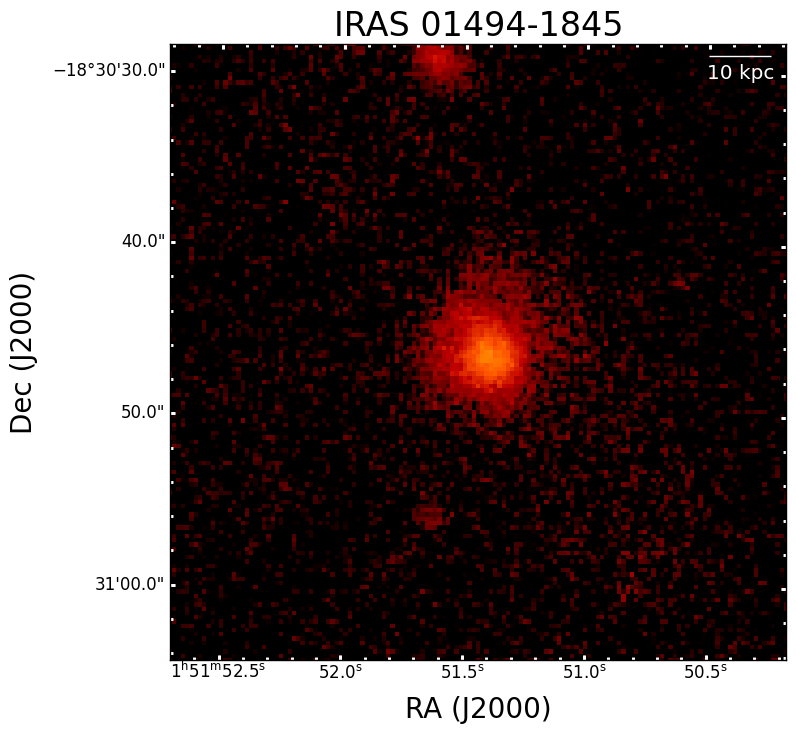} \\
    \vspace{0.1cm}
    \includegraphics[width=2.295in,height=1.95in]{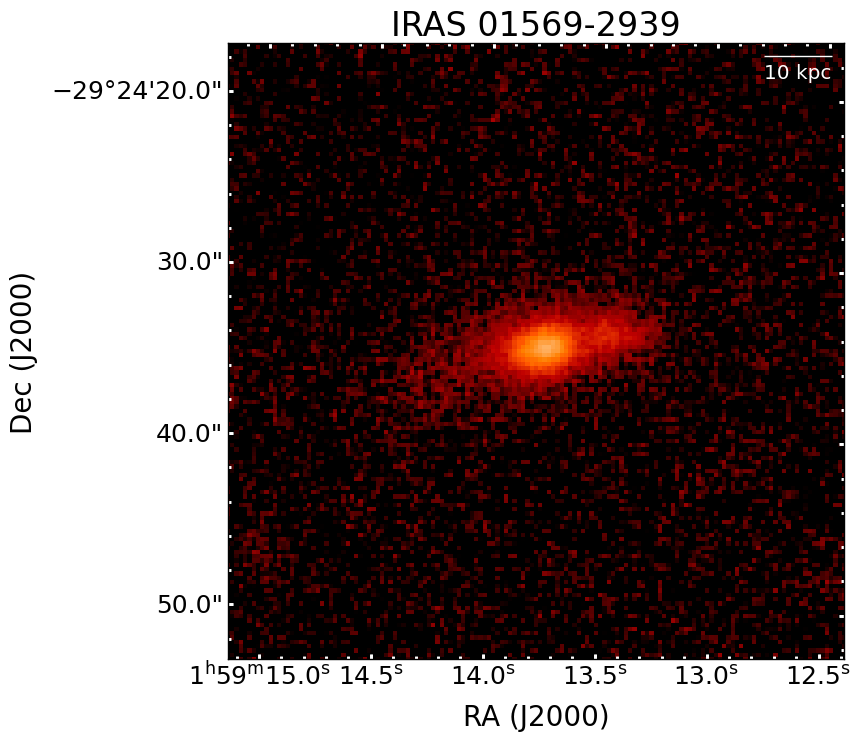} \hfill
    \includegraphics[width=2.295in,height=1.95in]{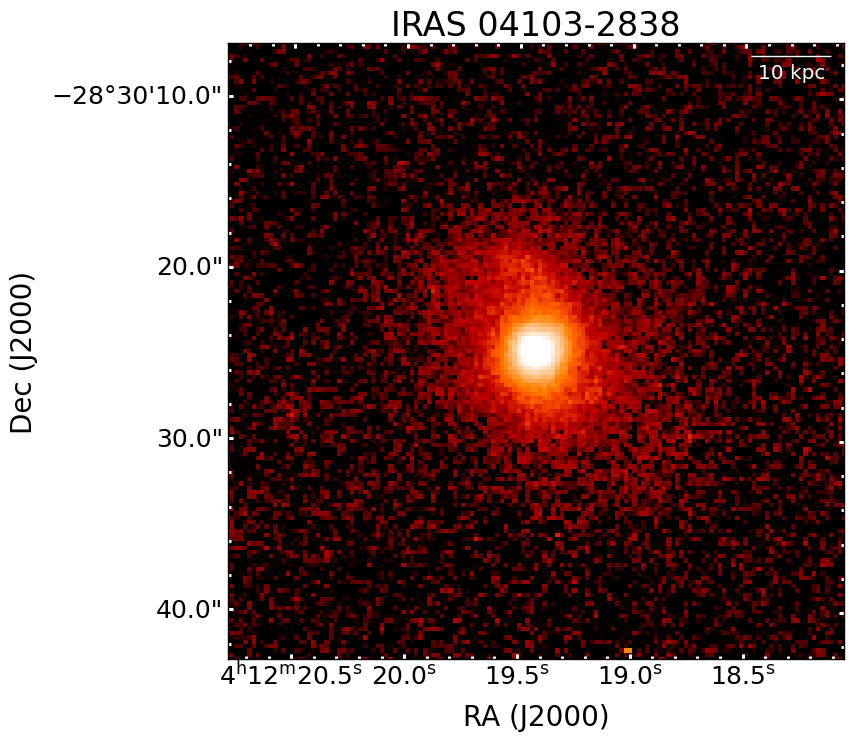} \hfill
    \includegraphics[width=2.295in,height=1.95in]{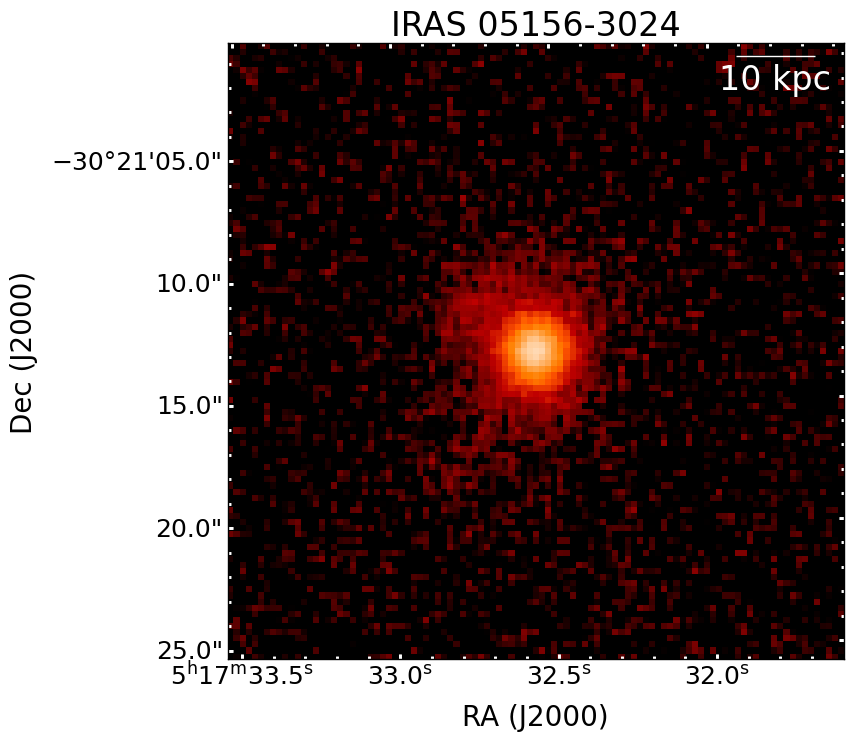}

    \vspace{0.1cm}
    \includegraphics[width=2.295in,height=1.95in]{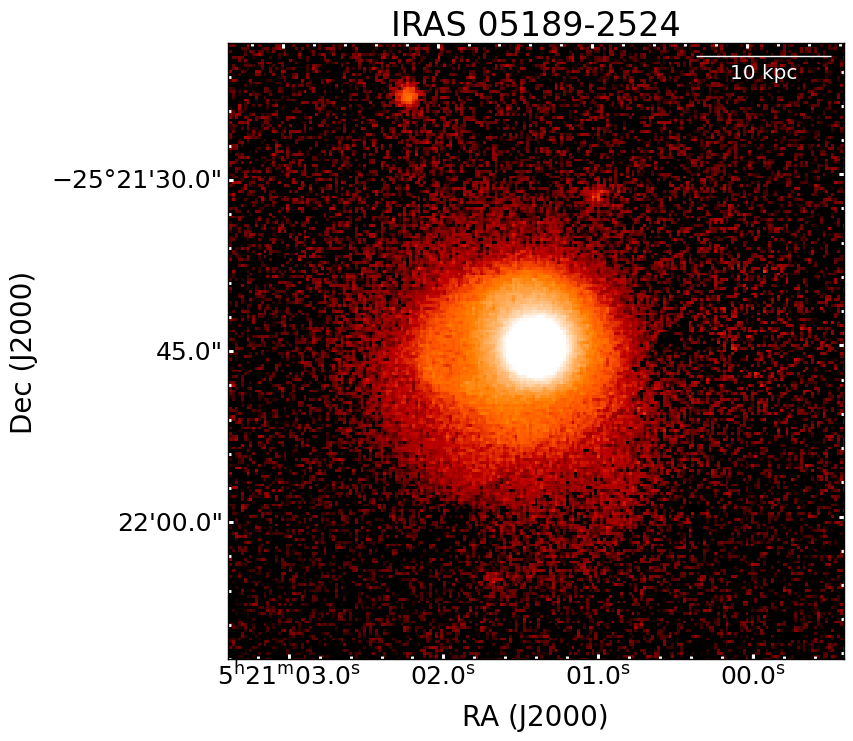} \hfill
    \includegraphics[width=2.295in,height=1.95in]{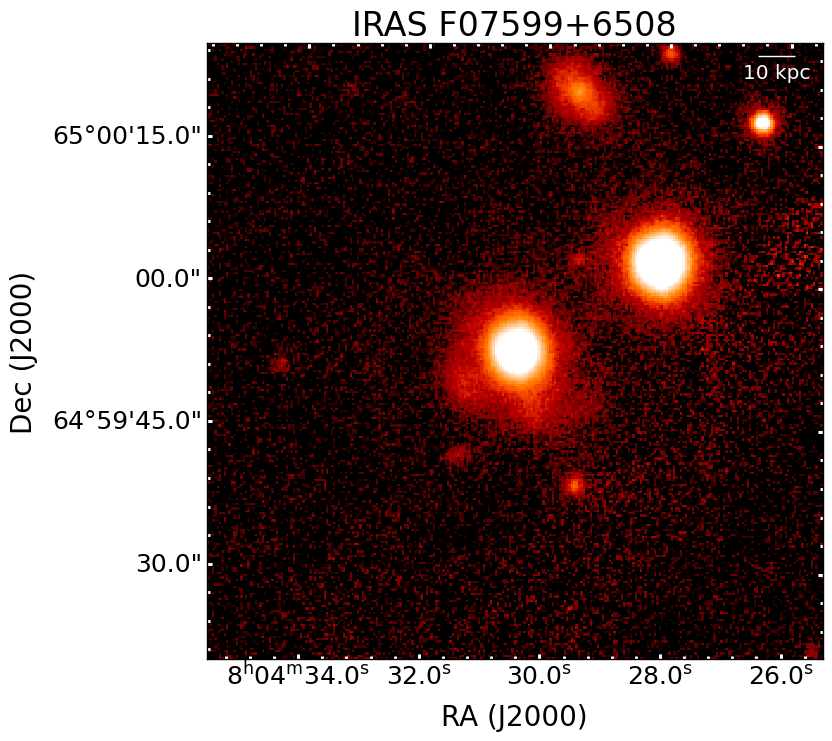} \hfill
    \includegraphics[width=2.295in,height=1.95in]{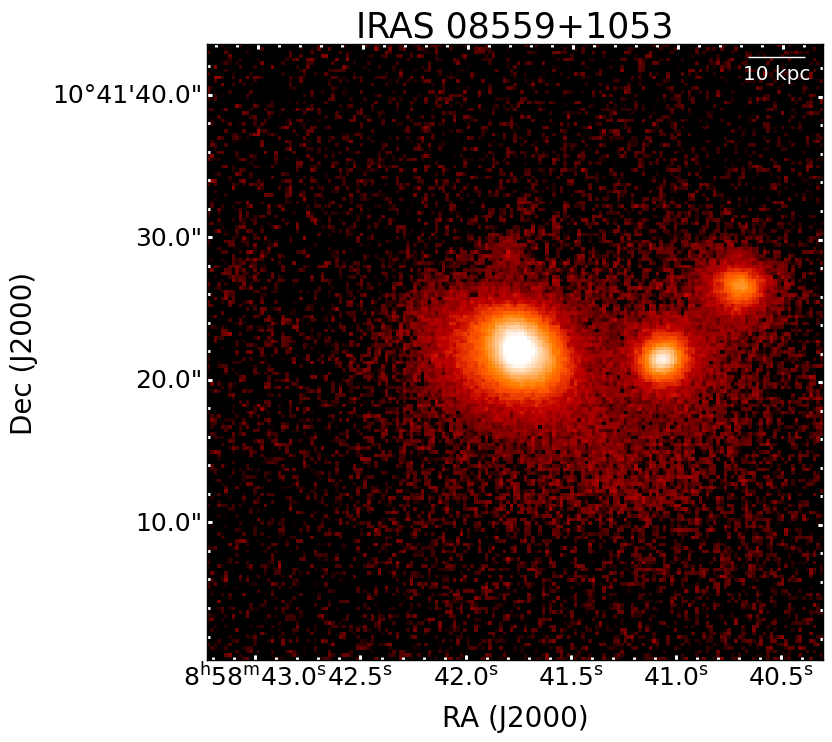}

    \vspace{0.1cm}
    \includegraphics[width=2.295in,height=1.95in]{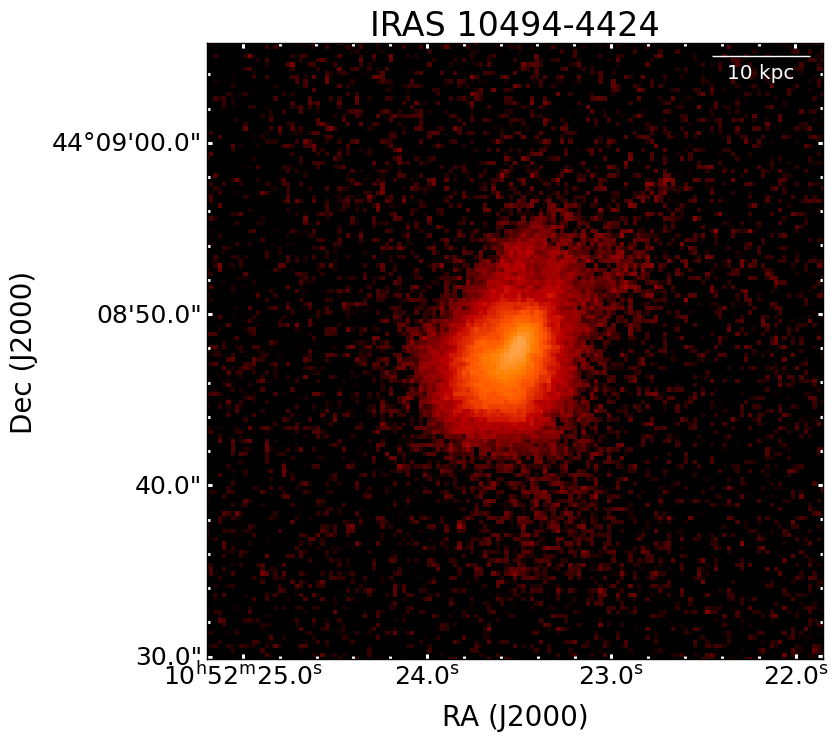} \hfill
    \includegraphics[width=2.295in,height=1.95in]{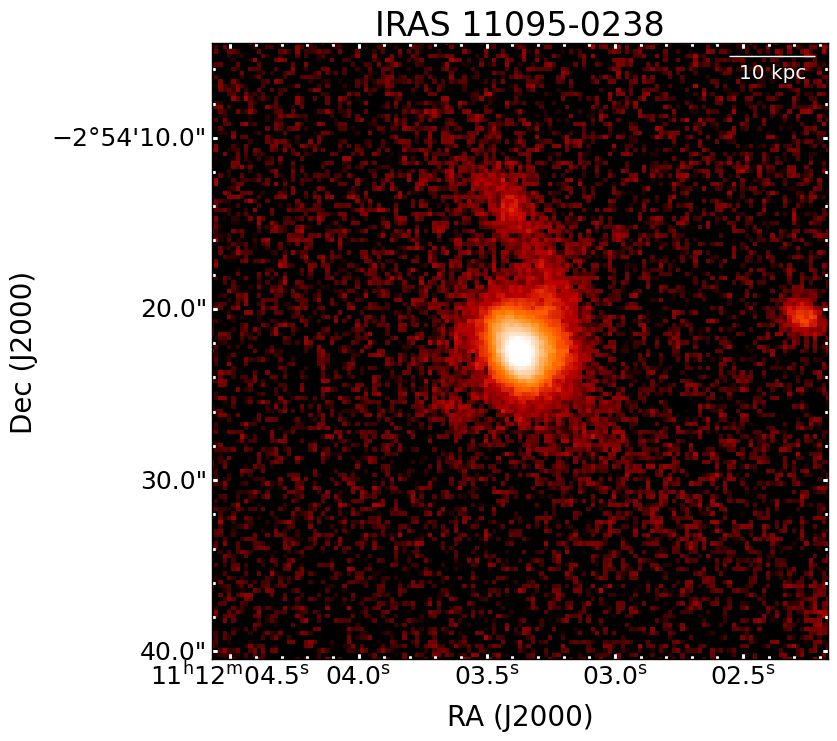} \hfill
    \includegraphics[width=2.295in,height=1.95in]{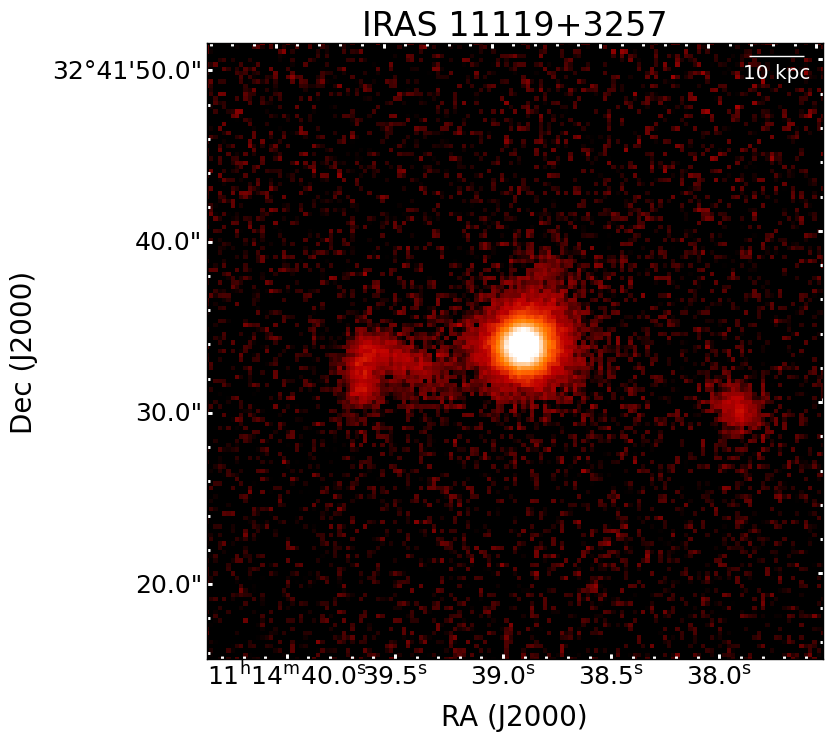}
    
    \caption{PANSTARRS images of host galaxy of source IRAS 00456$-$2904, IRAS F01004$-$2237, IRAS 01494$-$1845, IRAS 01569$-$2939, IRAS 04103$-$2838, IRAS 05156$-$3024, IRAS 05189$-$2524, IRAS F07599+6508, IRAS 08559+1053, IRAS 10494+4424, IRAS 11095$-$0238 and IRAS 11119+3257.}
\label{panstar1}

\end{figure*}

The source IRAS 11095$-$0238 is the late-stage merger system with two distinct nuclei at a separation of approximately 0.5 arcseconds. HST observations confirm that this is as a close binary system \citep{Bushouse_2002}. The WISE color \( W1 - W2 =1.720\) also indicates it is in a late merger stage. According to \citet{Veilleux_2002} its interaction class is merger and the nuclear structure is single core in optical images. The Pan-STARRS image of the host galaxy (Figure \ref{panstar1}) exhibits features consistent with this merger scenario. The IR and radio bands show a single compact component. This object exhibits two tidal tails. The northern tidal tail is more prominent than the southeast tail \citep{duc1997southern}. Optically the object is classified either as LINER \citep{Veron2010} or as an AGN/starburst composite system \citep{Yuan2010}. The spectral analysis shows a low-frequency turnover at 224.0 MHz and a high break frequency of  $3.0\times 10^{13}$MHz. To obtain the maximum spectral age, the highest frequency available for flux measurements is taken as the break frequency. The maximum spectral age of this system is <19 Myrs. The highest frequency radio data available for this galaxy is the 3 GHz VLASS image and the source is unresolved at this frequency. The low-frequency turn over can be due to two merging radio active nuclei, but it needs high resolution radio observations to detect the components.

\subsection*{IRAS 12265+0219}

This source is very bright and is a famous radio-loud quasar commonly known as 3C273 at a redshift of z = 0.158 \citep{schmidt19633}. The estimated \(q_\nu\) value of this source is -1.300 and it is also the most luminous radio source in our sample (log $L_{1.4\mathrm{GHz}}$ = 27.699 WHz$^{-1}$). This object is highly nucleated but shows a thick tidal feature toward the northeast and a very faint feature to the northwest \citep{Veilleux_2002}. Its host galaxy is an elliptical one (see in Figure \ref{panstar2}), and it shows all the characteristics that are typical of high luminosity quasars including a flat radio spectrum of the core, rapid and strong variability in all observed energy ranges, variable polarization, and a radio jet with superluminal motion \citep{bach2007multi}. 
According to WISE color and  \citet{Veilleux_2002} it is a merger system. The spectral analysis shows a high break frequency of $5.9773\times10{^6}$ MHz but no turnover frequency. To obtain the maximum spectral age, the highest frequency available for flux measurements is taken as the break frequency. The spectral age of this system is <0.0093 Myrs. The highest frequency data which we have used for this galaxy is the 22 GHz VLA image. This image does not show any resolved morphology. This source show a transformation the gas-rich merger remnant into a radio-loud quasar.

\subsection*{IRAS 13428+5608}

This source is known as Mrk 273, and exhibits a striking tidal tail towards the south (see in Figure \ref{panstar2}). This is a rare example of very extended radio emission. LOFAR detected 300 Kpc radio emission for this object \citep{Kukreti2022}.This is a late-stage merger with three nuclei at the center. Though \citet{Veilleux_2002} noted merger interaction class but WISE color represents its a evolved merger phase. Initially the north and south west nuclear components were detected through IR observations, while the south east component is detected through radio observations. Later high resolution observations indicate detection of north nucleus in radio also \citep{carilli2000}. The positions of these three AGNs are shown in \citet{Rodrigues2014}. The radio AGNs are not on the brightest region of the optical host galaxy. For radio spectrum fitting we used central core flux density but not the northern or southern arc. The GMRT detected extended emission in both bands, as seen in Figures~\ref{band3} and~\ref{band4}. The spectral fitting gives a low-frequency turnover at 115.7 MHz, a break frequency at 4891 MHz. The estimated spectral age is 3.5 Myrs. Although our study did not resolve the central part of the source, optical images indicate that the source is young with multiple nuclei.

\subsection*{IRAS 13443+0802}
IRAS 13443+0802 is clearly involved in interactions with two other objects (see in Figure \ref{panstar2}). All three sources within this field are classified as emission-line galaxies with an H II region-type spectrum, sharing similar redshifts (z = 0.135). This triple system comprising a close pair with a nuclear separation of $\sim$5 arcsec ($\sim$12 kpc) in the North South direction, along with a third galaxy located $\sim$15 arcsec ($\sim$35 kpc) to the South West of that pair \citep{Tadhunter}. According to the paper by \citet{Veilleux_2002}, there is a possibility that a bridge of material may be present between the eastern binary and the southwest component. Two detected WISE components show one AGN and one star forming region is interacting in this system. This system has a break frequency at 1997 MHz but no turnover frequency. The spectral age is estimated to be 0.43 Myrs, indicating this source is a relatively young radio source. Figure~\ref{kntr} shows the 3 GHz radio contours overlaid on the optical image. 
It clearly resolved two nuclei and indicates that the north is brighter than the south nucleus. Radio peak positions are separated by 7.552 arcseconds. The presence of binary AGNs is possible for this system but it needs to be confirmed by performing high resolution multi-frequency observations.

\subsection*{IRAS 13451+1232}
Another name of this target is PKS 1345+12. It is identified as a quasar and stands out as one of the nearest bright GPS/CSS radio sources. This quasar has been considered a prime candidate for acting as a link between ultraluminous infrared galaxies and young radio galaxies \citep{2004A&A...424..119M}. It is the second most luminous radio source in our sample (log $L_{1.4\mathrm{GHz}}$ = 26.305 WHz$^{-1}$) and the estimated \(q_\nu\) value is -0.361. The host galaxy of this source is undergoing a pre-merger process. This is indicated by the presence of a double nucleus, irregular isophotes, a tail, and the existence of several nearby smaller companions these features strongly suggest that the merging process has triggered the observed radio activity \citep{Heckman1986}. The optical image reveals two distinct nuclei, separated by approximately 2 arcseconds (4 kpc), as shown in Figure~\ref{panstar2} \citep{xiang2002evn}. The primary concentration of activity occurs within the western nucleus, identified as a Sy2 in the optical spectrum \citep{Veilleux1999} and as a compact source in radio wavelengths \citet{1999ApJ...521L.107E}. The VLBI observations revealed the inner structure into twin lobes and a core \citep{2001A&A...377..377S} with a projected linear size $<$ 150 pc. In our analysis the integrated radio spectrum has a low frequency turnover at 359.1 MHz. The break frequency of this target is at 7666 MHz which gives a spectral age estimate of 0.21 Myrs. Double nuclear components have been detected in optical, IR and radio.

\subsection*{IRAS 15327+2340}
This is a well known interacting system, Arp 220. The morphology of Arp 220, showing a double  nucleus structure in optical, radio and IR. The optical host galaxy is shown in Figure~\ref{panstar2}. Its tidal tails and dust lanes, indicate that it is likely the outcome of a recent merging phenomenon. \citet{2016A&A...593A..86V} reported steep spectrum outflows and extended emission at 150 MHz. The estimated \(q_\nu\) value is 2.593 which is above the mean value of 2.34 (\citet{2001ApJ...554..803Y}), indicates that most of its radio emission comes from strong starburst activity. According to \citet{rovilos2003continuum} this galaxy exhibits very powerful maser emission and does not belong to the category of Compact Symmetric Objects (CSOs) or Medium Symmetric Objects (MSOs). This source exhibits low  frequency turnover at 291.2 MHz and the break frequency is 7.0789 × 10${^5}$ MHz. To obtain the spectral age for this source we take the highest frequency available for flux measurements as a lower limit of the break frequency. Therefore, the lower limit of spectral age is estimated to be $<0.76$ Myr.

\subsection*{IRAS F23234+0946}

\begin{figure*}
    \centering
    \includegraphics[width=2.295in,height=1.95in]{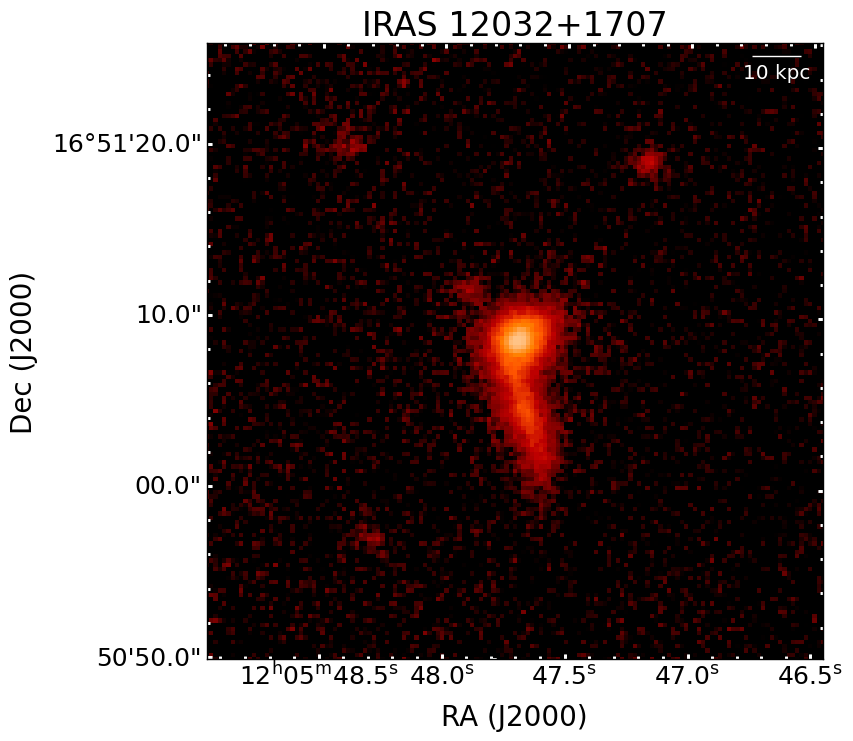} 
    \hfill
    \includegraphics[width=2.295in,height=1.95in]{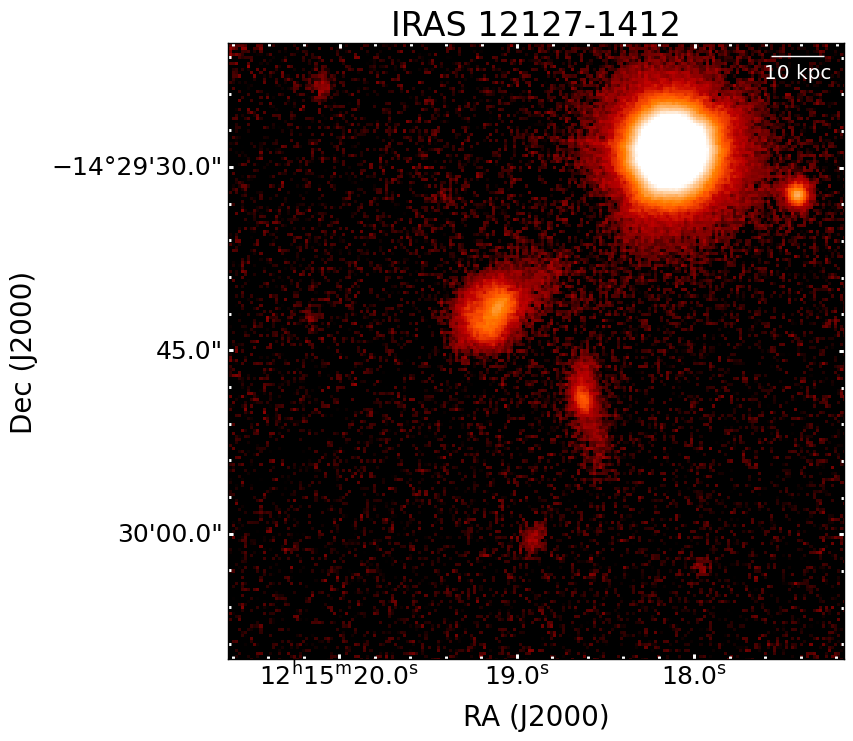} \hfill
    \includegraphics[width=2.2in,height=1.95in]{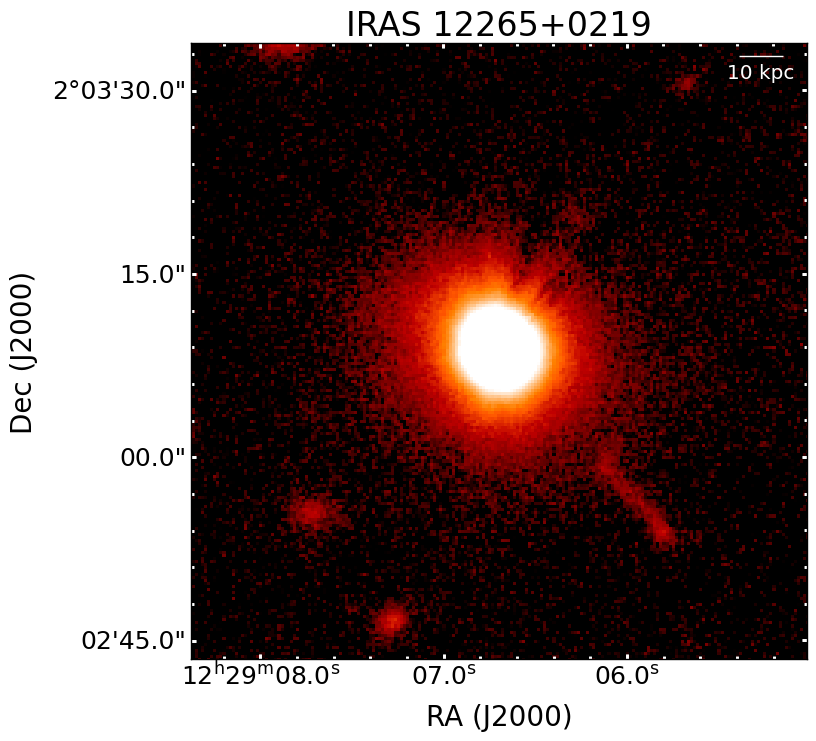} \\
    \vspace{0.1cm}
    \includegraphics[width=2.295in,height=1.95in]{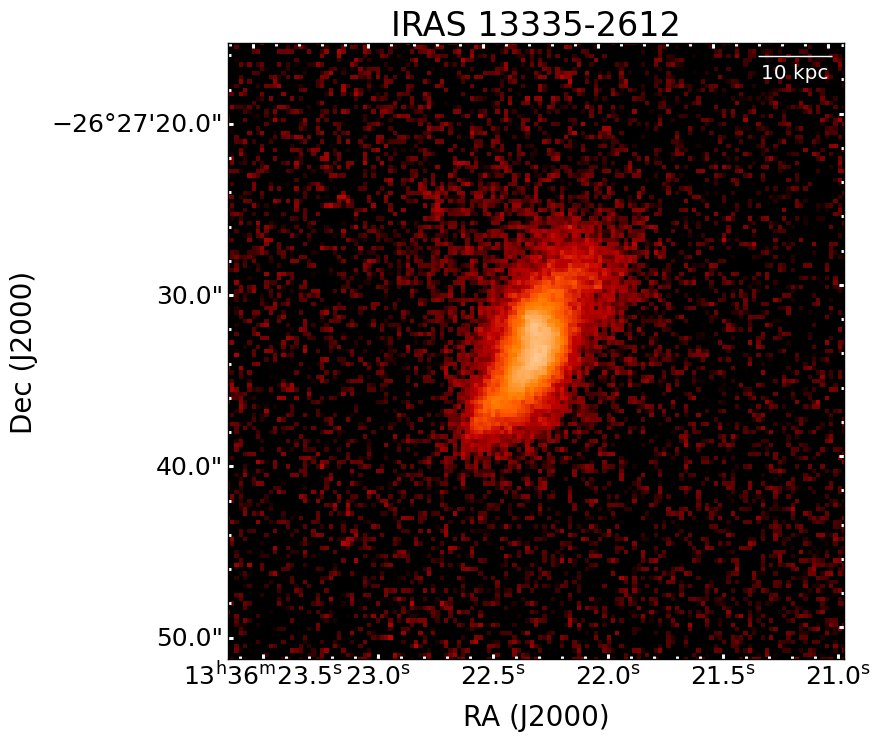} \hfill
    \includegraphics[width=2.295in,height=1.95in]{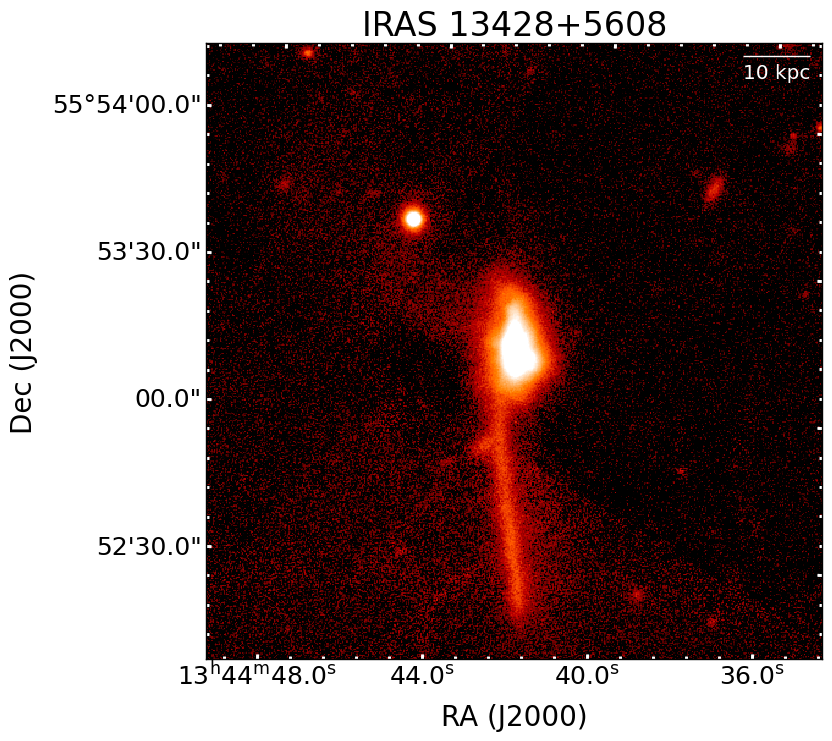} \hfill
    \includegraphics[width=2.295in,height=1.95in]{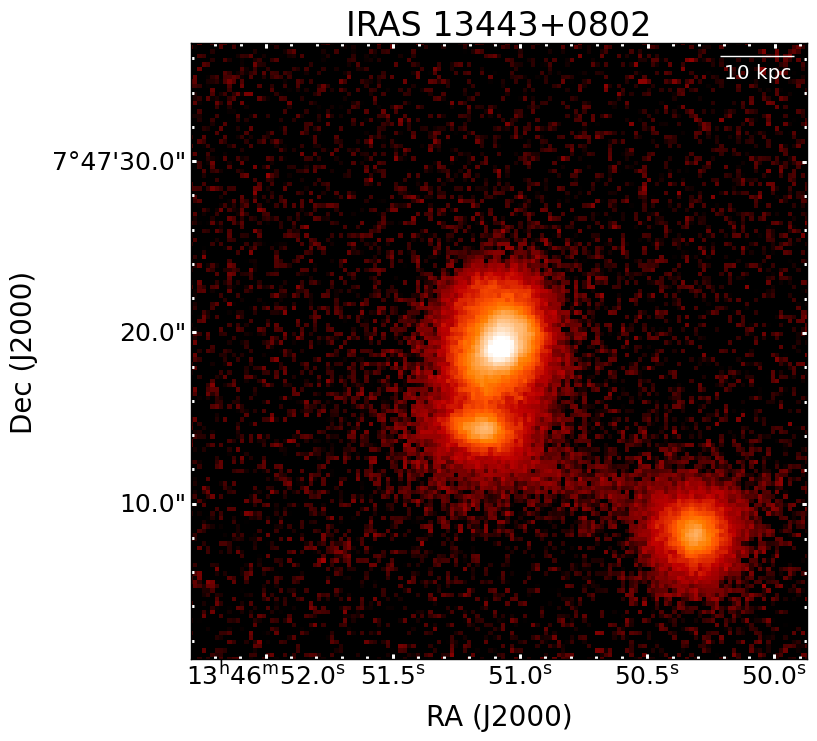}
    \vspace{0.1cm}
    \includegraphics[width=2.295in,height=1.95in]{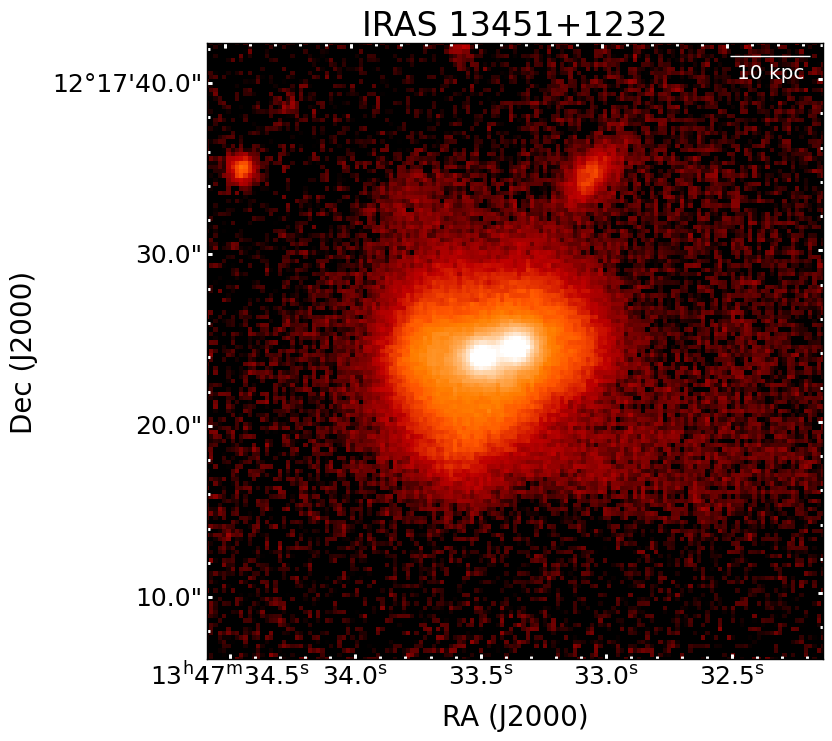} \hfill
    \includegraphics[width=2.295in,height=1.95in]{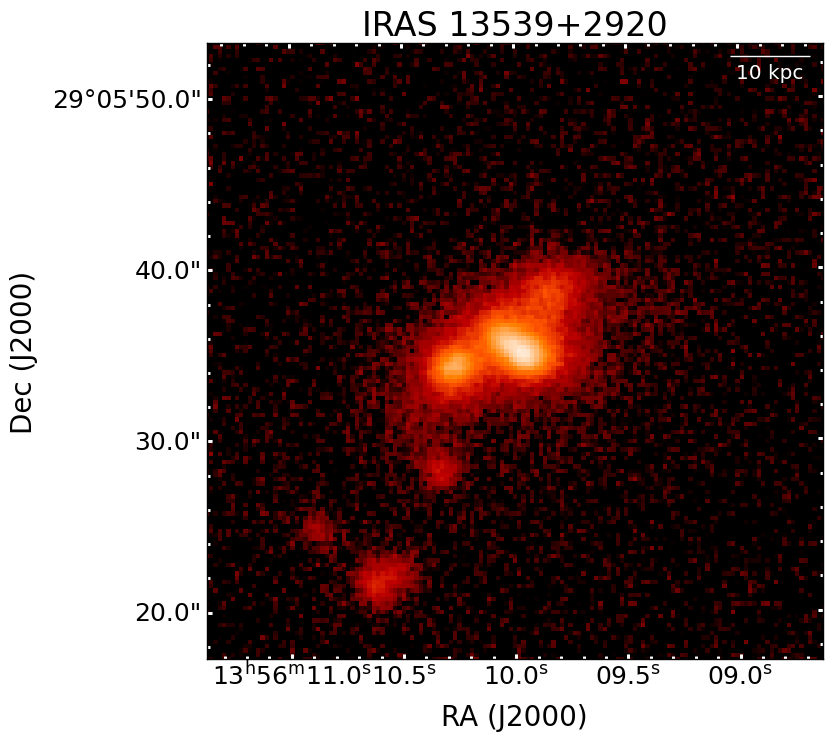} \hfill
    \includegraphics[width=2.295in,height=1.95in]{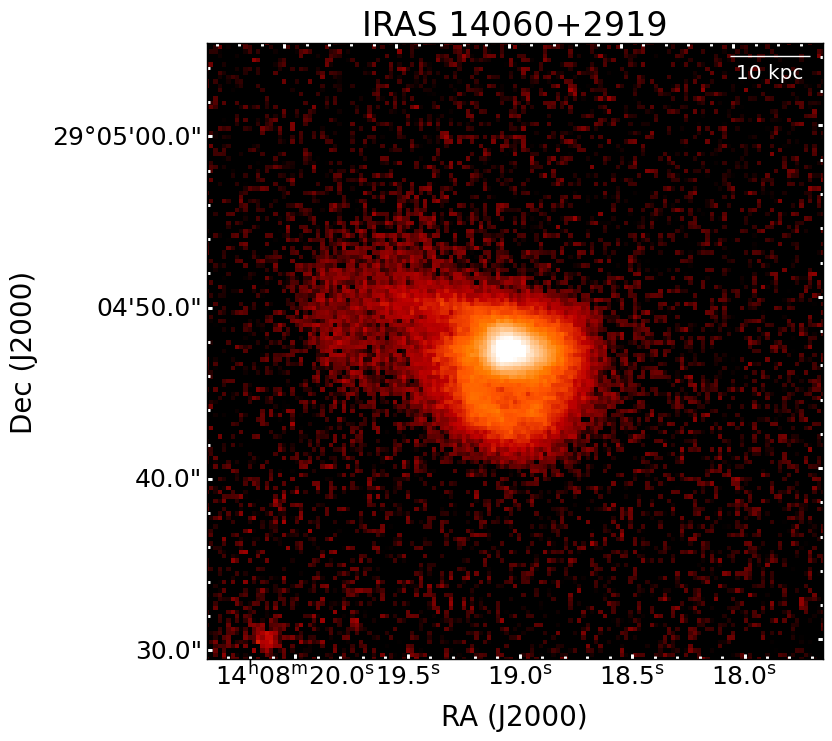}
\vspace{0.1cm}
    \includegraphics[width=2.295in,height=1.95in]{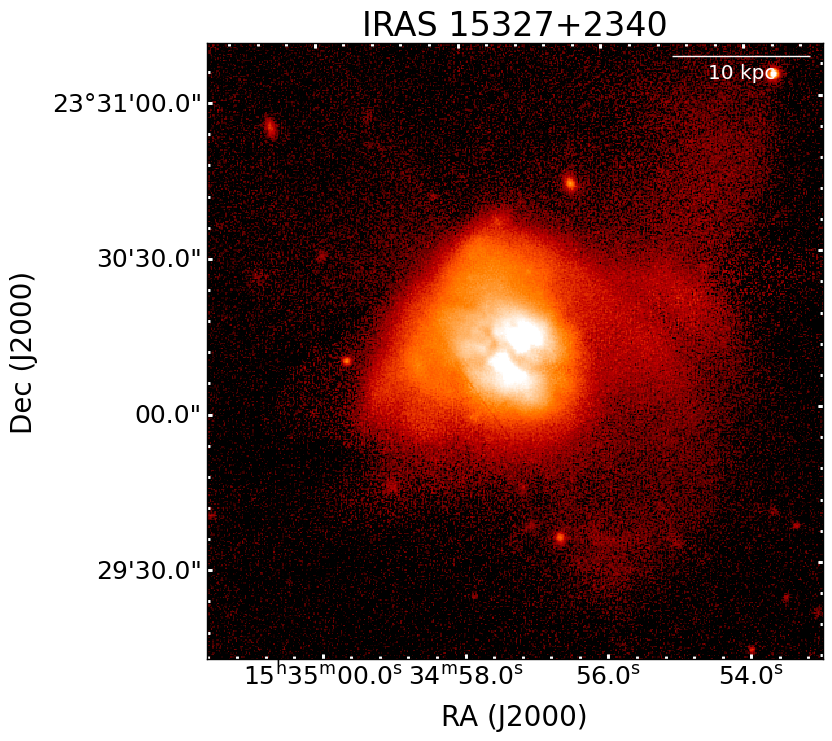} \hfill
    \includegraphics[width=2.295in,height=1.95in]{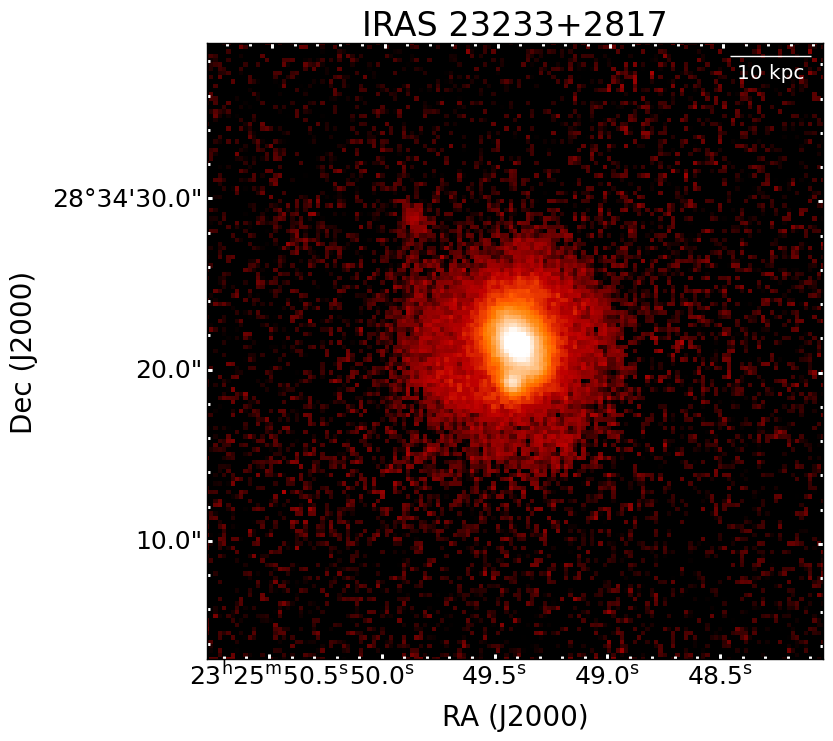} \hfill
    \includegraphics[width=2.295in,height=1.95in]{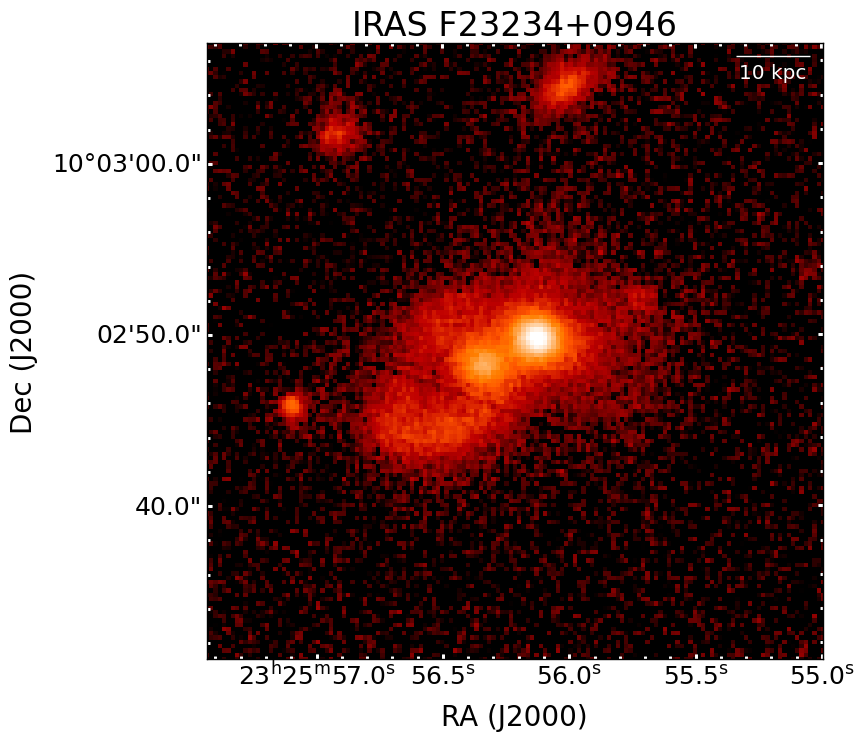}

\caption{PANSTARRS images of host galaxy of source IRAS 12032+1707, IRAS 12127$-$1412, IRAS 12265+0219, IRAS 13335$-$2612, IRAS 13428+5608, IRAS 13443+0802, IRAS 13451+1232, IRAS 13539+2920, IRAS 14060+2919, IRAS 15327+2340, IRAS 23233+2817 and IRAS F23234+0946. }
\label{panstar2}

\end{figure*}

This is a double nucleus system in both optical and IR. A hooklike tidal tail is present and it extends to the southeast as seen in Figure \ref{panstar2}. The two nuclei are situated in the east and west, with the western nucleus showing no sign of AGN activity in the mid-infrared spectrum and displaying the highest brightness at optical wavelengths \citet{imanishi2007spitzer}. According to \citet{Rodr2009}, the western nucleus has a much higher percentage contribution of old stellar population (60–82 percent) compared to the eastern nucleus when these two are compared. The estimated \(q_\nu\) value for this source is 2.250. To obtain the spectral age for this source we use the highest frequency available for flux measurements is taken as the break frequency. This source exhibits no turnover frequency. The break frequency is 1.6343 × 10${^4}$ MHz. The 3 GHz VLASS archival image provides the highest-frequency radio data for this source. The radio synchrotron age is estimated to be $<0.14$ Myrs. The western nucleus shows distinct radio emission (Figure \ref{kntr}). 
Hence we expect AGN activity in the western component also.

\section{Pan-STARRS Optical Images of ULIRGs in Our Sample}
\label{appendix:PANfigures}

This appendix shows the Pan-STARRS optical images of the ULIRGs studied in this work. 
These images are included to complement the discussion in Section~4.4, where we 
describe the optical appearance of the host galaxies. 
The Pan-STARRS images provide a view of the host galaxy morphologies and help 
illustrate features such as disturbed structures, tidal signatures, and possible 
multiple components, which are relevant for understanding the interaction and merger 
status of these systems.

\end{document}